\documentclass[oneside,10pt]{article}

\usepackage{times,url,mathrsfs}
\usepackage{amsmath}
\usepackage{amsfonts}
\usepackage{graphicx}
\usepackage{subfigure}
\newtheorem{theorem}{Theorem}
\newtheorem{lemma}{Lemma}

\begin{document}

\title{On Practical Algorithms for Entropy Estimation and the Improved Sample Complexity of Compressed Counting}

\author{ Ping Li \\
         Department of Statistical Science\\
         Faculty of Computing and Information Science\\
       Cornell University\\
         Ithaca, NY 14853\\
       pingli@cornell.edu
  }
%\date{}
\maketitle
%\vspace{-0.1in}
\begin{abstract}

The long-standing problem of Shannon entropy estimation in data streams (assuming the strict Turnstile model) is now an easy task by using the technique proposed in this paper. Essentially speaking, in order to estimate the Shannon entropy with a guaranteed $\nu$-additive accuracy, it suffices to estimate the $\alpha$th frequency moment, where $\alpha = 1-\Delta$, with a guaranteed  $\epsilon$-multiplicative accuracy, where $\epsilon = \nu\Delta$. Previous studies have shown that $\Delta$ has to be extremely small (e.g., $\Delta<10^{-4}$ or even much smaller). In other words, the sample complexity for entropy estimation is $O\left(\frac{V}{\epsilon^2}\right) = O\left(\frac{V}{\nu^2\Delta^2}\right)$, where $V$ is the coefficient essentially determined by the variance of the estimator of frequency moments. In this paper, the proposed algorithm achieves $V = O\left(\Delta^2\right)$ and hence is a practical technique with complexity $O\left(\frac{1}{\nu^2}\right)$ (which is essentially $O(1)$ if we consider $\nu=O(1)$). We provide the (small) complexity bound constants  numerically (for $0<\nu<1$) and analytically (for small $\nu$). \\

Prior well-known algorithms based on {\em symmetric stable random projections} could only achieve $V = O\left(1\right)$, meaning that the sample complexity would be $O\left(\frac{1}{\nu^2\Delta^2}\right)$, which will be extremely large. For example, if $\nu=O(1)$ and $\Delta=10^{-5}$, then $O\left(\frac{1}{\nu^2\Delta^2} \right)= O\left({10^{10}}\right)$. \\

\textbf{Compressed Counting (CC)}\cite{Proc:Li_SODA09}, based on {\em maximally skewed stable random projections}, was recently proposed for  estimating the $\alpha$th frequency moment of data streams. \cite{Proc:Li_SODA09} proposed algorithms for CC based on the {\em geometric mean} and {\em harmonic mean} estimators. It was proved that the {\em geometric mean} estimator could achieve $V = O\left(\Delta\right)$, leading to an $O\left(\frac{1}{\nu^2\Delta}\right)$ algorithm, which unfortunately could still be impractical.  In this paper, we prove that the {\em harmonic mean} estimator for CC also could only achieve $V=O\left(\Delta\right)$.\\

The proposed new estimator for CC  has a simple clean form:  $\frac{1}{\Delta^\Delta}\left[\frac{k}{\sum_{j=1}^k x_j^{-\alpha/\Delta}}\right]^\Delta$, where $x_j$'s are the projected data and $k$ is the sample size.  We prove that its variance achieves $V = O\left(\Delta^2\right)$, leading to a practical algorithm with complexity $O\left(\frac{1}{\nu^2}\right)$. In other words, if $\Delta=10^{-5}$, the new algorithm improves the prior algorithms based the {\em symmetric stable random projections} roughly by a factor of $10^{10}$; and it improves the {\em geometric/harmonic mean} algorithms for CC roughly by a factor of $10^{5}$. \\

Our extensive experiments (in the Appendix) verify that, using the proposed algorithm, $k\approx10$ samples could provide accurate estimates of the Shannon entropy. The proposed algorithm is also numerically very stable, even for $\Delta$ as small as $10^{-10}$.

\end{abstract}

\vspace{-0.1in}

\section{Introduction}

The problem of ``scaling up for high dimensional data and high speed data streams'' is among the  ``ten challenging problems in data mining research''\cite{Article:ICDM10}. This paper is devoted to estimating entropy of data streams. Mining data streams\cite{Book:Henzinger_99,Proc:Babcock_PODS02,Proc:Aggarwal_KDD04,Article:Muthukrishnan_05} in (e.g.,) 100 TB scale  databases has become an important area of research, e.g., \cite{Proc:Domeniconi_ICDM01,Proc:Aggarwal_KDD04}, as network data can easily reach that scale\cite{Article:ICDM10}. Search engines are a typical source of data streams\cite{Proc:Babcock_PODS02}.

Consider the  {\em Turnstile}  stream model\cite{Article:Muthukrishnan_05}. The input  stream $a_t = (i_t, I_t)$, $i_t\in [1,\ D]$ arriving sequentially describes the underlying signal $A$, meaning
\begin{align}\label{eqn_Turnstile}
A_t[i_t] = A_{t-1}[i_t] + I_t,
\end{align} where the increment $I_t$ can be either positive (insertion) or negative (deletion).  Restricting $A_t[i]\geq 0$ results in the {\em strict-Turnstile} model, which suffices for describing almost all natural phenomena\cite{Article:Muthukrishnan_05}.

This study focuses on the {\em relaxed strict-Turnstile} model and  studies efficient algorithms for estimating the {\em $\alpha$th frequency moment} of data streams
\begin{align}\label{eqn_moment}
F_{(\alpha)} = \sum_{i=1}^D A_t[i]^\alpha.
\end{align}
We are particularly interested in the case of $\alpha\rightarrow 1$, which is very important for estimating {\em Shannon entropy}.

 The {\em relaxed strict-Turnstile model} only requires $A_t[i]\geq 0$ at the time $t$ one cares about (e.g., the end of streams); and hence it is considerably more flexible than the {\em strict-Turnstile} model.

\subsection{Entropy, Moments, and Estimation Complexity}

A very useful (e.g.,  in Web and networks\cite{Proc:Feinstein_DARPA03,Proc:Lall_SIGMETRICS06,Proc:Zhao_IMC07,Proc:Mei_WSDM08} and neural comptutations\cite{Article:Paninski_NC03}) summary statistic is  the {\em Shannon entropy}
\begin{align}\label{eqn_Shannon}
H = -\sum_{i=1}^D\frac{A_t[i]}{F_{(1)}}\log \frac{A_t[i]}{F_{(1)}}.
\end{align}
Various generalizations of the Shannon entropy have been proposed. The R\'enyi entropy\cite{Proc:Renyi_61}, denoted by $H_\alpha$, and the Tsallis entropy\cite{Article:Havrda_67,Article:Tsallis_88}, denoted by $T_\alpha$,   are respectively defined as
\begin{align}\label{eqn_Renyi}
&H_\alpha =\frac{1}{1-\alpha} \log \frac{\sum_{i=1}^D A_t[i]^\alpha}{\left(\sum_{i=1}^D A_t[i]\right)^\alpha} = \frac{1}{1-\alpha} \log \frac{F_{(\alpha)}}{F_{(1)}^\alpha}, \\ &T_\alpha = \frac{1}{1-\alpha} \left( \frac{F_{(\alpha)}}{F_{(1)}^\alpha}-1\right).
\end{align}

As $\alpha\rightarrow 1$, both R\'enyi entropy and Tsallis entropy converge to Shannon entropy:
$\lim_{\alpha\rightarrow 1} H_\alpha  = \lim_{\alpha\rightarrow 1}T_\alpha = H$.  Thus, both R\'enyi entropy and Tsallis entropy can be computed from the $\alpha$th frequency moment; and one can approximate Shannon entropy from either $H_\alpha$ or $T_\alpha$ by letting $\alpha\approx 1$. Several studies\cite{Proc:Zhao_IMC07,Proc:Harvey_ITW08,Proc:Harvey_FOCS08})   used this idea to approximate  Shannon entropy, all of which relied critically on efficient algorithms for estimating the $\alpha$th  frequency moments (\ref{eqn_moment}) near $\alpha =1$. In fact, one can numerically verify that the $\alpha$ values proposed in \cite{Proc:Harvey_ITW08,Proc:Harvey_FOCS08} are extremely close to 1,  for example, $\Delta = |1-\alpha|<10^{-7}$ \cite[Alg. 1]{Proc:Harvey_ITW08} or $\Delta<10^{-4}$\cite{Proc:Harvey_FOCS08} are quite likely.\footnote{
In \cite[Alg. 1]{Proc:Harvey_ITW08}, $\Delta = \frac{c}{16\log(1/c)}$, $c = \frac{\nu}{4\log(D)\log(m)}$, where $m$ is the number of streaming updates. If we let $D=2^{64}$, $m=2^{64}$, $\nu=0.01$, then $\Delta \approx 6\times 10^{-9}$. If we let $m=10^6$, $\nu=0.1$, then $\Delta \approx 2\times 10^{-7}$. \\
\cite[Sec. 4.2.2 and 5.2]{Proc:Harvey_FOCS08} provides some improvements, to allow larger $\Delta$. If $m=2^{64}$ and $\nu=0.01$, then $\Delta \approx 7\times 10^{-6}$. If $m=10^{6}$ and $\nu=0.1$, then $\Delta\approx 10^{-4}$.}\\

From the definition of the R\'enyi and Tsallis entropies, it is clear that, in order to achieve a $\nu$-additive guarantee for the Shannon entropy, it suffices to estimate the $\alpha$th frequency moment with an $\epsilon=\nu\Delta$ guarantee (for sufficiently small $\Delta$). For example, suppose an estimator $\hat{F}_{(\alpha)}$ guarantees (with high probability) that $(1-\epsilon)F_{(\alpha)} \leq \hat{F}_{(\alpha)} \leq (1+\epsilon)F_{(\alpha)}$, then the estimated R\'enyi entropy, denoted by $\hat{H}_{\alpha}$ would satisfy  $H_{\alpha} - \nu \leq \hat{H}_{\alpha} \leq H_{\alpha} + \nu$, assuming $\Delta$ is sufficiently small.

Another perspective is from the estimation variances. From the definitions of the R\'enyi and Tsallis entropies, it is clear that we need estimators of the frequency moments with variances proportional to $O\left(\Delta^2\right)$ in order to cancel the term $\frac{1}{(1-\alpha)^2}$. The estimation variance, of course, is also closely related to the sample complexity. \\

Suppose we have an unbiased estimator of $F_{(\alpha)}$ whose variance is $\frac{V}{k}F^2_{(\alpha)}$, where $k$ is the sample size. Then the sample complexity is essentially $O\left( (VF^2_{(\alpha)})/(\epsilon^2 F^2_{(\alpha)})\right) = O\left(V/\epsilon^2\right)$, using the standard argument popular in the theory literature, e.g., \cite{Article:Karmarkar_SJOC93}. The space complexity (in terms of bits) will be $O\left(V/\epsilon^2 \log \sum_{s=1}^t |I_s|\right)$.
The drawback of this argument is that it does not fully specify the constants. \\

In a summary, in order to provide a $\nu$ (e.g., 0.1) additive approximation of the Shannon entropy, one should use $O\left(V/\epsilon^2\right)  = O\left(V/(\nu\Delta)^2\right)$ samples for estimating the $(1\pm\Delta)$th frequency moments.  This bound  initially appears disappointing, because, if for example, $V=O(1)$, $\nu = 0.1$, $\Delta =10^{-5}$, then it requires $O\left(10^{12}\right)$ samples, which is very likely impractical. Well-known algorithms based on {\em symmetric stable random projections}\cite{Article:Indyk_JACM06,Proc:Li_SODA08} indeed exhibit $V=O(1)$.

\subsection{Some Applications of Shannon Entropy}

\subsubsection{Real-Time Network Anomaly Detection}

Network traffic is a typical example of high-rate data streams. An effective and reliable measurement of network traffic in real-time is crucial for anomaly detection and network diagnosis; and one such measurement metric is Shannon entropy\cite{Proc:Feinstein_DARPA03,Proc:Lakhina_SIGCOMM05,Proc:Xu_SIGCOMM05,Proc:Brauckhoff_IMC06,Proc:Lall_SIGMETRICS06,Proc:Zhao_IMC07}. The {\em Turnstile} data stream model (\ref{eqn_Turnstile}) is naturally suitable for describing network traffic, especially when the goal is to characterize the statistical distribution of the traffic. In its empirical form, a statistical distribution is described by histograms, $A_t[i]$, $i=1$ to $D$. It is possible that $D=2^{64}$ (IPV6) if one is interested in measuring the traffic streams of unique source or destination.

The Distributed Denial of Service (\textbf{DDoS}) attack is a representative example of network anomalies. A DDoS attack attempts to make computers unavailable to intended users, either by forcing users to reset the computers or by exhausting the resources of  service-hosting sites. For example, hackers may maliciously saturate the victim machines by sending many external communication requests. DDoS attacks typically target sites such as banks, credit card payment gateways, or military sites.

A DDoS attack changes the statistical distribution of network traffic. Therefore, a common practice to detect an attack is to monitor the network traffic using certain summary statistics. Since  Shannon entropy is a well-suited for characterizing a distribution, a popular detection method is to measure the time-history of entropy and alarm anomalies when the entropy becomes abnormal\cite{Proc:Feinstein_DARPA03,Proc:Lall_SIGMETRICS06}.

Entropy measurements do not have to be ``perfect'' for detecting attacks. It is however crucial that the  algorithm should be computationally efficient at  low memory cost, because the traffic data generated by large high-speed networks are enormous and transient (e.g., 1 Gbits/second). Algorithms should be real-time and one-pass, as the traffic data will not be stored\cite{Proc:Babcock_PODS02}. Many algorithms have been proposed for ``sampling'' the traffic data and estimating entropy over data streams\cite{Proc:Lall_SIGMETRICS06,Proc:Zhao_IMC07,Proc:Bhuvanagiri_ESA06,Proc:Guha_SODA06,Article:Chakrabarti_06,Proc:Chakrabarti_SODA07,Proc:Harvey_ITW08,Proc:Harvey_FOCS08},

\subsubsection{Entropy of Query Logs in Web Search}
The recent work\cite{Proc:Mei_WSDM08} was devoted to estimating the Shannon entropy of MSN search logs, to help answer some basic problems in Web search, such as,  {\em how big is the web?}

The search logs can be viewed as data streams, and \cite{Proc:Mei_WSDM08}  analyzed several ``snapshots'' of a sample of MSN search logs.  The sample used in \cite{Proc:Mei_WSDM08} contained 10 million $<$Query, URL,IP$>$ triples; each triple corresponded to a click from a particular IP address on a particular URL for a particular query.  \cite{Proc:Mei_WSDM08} drew their important conclusions on this (hopefully) representative sample. Alternatively, one could apply data  stream algorithms such as CC on the whole history of MSN (or other search engines).

\vspace{-0.05in}
\subsubsection{Entropy in Neural Computations}

A workshop in NIPS'03 was devoted to entropy estimation, owing to the wide-spread use of Shannon entropy in Neural Computations\cite{Article:Paninski_NC03}. (\url{http://www.menem.com/~ilya/pages/NIPS03}) For example, one application of entropy is to study the underlying structure of  spike trains.

\subsection{Previous Algorithms for Estimating Frequency Moments}

The problem of approximating $F_{(\alpha)}$ has been very heavily studied in theoretical computer science and databases, since the pioneering work of \cite{Proc:Alon_STOC96}, which studied $\alpha = 0$,  2, and $\alpha>2$. \cite{Proc:Feigenbaum_FOCS99,Article:Indyk_JACM06,Proc:Li_SODA08} provided improved algorithms for $0<\alpha\leq 2$. \cite{Proc:Indyk_STOC05} provided algorithms for $\alpha >2$ to  achieve the lower bounds proved by
 \cite{Proc:Saks_STOC02,Proc:Kumar_FOCS02,Proc:Woodruff_SODA04}. \cite{Proc:Ganguly_RANDOM07} suggested using even more space to trade for some speedup in the processing time.\\

Note that the first moment (i.e., the sum), $F_{(1)}$, can be computed easily with a simple counter\cite{Article:Morris_CACM78,Article:Flajolet_BIT85,Proc:Alon_STOC96}. This important property was recently  captured by the method of \textbf{Compressed Counting (CC)}\cite{Proc:Li_SODA09}, which was based on the {\em maximally-skewed stable random projections}.
\cite{Proc:Li_SODA09} provided two algorithms, based on the {\em geometric mean} and {\em harmonic mean},\footnote{
The {\em geometric mean} and {\em harmonic mean} algorithms could be empirically improved using another algorithm based on numerical optimizations\cite{Proc:Li_UAI09}, which is very difficult for precise theoretical analysis (variances and bounds).
}
and proved some important theoretical results:
\begin{itemize}
\item The {\em geometric mean} algorithm has the variance proportional to $O(\Delta)$ in the neighborhood of $\alpha=1$, where $\Delta = |1-\alpha|$. This is the first algorithm that captured the intuition that, in the neighborhood of $\alpha=1$, the moment estimation algorithms should work better and better as $\alpha\rightarrow 1$, in a continuous fashion.

    \textbf{Our comments}: The {\em geometric mean} algorithm, unfortunately, did not provide an adequate mechanism for entropy estimation. As previously discussed, this methods leads to an entropy estimation algorithm with complexity $O\left(1/(\nu^2 \Delta)\right)$, which is actually quite intuitive from the definitions of the Tsallis entropy and  R\'enyi entropy. Both entropies contain the $\frac{1}{1-\alpha}=\frac{1}{\Delta}$ terms, meaning that the variance will blow up as $O\left(1/\Delta^2\right)$, which can not be canceled by $O\left(\Delta\right)$. Note that \cite{Proc:Li_SODA09} did not show the variance of the {\em harmonic mean} algorithm is also proportional to $O\left(\Delta\right)$; this paper will provide the proof.

\item For fixed $\epsilon$, as $\Delta\rightarrow 0$, the sample complexity bound of the {\em geometric mean} algorithm is $O\left(1/\epsilon\right)$ with all constants specified. This result was a major improvement over the well-known $O\left(1/\epsilon^2\right)$ bound\cite{Proc:Woodruff_SODA04,Article:Indyk_JACM06,Proc:Li_SODA08}. Note that the assumption of fixing $\epsilon$ and letting $\Delta\rightarrow 0$ is needed for theoretical convenience in order to derive bounds with no unspecified constants. This study will continue to use this assumption.

    \textbf{Our comments}: When $\alpha=1$, the moment estimation problem is trivial and only requires one simple counter. Therefore, even intuitively,  $O\left(1/\epsilon\right)$ can not possibly be the true complexity bound. %One might speculate that some sort of ``logarithmic'' bound could be the answer. Our study verifies this natural conjecture.

\end{itemize}

\section{The Proposed Algorithm}

We consider the {\em relaxed strict-Turnstile} model (\ref{eqn_Turnstile}). Conceptually, we  multiply the data stream vector $A_t\in\mathbb{R}^{1\times D}$ by a random projection matrix $\mathbf{R}\in\mathbb{R}^{D\times k}$. The resultant vector $X = A_t \times \mathbf{R}\in\mathbb{R}^{k\times 1}$ is only of length $k$.  More specifically, the entries of the projected vector $X$ are
\begin{align}\notag
x_j =\left[A_t\times\mathbf{R}\right]_j= \sum_{i=1}^D r_{ij} A_t[i], \ \ j = 1, 2, ..., k
\end{align}

$r_{ij}$'s are random variables generated from the following (\textbf{non-standard}) skewed stable distribution\cite{Book:Zolotarev_86}:
\begin{align}\label{eqn_r_ij}
r_{ij} =  \frac{  \sin\left(\alpha v_{ij} \right)}{\left[\sin v_{ij}
\right]^{1/\alpha}} \left[\frac{\sin\left( v_{ij}\Delta\right)}{w_{ij}}
\right]^{\frac{\Delta}{\alpha}}, \ \ \ \ \Delta = 1-\alpha>0,
\end{align}
where $v_{ij} \sim Uniform(0,\pi)$ (i.i.d.) and $w_{ij}\sim Exp(1)$ (i.i.d.), an exponential distribution with mean 1. We use this formulation to avoid numerical problems and simplify the analysis. \\

Of course, in data stream computations, the matrix $\mathbf{R}$ is never fully materialized. The standard procedure in data stream computations is to generate entries of $\mathbf{R}$ on-demand\cite{Article:Indyk_JACM06}. In other words, whenever an stream element $a_t = (i_t, I_t)$ arrives, one updates entries of $X$ as
\begin{align}\notag
x_j \leftarrow x_j + I_t r_{i_tj},  \ \ \ j = 1, 2, ..., k.
\end{align}

The proposed algorithm is defined as follows:
\begin{align}\label{eqn_F}
\hat{F}_{(\alpha)} = \frac{1}{\Delta^\Delta}\left[\frac{k}{\sum_{j=1}^k x_j^{-\alpha/\Delta}}\right]^\Delta
\end{align}

The following Theorem proves that this new estimator is (asymptotically) unbiased with the variance proportional to $O\left(\Delta^2\right)$. Note that $\Delta^\Delta\rightarrow 1$  as $\Delta\rightarrow 0$.

\begin{theorem}\label{thm_F_var}
\begin{align}
&E\left(\hat{F}_{(\alpha)}\right) = F_{(\alpha)}\left(1 + O\left(\frac{\Delta}{k}\right)\right),\\\label{eqn_Var_F}
&Var\left(\hat{F}_{(\alpha)}\right) = \frac{\Delta^2}{k}F_{(\alpha)}^2\left(3-2\Delta+O\left(\frac{1}{k}\right)\right).
\end{align}
\textbf{Proof:} \hspace{0.2in} See Appendix \ref{proof_thm_F_var}.
\end{theorem}

In this paper, we only consider $\alpha = 1-\Delta<1$. This is because the maximally-skewed stable distributions have good theoretical properties when $\alpha<1$\cite{Proc:Li_SODA09}; for example, all negative moments exist; see Lemma \ref{lem_moments}.

\subsection{Review Maximally-Skewed Stable Random Projections and Estimators}

The standard procedure for sampling from skewed stable distributions is based on the Chambers-Mallows-Stuck method\cite{Article:Chambers_JASA76}. To generate a sample from $S(\alpha,\beta=1,1)$, i.e., $\alpha$-stable, maximally-skewed ($\beta=1$), with unit scale,  one first generates an exponential random variable with mean 1, $W \sim Exp(1)$,  and a uniform random variable $U \sim Uniform \left(-\frac{\pi}{2}, \frac{\pi}{2}\right)$, then,
\begin{align}\label{eqn_sampling_skewed}
Z^\prime &= \frac{\sin\left(\alpha(U+\rho)\right)}{\left[\cos U \cos\left(\rho \alpha\right)
\right]^{1/\alpha}} \left[\frac{\cos\left( U - \alpha(U + \rho)\right)}{W}
\right]^{\frac{1-\alpha}{\alpha}} \sim S(\alpha,\beta=1,1),
\end{align}
where $\rho = \frac{\pi}{2}$ when $\alpha<1$ and $\rho = \frac{\pi}{2}\frac{2-\alpha}{\alpha}$ when $\alpha>1$. \\

Note that $\cos\left(\frac{\pi}{2}\alpha\right)\rightarrow 0$ as $\alpha\rightarrow 1$. For convenience (and avoiding numerical problems), we will use
\begin{align}\notag
Z = Z^\prime\cos^{1/\alpha}\left(\rho\alpha\right) \sim S\left(\alpha,\beta=1,\cos\left(\rho\alpha\right)\right).
\end{align}

In this study, we will only consider\ \  $\alpha = 1-\Delta<1$, i.e, $\rho=\frac{\pi}{2}$. After simplification, we obtain
\begin{align}
Z &= \frac{  \sin\left(\alpha V \right)}{\left[\sin V
\right]^{1/\alpha}} \left[\frac{\sin\left( V\Delta\right)}{W}
\right]^{\frac{\Delta}{\alpha}},
\end{align}
where $V = \frac{\pi}{2} + U\sim Uniform(0,\pi)$. This explains (\ref{eqn_r_ij}).\\

Lemma \ref{lem_Z_order} shows $\log Z = O\left(|\Delta\log\Delta|\right)$, which can be accurately represented using $O\left(\log 1/\Delta\right)$ bits. The proof is omitted since it is straightforward.

\begin{lemma}\label{lem_Z_order}
For any given $V\neq 0$, and $W\neq 0$, as $\Delta\rightarrow 0$,
\begin{align}\notag
Z = 1+O\left(|\Delta\log\Delta|\right), \ \ \ \text{i.e.,} \ \ \log Z = O\left(|\Delta\log\Delta|\right).
\end{align}
\end{lemma}

\vspace{0.2in}

Let $X = A_t\times \mathbf{R}$, where entries of $\mathbf{R}$  are i.i.d. samples of $S\left(\alpha,\beta=1,\cos\left(\frac{\pi}{2}\alpha\right)\right)$. Then by properties of stable distributions, entries of $X$ are
\begin{align}\notag
x_j = \left[A_t\times \mathbf{R}\right]_j = \sum_{i=1}^D r_{i,j}A_t[i] \sim S \left(\alpha,\beta=1, \cos\left(\frac{\pi}{2}\alpha\right)F_{(\alpha)}\right),
\end{align}
where $F_{(\alpha)}=\sum_{i=1}^D A_t[i]^\alpha$ as defined in (\ref{eqn_moment}).\\

Therefore, CC boils down to estimating  $F_{(\alpha)}$ from $k$ i.i.d. stable samples. \cite{Proc:Li_SODA09} provided two statistical estimators, the {\em geometric mean} and {\em harmonic mean} estimators, which are derived based on the following basic moment formula.

%\subsubsection{The Basic Moment Formula}

\begin{lemma}\label{lem_moments}
\cite{Proc:Li_SODA09}.   \hspace{0.2in}
If {\small$X \sim S(\alpha,\beta=1,F_{(\alpha)}\cos\left(\frac{\alpha\pi}{2}\right))$}, then $X>0$, and  for any {\small$-\infty<\lambda<\alpha<1$},
\begin{align}\notag%\label{eqn_moment}%\notag
{E}\left(X^\lambda\right) = {F}_{(\alpha)}^{\lambda/\alpha} \frac{ \Gamma\left(1-\frac{\lambda}{\alpha}\right) }
{\Gamma\left(1-\lambda\right)}.
\end{align}
\end{lemma}
\subsubsection{The  Geometric Mean Estimator}

Assume $x_j$, $j=1$ to $k$, are i.i.d. samples from $S(\alpha,\beta=1,F_{(\alpha)}\cos\left(\frac{\alpha\pi}{2}\right))$. After simplifying the corresponding expression in \cite{Proc:Li_SODA09}, we obtain
%\begin{align}\notag
%&\hat{F}_{(\alpha),gm} = \frac{\prod_{j=1}^k |x_j|^{\alpha/k}} { \cos^k\left(\frac{\alpha\pi}{2k}\right)
%\left[\frac{2}{\pi}\sin\left(\frac{\pi\alpha}{2k}\right)
%\Gamma\left(1-\frac{1}{k}\right)\Gamma\left(\frac{\alpha}{k}\right)\right]^k}
%\end{align}
\begin{align}%\notag
&\hat{F}_{(\alpha),gm} = \left[\frac{\Gamma\left(1-\frac{\alpha}{k}\right)}{\Gamma\left(1-\frac{1}{k}\right)}\right]^k \prod_{j=1}^k x_j^{\alpha/k},
\end{align}
which is unbiased and has asymptotic  variance
\begin{align}\label{eqn_Var_gm}
\text{Var}\left(\hat{F}_{(\alpha),gm}\right)=
\frac{F_{(\alpha)}^2}{k}\frac{\pi^2}{6}\Delta\left(1+\alpha\right)+O\left(\frac{1}{k^2}\right)
\end{align}
\noindent As $\alpha\rightarrow 1$, the asymptotic variance  approaches zero at the rate of only $O\left(\Delta\right)$, which is not adequate.

\subsubsection{The Harmonic Mean Estimator}
\begin{align}%\notag
\hat{F}_{(\alpha),hm} = \frac{k\frac{1}{\Gamma(1+\alpha)}}{\sum_{j=1}^k|x_j|^{-\alpha}}
\left(1- \frac{1}{k}\left(\frac{2\Gamma^2(1+\alpha)}{\Gamma(1+2\alpha)}-1\right) \right),
\end{align}
which is asymptotically unbiased and has variance
\begin{align}%\notag
\text{Var}\left(\hat{F}_{(\alpha),hm}\right) = \frac{F^{2}_{(\alpha)}}{k}\left(\frac{2\Gamma^2(1+\alpha)}{\Gamma(1+2\alpha)}-1\right) + O\left(\frac{1}{k^2}\right).
\end{align}

\cite{Proc:Li_SODA09} only graphically showed that the {\em harmonic mean} estimator is noticeably better than the {\em geometric mean} estimator. We prove the following Lemma, which says the variance of the {\em harmonic mean} is also proportional to $O\left(\Delta\right)$. Thus, the {\em harmonic mean} estimator is  not adequate for entropy estimation either.

\begin{lemma}\label{lem_hm}
As $\Delta = 1-\alpha\rightarrow 0$,
\begin{align}
\frac{2\Gamma^2(1+\alpha)}{\Gamma(1+2\alpha)}-1 = \Delta + \Delta^2\left(2-\frac{\pi^2}{6}\right)+O\left(\Delta^3\right).
\end{align}
\textbf{Proof:} \hspace{0.2in} See Appendix \ref{proof_lem_hm}.

\end{lemma}

%
%The proposed estimator of $F_{(\alpha)}$ is based on the {\em sample minimum}:
%\begin{align}\notag
%\hat{F}_{(\alpha),\min} = \left[\min\left\{x_j, j = 1, 2, ..., k\right\}\right]^\alpha
%\end{align}

\subsection{The Distribution Function}

This section provides the distribution function of $Z \sim S\left(\alpha<1,\beta=1,\cos\left(\frac{\pi}{2}\alpha\right)\right)$, which will be needed in deriving  the
proposed estimator (\ref{eqn_F}).

%and later in analyzing the complexity bound of the simplified estimator.

\begin{lemma}\label{lem_CDF}
Suppose a random variable $Z \sim S\left(\alpha<1,\beta=1,\cos\left(\frac{\pi}{2}\alpha\right)\right)$.
The cumulative distribution function (CDF) is
\begin{align}\notag
&F_Z(t) =  \mathbf{Pr}\left(Z\leq t\right) =\frac{1}{\pi}\int_0^\pi  \exp\left(- t^{-\alpha/\Delta} g\left(\theta;\Delta\right)\right) d\theta.
\end{align}
where
\begin{align}\notag
g(\theta;\Delta) = \frac{  \left[\sin\left(\alpha \theta \right)\right]^{\alpha/\Delta} }{\left[\sin \theta \right]^{1/\Delta}} \sin\left( \theta\Delta\right), \hspace{0.5in} \theta\in(0,\pi)
\end{align}

Assume $\Delta = 1-\alpha<0.5$, then
$g(\theta;\Delta)$ is monotonically increasing in $(0,\pi)$, with
\begin{align}\notag
\lim_{\theta\rightarrow0+}g(\theta;\Delta) = g\left(0+;\Delta\right)=\Delta\alpha^{\alpha/\Delta}.
\end{align}
Moreover, $g(\theta;\Delta)$ is a convex function of $\theta$.

\textbf{Proof:}\hspace{0.2in} See Appendix \ref{proof_lem_CDF}.\\
\end{lemma}

\begin{figure}[h]
\begin{center}\mbox{
\includegraphics[width=2.25in]{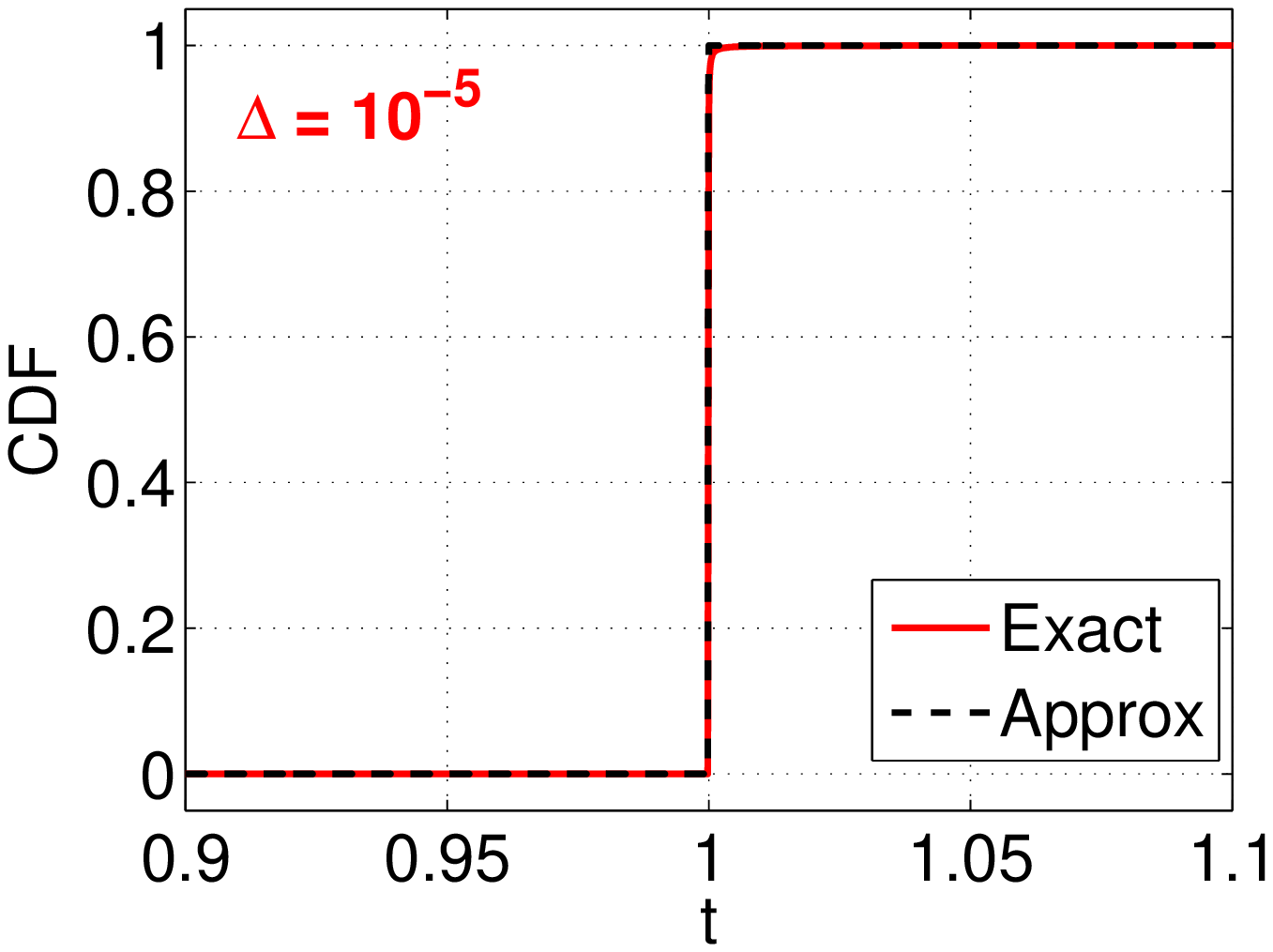}\hspace{-0.16in}
\includegraphics[width=2.25in]{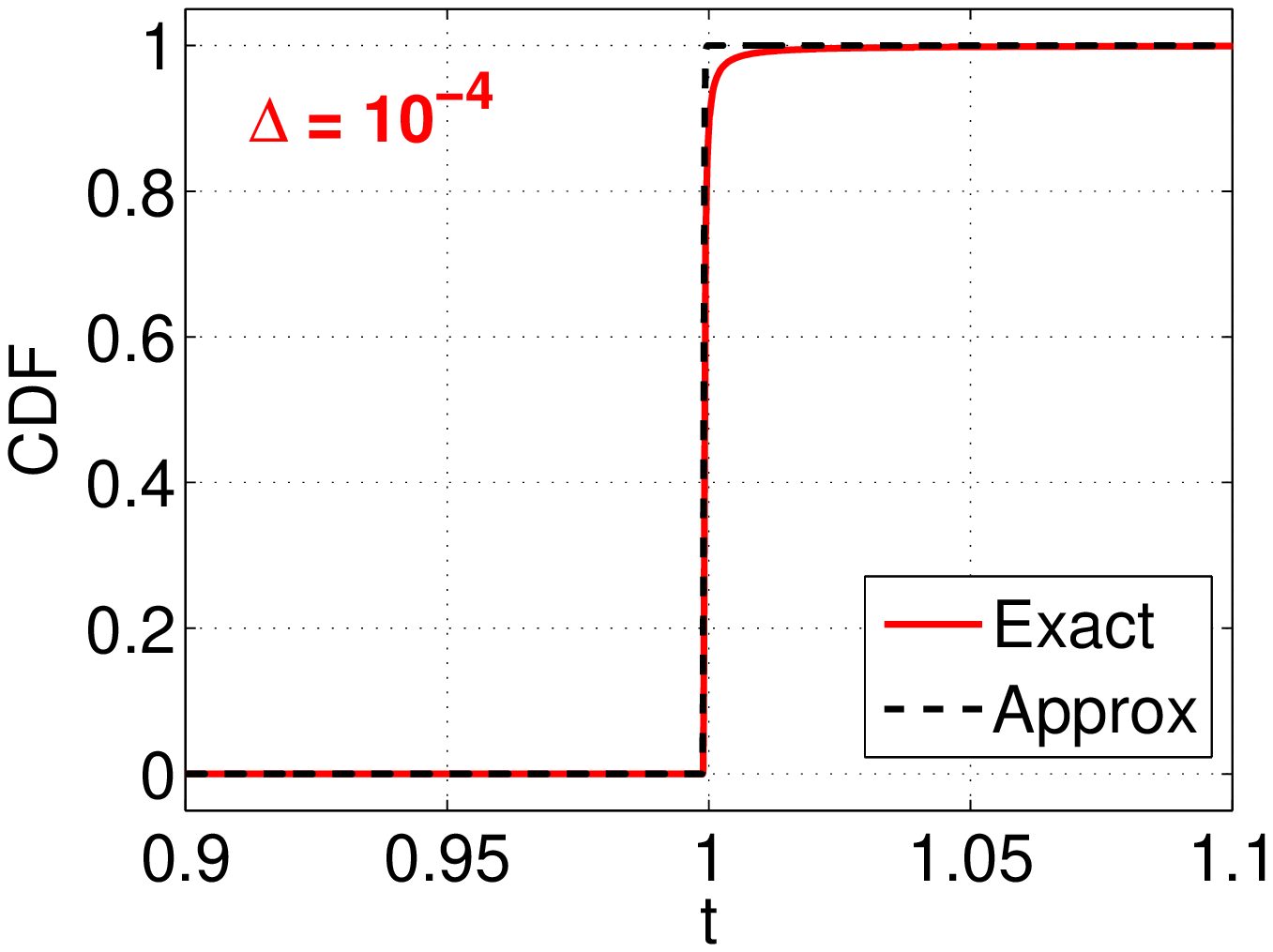}\hspace{-0.16in}
\includegraphics[width=2.25in]{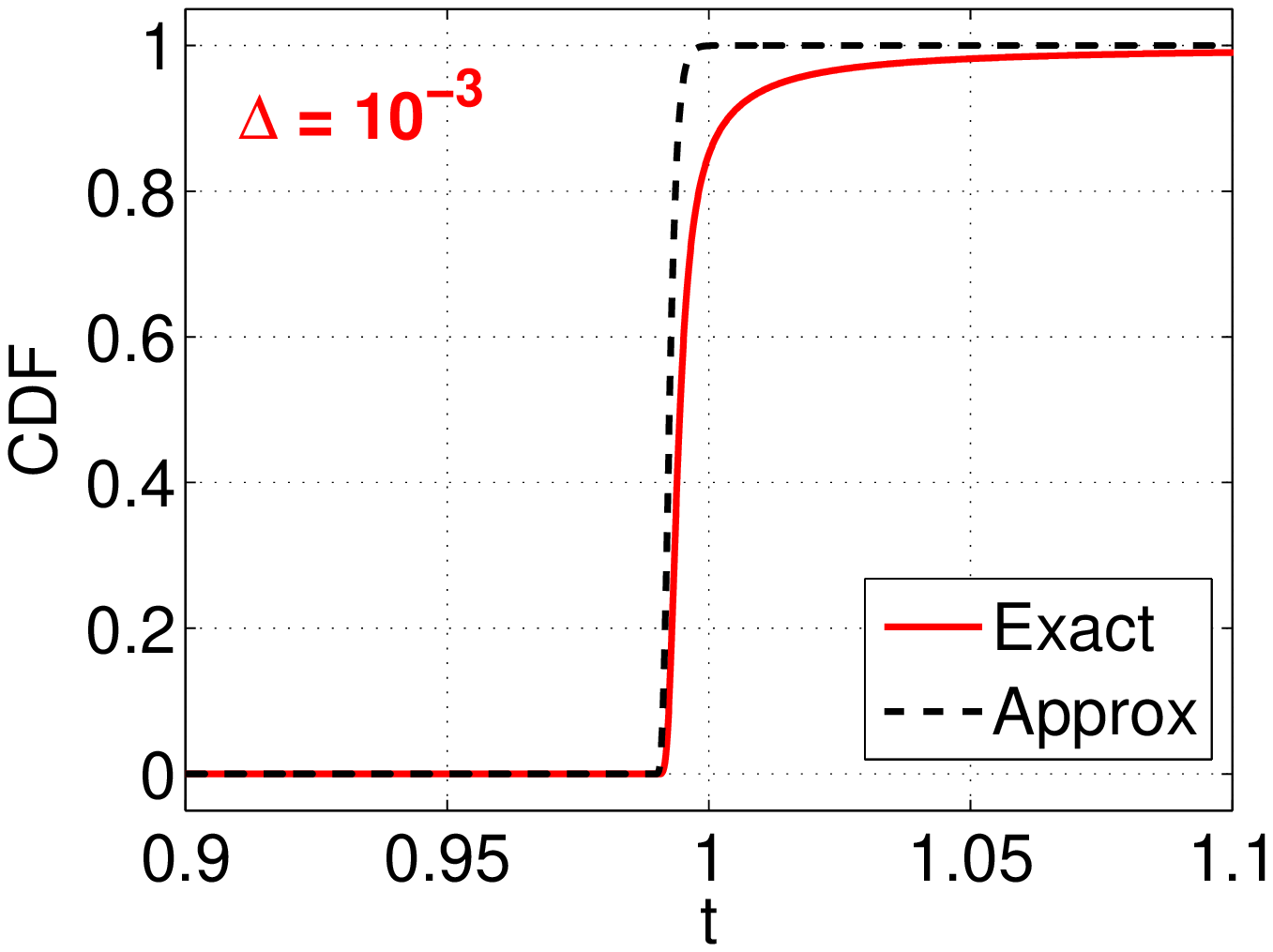}
}
\end{center}
\vspace{-0.2in}
\caption{We plot the CDF curves as derived in Lemma \ref{lem_CDF}, for $\Delta=10^{-5}$, $10^{-4}$, and $10^{-3}$. As $\Delta\rightarrow 0$, the exact CDF (solid curves) is very close to the approximate CDF (dashed curves), which we obtain by replacing the exact $g(\theta;\Delta)$ function in Lemma \ref{lem_CDF} with the limit $g(0+;\Delta)$.
 }\label{fig_CDF}
\end{figure}

Note that $g\left(0+;\Delta\right) = \Delta\alpha^{\alpha/\Delta} \approx \Delta e^{-1}$ approaches zero as $\Delta\rightarrow 0$. Thus, one might be wondering if we replace $g\left( \theta;\Delta\right)$ by $g\left(0+;\Delta\right)$, the errors may be quite small. This conjecture is verified in Figure \ref{fig_CDF}.

\subsection{The Intuition Behind the Proposed Algorithm}

Basically, we derive the proposed estimator by ``guessing.'' We first derive a maximum likelihood estimator (MLE) for a slightly different distribution based on the intuition from Lemma \ref{lem_CDF} and Figure \ref{fig_CDF}. Then we verify that this MLE is actually a very good estimator (in terms of both the variances and tail bounds) for the stable distribution we care about.

Here, we consider a random variable $Y$ whose cumulative distribution function (CDF) is
\begin{align}\label{eqn_CDF_Y}
F_Y(t) = \mathbf{Pr}\left(Y\leq t\right) = \exp\left(-t^{-\alpha/\Delta}\Delta\alpha^{\alpha/\Delta}\right),  \hspace{0.2in} t \in [0, \infty).
\end{align}
It is indeed a CDF because it is an increasing function of $t\in[0,\infty)$, $F_Y(0) = 0$, and $F_Y(\infty) = 1$.

Similar to stable random projections, we are interested in estimating $c^\alpha$ from $k$ i.i.d. samples $x_j = cY_j$, $j=1$ to $k$. Statistics theory tells us that the maximum likelihood estimator (MLE) has the (asymptotic) optimality. Because the distribution function of $Y_j$ is known, we can actually compute the MLE in this case.

\begin{theorem}\label{thm_MLE}
Suppose $Y_j$, $j = 1$ to $k$, are i.i.d. samples from a distribution whose CDF is given by (\ref{eqn_CDF_Y}). Let $x_j = cY_j$, where $c>0$. Then the maximum likelihood estimator of $c^\alpha$ is given by
\begin{align}\label{eqn_MLE}
\frac{1}{\Delta^\Delta\alpha^\alpha} \left[\frac{k}{\sum_{j=1}^k x_j^{-\alpha/\Delta}}\right]^\Delta
\end{align}
\textbf{Proof:}\hspace{0.2in} See Appendix \ref{proof_thm_MLE}.
\end{theorem}

Compared with the proposed estimator $\hat{F}_{(\alpha)}$ in (\ref{eqn_F}),  the MLE solution has the addition term of $\frac{1}{\alpha^\alpha}$. Note that, while both $\Delta^\Delta$ and $\alpha^\alpha$  approach 1, $\Delta^\Delta\rightarrow 1$ considerably slower than $\alpha^\alpha\rightarrow 1$, because
\begin{align}\notag
&\alpha^\alpha = \exp\left(\alpha\log \alpha\right) = \exp\left(-\Delta + O\left(\Delta^2\right)\right)\\\notag
&\Delta^\Delta = \exp\left(\Delta\log\Delta\right)
\end{align}
For example, when $\Delta=0.1$, $\Delta^\Delta = 0.7943$, $\alpha^\alpha = 0.9095$;  when $\Delta=0.01$, $\Delta^\Delta=0.9550$, $\alpha^\alpha = 0.9901$.

Therefore, while $\alpha^\alpha$ may be considered negligible, it may be preferable to keep $\Delta^\Delta$. In fact, when proving that the proposed estimator $\hat{F}_{(\alpha)}$ is (asymptotically) unbiased (see Appendix \ref{proof_thm_F_var}), we do need  the $\Delta^\Delta$ term.

\section{The Tail Bounds of the Proposed Estimator}

Theorem \ref{thm_F_var} has proved that the proposed estimator
\begin{align}\notag
\hat{F}_{(\alpha)} = \frac{1}{\Delta^\Delta}\left[\frac{k}{\sum_{j=1}^k x_j^{-\alpha/\Delta}}\right]^\Delta
\end{align}
is asymptotically unbiased with variance proportional to $O(\Delta^2)$. Using the standard argument, we know that the sample complexity bound must be $O\left(\frac{\Delta^2}{\epsilon^2}\right) = O\left(\frac{\Delta^2}{\nu^2\Delta^2}\right) = O\left(\frac{1}{\nu^2}\right)$. We are, however, very interested in the precise complexity bounds, not just the orders.

Normally, we would like to present the tail bounds as, e.g., $\mathbf{Pr}\left(\hat{F}_{(\alpha)} \geq \left(1+\epsilon\right)F_{(\alpha)}\right) \leq \exp\left(-k\frac{\epsilon^2}{G_R}\right)$, which immediately leads to the statement that:

{\em With probability at least $1-\delta$, it suffices to use $k \geq \frac{G_R}{\epsilon^2}\log1/\delta$ to guarantee $\hat{F}_{(\alpha)} \leq (1+\epsilon)F_{(\alpha)}$.}\\

Ideally, we hope $G_R$ will be as small as possible. In fact, in order to achieve a $\nu$-additive algorithm for entropy estimation, we need $\epsilon = \nu\Delta$ (where $\Delta<10^{-4}$ or even much smaller). Therefore, we really need $G_R = O\left(\Delta^2\right)$. In this sense, it is no longer appropriate to treat $G_R$ as a ``constant.''\\

Theorem \ref{thm_tail} presents the tail bounds for $\hat{F}_{(\alpha)}$.
\begin{theorem}\label{thm_tail}
For any $\epsilon >0$ and $0< \Delta=1-\alpha<1$, we have the right tail bound for the proposed estimator:
\begin{align}\label{eqn_G_R}
&\mathbf{Pr}\left(\hat{F}_{(\alpha)} \geq (1+\epsilon)F_{(\alpha)}\right) \leq \exp\left(-k\frac{\epsilon^2}{G_R}\right)\\\notag
&\frac{\epsilon^2}{G_R} = - \left( \log \sum_{n=0}^\infty \frac{(-t_R)^n}{n!} \frac{\Gamma\left(1+\frac{n}{\Delta}\right)}{\Gamma\left(1+\frac{n\alpha}{\Delta}\right)} +
\frac{t_R}{(1+\epsilon)^{1/\Delta} \Delta }\right)
\end{align}
where $t_R$ is the solution to
\begin{align}\label{eqn_t_R}
\frac{\sum_{n=1}^\infty \frac{(-1)^n(t_R)^{n-1}}{(n-1)!} \frac{\Gamma\left(1+\frac{n}{\Delta}\right)}{\Gamma\left(1+\frac{n\alpha}{\Delta}\right)}}
{\sum_{n=0}^\infty \frac{(-t_R)^n}{n!} \frac{\Gamma\left(1+\frac{n}{\Delta}\right)}{\Gamma\left(1+\frac{n\alpha}{\Delta}\right)}
} +
\frac{1}{(1+\epsilon)^{1/\Delta} \Delta } = 0
\end{align}

For any $0<\epsilon <1$ and $0< \Delta=1-\alpha<1$, we have the left tail bound:
\begin{align}\label{eqn_G_L}
&\mathbf{Pr}\left(\hat{F}_{(\alpha)} \leq (1-\epsilon)F_{(\alpha)}\right) \leq \exp\left(-k\frac{\epsilon^2}{G_L}\right)\\\notag
&\frac{\epsilon^2}{G_L} = -  \log \sum_{n=0}^\infty \frac{(t_L)^n}{n!} \frac{\Gamma\left(1+\frac{n}{\Delta}\right)}{\Gamma\left(1+\frac{n\alpha}{\Delta}\right)}+
\frac{t_L}{(1-\epsilon)^{1/\Delta} \Delta}
\end{align}
where $t_L$ is the solution to
\begin{align}\label{eqn_t_L}
-\frac{\sum_{n=1}^\infty \frac{(t_L)^{n-1}}{(n-1)!} \frac{\Gamma\left(1+\frac{n}{\Delta}\right)}{\Gamma\left(1+\frac{n\alpha}{\Delta}\right)}}
{\sum_{n=0}^\infty \frac{(t_L)^n}{n!} \frac{\Gamma\left(1+\frac{n}{\Delta}\right)}{\Gamma\left(1+\frac{n\alpha}{\Delta}\right)}
} +
\frac{1}{(1-\epsilon)^{1/\Delta} \Delta } = 0
\end{align}
\textbf{Proof:} \ See Appendix \ref{proof_thm_tail}.\\

\end{theorem}

These bounds appear  to be too complicated to gain insightful information. People may be even wondering about numerical stability  of the infinite sums.

First of all, we notice that when $\Delta = 1$ (i.e., $\alpha = 0$), we can compute the tail bounds exactly, as presented in Lemma \ref{lem_tail_d=1}.
\begin{lemma}\label{lem_tail_d=1}
When $\Delta = 1$, i.e., $\alpha = 0$,
\begin{align}
&\frac{\epsilon^2}{G_R} = \log (1+\epsilon) - \frac{\epsilon}{1+\epsilon}, \hspace{0.2in} \epsilon>0\\
&\frac{\epsilon^2}{G_L} = \log (1-\epsilon) + \frac{\epsilon}{1-\epsilon}, \hspace{0.2in} 0<\epsilon<1.
\end{align}
\textbf{Proof:} \hspace{0.2in} When $\Delta=1$ ($\alpha=0$), we have $\Gamma\left(1+\frac{n}{\Delta}\right)=n!$, $\Gamma\left(1+\frac{n\alpha}{\Delta}\right)=1$, $\sum_{n=0}^\infty t^n = \frac{1}{1-t}$, and $\sum_{n=0}^\infty (-t)^n = \frac{1}{1+t}$. The conclusions follow easily. $\Box$
\end{lemma}

\vspace{0.3in}

Next, we re-formulate the tail bounds to facilitate numerical evaluations. Our numerical results show that, when $\Delta$ is small, $G_R \approx (6\sim 9) \Delta^2$ and $G_L \approx (4\sim 6) \Delta^2$, for $0<\nu<1$. Thus, we indeed have an algorithm for entropy estimation with complexity $O\left(\frac{1}{\nu^2}\right)$. \\

%\subsection{The Numerical Evaluations of the Tail Bounds}

The tail bounds (\ref{eqn_G_R}) and (\ref{eqn_G_L}) contain $\frac{\Gamma\left(1+\frac{n}{\Delta}\right)}{\Gamma\left(1+\frac{n\alpha}{\Delta}\right)}$, which can be written as
\begin{align}\notag
\frac{\Gamma\left(1+\frac{n}{\Delta}\right)}{\Gamma\left(1+\frac{n\alpha}{\Delta}\right)}
=\frac{\Gamma\left(1+\frac{n}{\Delta}\right)}{\Gamma\left(1+\frac{n}{\Delta} - n\right)} = \frac{n}{\Delta} \left(\frac{n}{\Delta}-1\right)...\left(\frac{n}{\Delta}-n+1\right) = \frac{1}{\Delta^n} n\left(n-\Delta\right)\left(n-2\Delta\right)...\left(n-(n-1)\Delta\right).
\end{align}
Therefore,
\begin{align}\notag
\frac{1}{n!}\frac{\Gamma\left(1+\frac{n}{\Delta}\right)}{\Gamma\left(1+\frac{n\alpha}{\Delta}\right)}
=\frac{1}{\Delta^n}\prod_{j=0}^{n-1}\frac{n-j\Delta}{n-j} \leq \frac{1}{\Delta^n} \frac{n^n}{n!}
%\leq \frac{1}{\Delta^n} \frac{n^n}{\sqrt{2\pi(n+1)}\left(\frac{n+1}{e}\right)^{n+1}}
\leq \frac{1}{\Delta^n}\frac{n^n}{(n-1)!} \leq \frac{1}{\Delta^n} \frac{e^n}{\sqrt{2\pi n}}
\end{align}
according to the Stirling's series \cite[8.327]{Book:Gradshteyn_94}
\begin{align}\notag
\Gamma(n)=(n-1)! = \sqrt{2\pi n}\left(\frac{n}{e}\right)^n \left[1 + \frac{1}{12n}+\frac{1}{288n^2}-\frac{139}{51840n^3}-...\right].
\end{align}

Thus, for numerical reasons, we can rewrite (\ref{eqn_G_R}) and (\ref{eqn_G_L}) as
\begin{align}\notag
&\frac{\epsilon^2}{G_R} = - \log\left(1+\sum_{n=1}^\infty \left(-t_R \frac{e}{\Delta}\right)^n \prod_{j=0}^{n-1} \frac{n-j\Delta}{(n-j)e}\right) - \left(t_R\frac{e}{\Delta}\right)\frac{1}{e(1+\epsilon)^{1/\Delta}}\\\notag
&\frac{\epsilon^2}{G_L} = - \log\left(1+\sum_{n=1}^\infty \left(t_L \frac{e}{\Delta}\right)^n \prod_{j=0}^{n-1} \frac{n-j\Delta}{(n-j)e}\right) + \left(t_L\frac{e}{\Delta}\right)\frac{1}{e(1-\epsilon)^{1/\Delta}}.
\end{align}
The infinite series always converge provided $t_R\leq \frac{\Delta}{e}$ and $t_L\leq \frac{\Delta}{e}$.  In fact, because the bounds hold for any $t>0$ (not necessarily the optimal values, $t_R$ and $t_L$), we know $\frac{\epsilon^2}{G_R} = O(1)$ and $\frac{\epsilon^2}{G_L} = O(1)$ if using (e.g.,) $t = 0.5\frac{\Delta}{e}$. In other words, $G_R = \left(\Delta^2\right)$ and $G_L = \left(\Delta^2\right)$, as desired. We state this as a Lemma.\\

\begin{lemma}\label{lem_G}
The tail bound constants (\ref{eqn_G_R}) and (\ref{eqn_G_L})
\begin{align}\notag
\frac{\epsilon^2}{G_R} = O(1), \hspace{0.5in} \frac{\epsilon^2}{G_L} = O(1),  \hspace{0.5in} (\epsilon = \nu\Delta).
\end{align}
In other words
\begin{align}\notag
G_R = O\left(\Delta^2\right), \hspace{0.5in} G_L = O\left(\Delta^2\right).
\end{align}
Therefore, to estimate $F_{(\alpha)}$ within a $(1\pm\nu\Delta)$ factor, it suffices to let the sample size $k  = O\left(\frac{1}{\nu^2}\right)$, using the proposed estimator $\hat{F}_{(\alpha)}$. \\

\end{lemma}

\begin{figure}[h]
\begin{center}\mbox{
\includegraphics[width=3in]{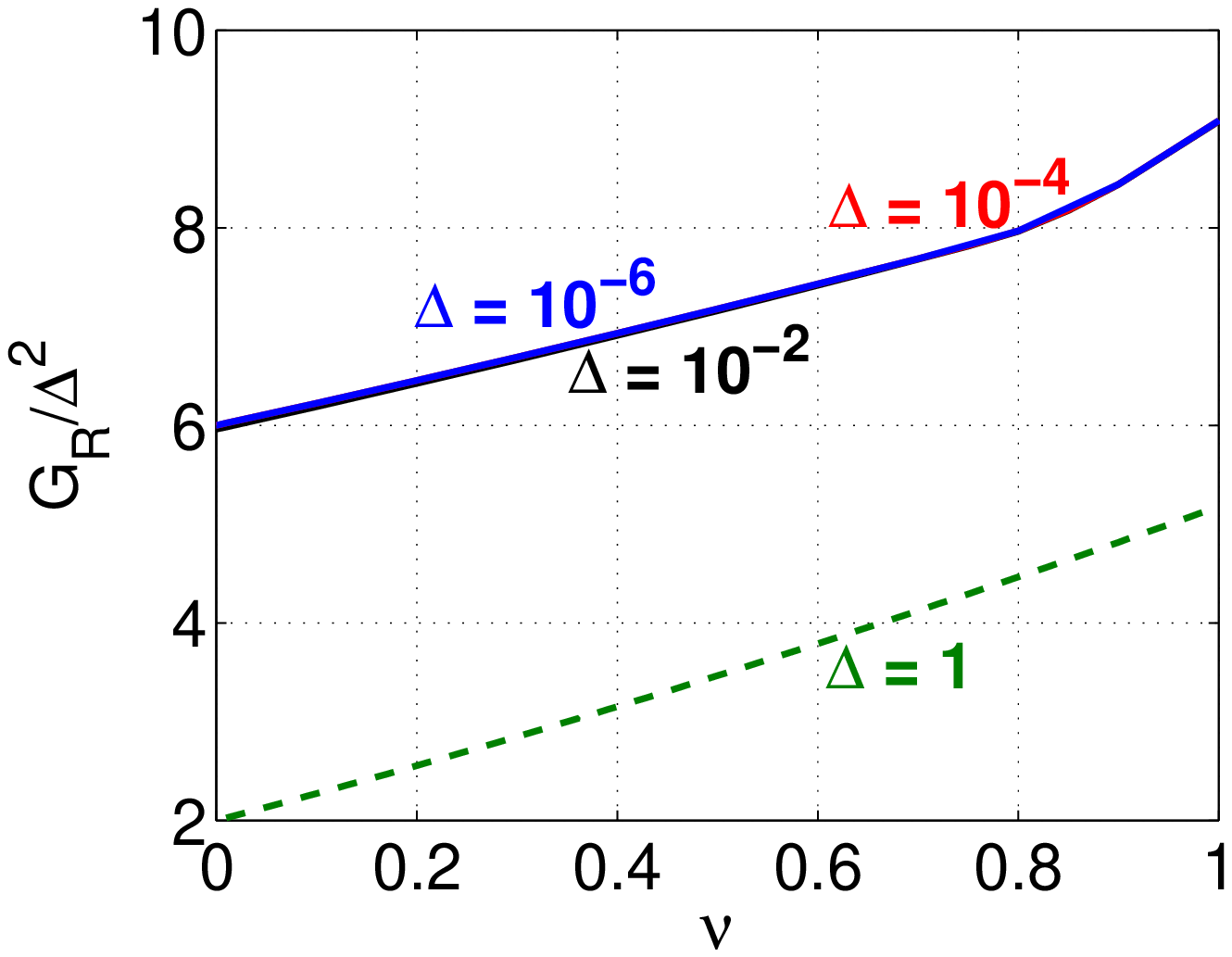}
\includegraphics[width=3in]{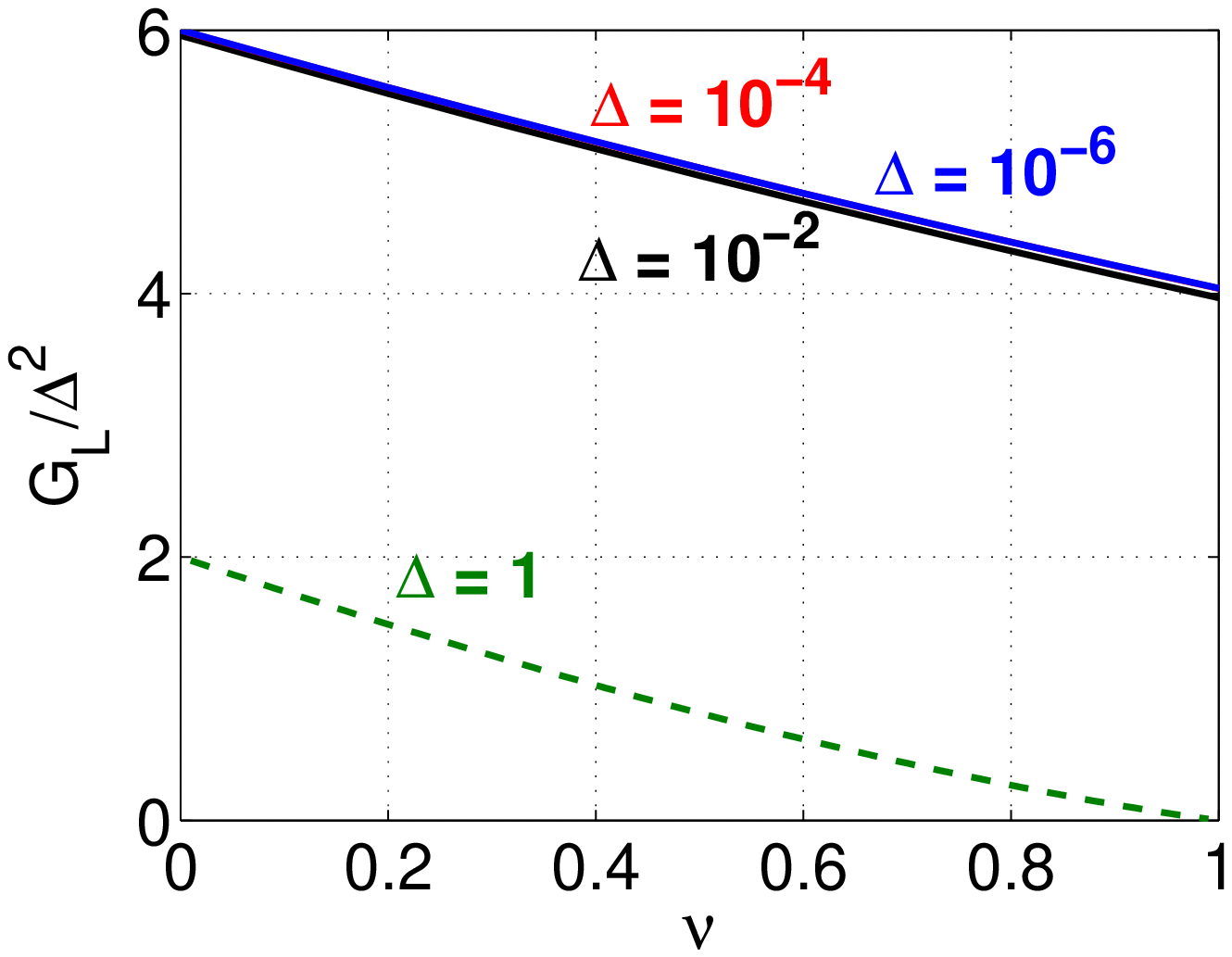}}
\end{center}
\vspace{-0.2in}
\caption{Numerical values of $G_R$ and $G_L$ in the tail bounds (\ref{eqn_G_R}) and (\ref{eqn_G_L}), for $\Delta=10^{-2}$, $\Delta=10^{-4}$, and $\Delta=10^{-6}$, together with the closed-form expressions for $\Delta=1$ as obtained in Lemma \ref{lem_tail_d=1}.\hspace{0.3in} Because $G_R =O\left(\Delta^2\right)$ and $G_L =O\left(\Delta^2\right)$, we present the results in terms of $\frac{G_R}{\Delta^2}$ and $\frac{G_L}{\Delta^2}$. Note that as $\nu\rightarrow0$, both   $\frac{G_R}{\Delta^2}$ and $\frac{G_L}{\Delta^2}$ approach $6-4\Delta$, as proved in Lemma \ref{lem_G->0}. Also, note that the curves for $\Delta=10^{-2}$, $\Delta=10^{-4}$, and $\Delta=10^{-6}$ largely overlap.}\label{fig_G1e246}
\end{figure}

Figure \ref{fig_G1e246} presents the values in terms of $\frac{G_R}{\Delta^2}$ and $\frac{G_L}{\Delta^2}$ for $0<\nu<1$ and $\Delta = 10^{-2}$, $10^{-4}$, $10^{-6}$, together with the closed-form expressions for $\Delta=1$ as obtained in Lemma \ref{lem_tail_d=1}. The values are pleasantly small. Thus, at least numerically, we can say, for example, when $\Delta$ is small,
\begin{align}\notag
&\mathbf{Pr}\left(\hat{F}_{(\alpha)} \geq (1+\epsilon)F_{(\alpha)}\right) \leq \exp\left(-k\frac{\nu^2}{6\sim 9}\right),\\\notag
&\mathbf{Pr}\left(\hat{F}_{(\alpha)} \leq (1-\epsilon)F_{(\alpha)}\right) \leq \exp\left(-k\frac{\nu^2}{4\sim 6}\right).
%,\\\notag
%&\mathbf{Pr}\left(|\hat{F}_{(\alpha)} - F_{(\alpha)}| \geq \epsilon F_{(\alpha)}\right) \leq 2\exp\left(-k\frac{\nu^2}{9}\right).
\end{align}
In other words, with a probability at least $1-\delta$, using the proposed estimator, one can achieve $|\hat{F}_{(\alpha)} - F_{(\alpha)}| \leq (\nu\Delta) F_{(\alpha)}$ by using $k \geq 9\frac{\log2/\delta}{\nu^2}$ samples. And we know the constant 9 could be replaced by 6 if $\nu$ is small. \\

\vspace{0.2in}

Whenever possible, analytical expressions are always more desirable. In fact, when $\nu\rightarrow 0$, we can actually obtain the analytical expressions for $G_{R}$ and $G_L$.

\begin{lemma}\label{lem_G->0}
As $\nu\rightarrow 0$,
\begin{align}\label{qen_v->0}
\frac{G_R}{\Delta^2} \rightarrow 6-4\Delta, \hspace{0.5in} \ \  \frac{G_L}{\Delta^2} \rightarrow 6-4\Delta.
\end{align}
\textbf{Proof}: See Appendix \ref{proof_lem_G->0}.
\end{lemma}

\section{The Connection to the Sample Minimum Estimator}

In the previous (unpublished) work\cite{Report:Li_CC_min09}, we proposed the \textbf{\em sample minimum} estimator, which allowed us to prove a much improved  sample complexity bound than that in \cite{Proc:Li_SODA09}. Interestingly, the proposed estimator $\hat{F}_{(\alpha)}$ in this paper actually converges to the  {\em sample minimum} estimator, denoted by $\hat{F}_{(\alpha),\min}$,
\begin{align}\label{eqn_MLE}
\hat{F}_{(\alpha)} = \frac{1}{\Delta^\Delta} \left[\frac{k}{\sum_{j=1}^k x_j^{-\alpha/\Delta}}\right]^\Delta\rightarrow
\hat{F}_{(\alpha),\min} =
 \min\{x_j^\alpha, j =1, 2, ..., k\}.
\end{align}

This fact is quite intuitive. As $\Delta\rightarrow 0$, the smallest one of $x_j$'s is amplified the most by $x_j^{-\alpha/\Delta}$. This is analogous to the well-know fact that, the $l_p$ norm approaches the $l_\infty$ norm (which is the maximum element of the vector), as $p\rightarrow\infty$.\\

In \cite{Report:Li_CC_min09}, we  proved the  following (closed-form) sample complexity bound  for $\hat{F}_{(\alpha),\min}$:
\begin{theorem} \cite{Report:Li_CC_min09} \label{thm_right_bound_min}
As $\Delta=1-\alpha \rightarrow 0+$, for any fixed $\epsilon >0$,
\begin{align}\label{eqn_right_bound_min}
&\mathbf{Pr}\left(\hat{F}_{(\alpha),\min} \geq (1+\epsilon)F_{(\alpha)}\right)\leq
\exp\left(k \log\frac{1}{2}\left[
\Delta+\frac{\Delta}{\log(1+\epsilon)}+\frac{\Delta}{\Delta\log\Delta+\log(1+\epsilon)}+O\left(\Delta^2\right)
 \right]\right).
\end{align}
\end{theorem}

\vspace{0.2in}

Basically, in terms of $\epsilon = \nu\Delta$, Theorem \ref{thm_right_bound_min} is applicable when $\nu$ is large ($\nu\gg1$) and $\Delta$ is small. A simulation study in \cite{Report:Li_CC_min09} demonstrated that the bound in Theorem \ref{thm_right_bound_min} can be very sharp.\\

\section{Conclusion}

Real-world data are often dynamic and can be modeled as data streams. Measuring  summary statistics of data streams such as the Shannon entropy has become an important task in many applications, for example, detecting anomaly events in large-scale networks. One line of active research is to approximate the Shannon entropy using the $\alpha$th frequency moments of the stream with $\alpha$ very close to 1 (e.g., $\Delta = 1-\alpha<10^{-4}$ or even much smaller).

Efficiently approximating the $\alpha$th frequency moments of data streams has been very heavily studied in theoretical computer science and databases. When $0<\alpha\leq2$, it is well-known that efficient $O\left(1/\epsilon^2\right)$-space algorithms exist, for example, {\em symmetric stable random projections}\cite{Article:Indyk_JACM06,Proc:Li_SODA08}, which however are impractical for estimating Shannon entropy using $\alpha$ extremely close to 1. Recently, \cite{Proc:Li_SODA09} provided an algorithm to achieve the $O\left(1/\epsilon\right)$  bound in the neighborhood of $\alpha=1$, based on the idea of {\em maximally-skewed stable random projections} (also called {\em Compressed Counting (CC)}). The algorithms provided in \cite{Proc:Li_SODA09}, however, are still impractical.\\

 In this paper, we provide a truly practical algorithm for entropy estimation. We prove that its variance is proportional to $O\left(\Delta^2\right)$ whereas previous algorithms for CC developed in \cite{Proc:Li_SODA09} have variances proportional only to $O\left(\Delta\right)$. This new algorithm leads to an $O\left(1/\nu^2\right)$ algorithm for entropy estimation to achieve $\nu$-additive accuracy, while previous algorithms must use  $O\left(1/(\nu^2\Delta^2)\right)$ samples \cite{Article:Indyk_JACM06,Proc:Li_SODA08}, or $O\left(1/(\nu^2\Delta)\right)$ samples \cite{Proc:Li_SODA09}. Note that because $\Delta$ is so small, it is no longer appropriate to treat it as ``constant.''

We also analyze the precise sample complexity bound of the proposed new estimator, both numerically (for general $0<\nu<1$) and analytically (for small $\nu$), to demonstrate that the sample complexity bound of the new estimator is free of  large constants. This further confirms that our proposed new estimator is practical.

\appendix

\section{Proof of Theorem \ref{thm_F_var}}\label{proof_thm_F_var}

As defined in (\ref{eqn_F}), the proposed estimator
\begin{align}\notag
\hat{F}_{(\alpha)} = \frac{1}{\Delta^\Delta}\left[\frac{k}{\sum_{j=1}^k x_j^{-\alpha/\Delta}}\right]^\Delta = \left[\frac{1}{\hat{J}}\right]^\Delta,  \hspace{0.5in} \hat{J} = \Delta\frac{1}{k}\sum_{j=1}^k x_j^{-\alpha/\Delta}
\end{align}
where $x_j \sim S(\alpha,\beta=1,F_{(\alpha)}\cos\left(\frac{\alpha\pi}{2}\right))$, i.i.d. Denote $J = F_{(\alpha)}^{-1/\Delta}$. According to Lemma \ref{lem_moments},
\begin{align}\notag
E(\hat{J}) = E\left(\Delta x_j^{-\alpha/\Delta}\right) = \Delta F_{(\alpha)}^{-1/\Delta}\frac{\Gamma \left(1+\frac{1}{\Delta}\right)} {\Gamma \left(1+\frac{\alpha}{\Delta}\right)} = \Delta F_{(\alpha)}^{-1/\Delta}\frac{\Gamma \left(1+\frac{1}{\Delta}\right)} {\Gamma \left(\frac{1}{\Delta}\right)} = F_{(\alpha)}^{-1/\Delta} = J.
\end{align}
%\begin{align}\notag
%E(J^n) = E\left(\Delta x_j^{-\alpha/\Delta}\right) = \Delta F_{(\alpha)}^{-1/\Delta}\frac{\Gamma \left(1+\frac{1}{\Delta}\right)} {\Gamma \left(1+\frac{\alpha}{\Delta}\right)} = \Delta F_{(\alpha)}^{-1/\Delta}\frac{\Gamma \left(1+\frac{1}{\Delta}\right)} {\Gamma \left(\frac{1}{\Delta}\right)} = F_{(\alpha)}^{-1/\Delta} = J.
%\end{align}

\begin{align}\notag
Var\left(\Delta x_j^{-\alpha/\Delta}\right) =& E\left(\Delta^2 x_j^{-2\alpha/\Delta}\right) - F_{(\alpha)}^{-2/\Delta}\\\notag
=&\Delta^2 F_{(\alpha)}^{-2/\Delta}\frac{\Gamma \left(1+\frac{2}{\Delta}\right)} {\Gamma \left(1+\frac{2\alpha}{\Delta}\right)}  - F_{(\alpha)}^{-2/\Delta}\\\notag
=&F^{-2/\Delta}_{(\alpha)} \left(\Delta^2 \frac{2}{\Delta}\left(\frac{2}{\Delta}-1\right)-1\right)\\\notag
 =& F^{-2/\Delta}_{(\alpha)} \left(3-2\Delta\right)
\end{align}
\begin{align}\notag
Var(\hat{J}) = \frac{1}{k} Var\left(\Delta x_j^{-\alpha/\Delta}\right)  = \frac{1}{k}F^{-2/\Delta}_{(\alpha)} \left(3-2\Delta\right) = \frac{1}{k}J^2(3-2\Delta).
\end{align}
A bit more algebra can show
\begin{align}\notag
E\left(\hat{J}-J\right)^3 = \frac{J^3}{k^2}\left(17-21\Delta+6\Delta^2\right).
\end{align}

Recall  $\hat{F}_{(\alpha)}  = \hat{J}^{-\Delta}$. We will basically proceed by using the ``delta'' method popular in statistics. We need to be a bit careful here as $\Delta$ is small. Just to make sure the resultant higher-order terms are indeed negligible, we carry out the algebra.

By the Taylor expansion about $J$, we obtain
\begin{align}\notag
\hat{F}_{(\alpha)} = J^{-\Delta} - \left(\hat{J}-J\right)\left(\Delta J^{-\Delta-1}\right) + \frac{(\hat{J}-J)^2}{2}\Delta(\Delta+1)J^{-\Delta-2}+...
%-\frac{(\hat{J}-J)^3}{6}\Delta(\Delta+1)(\Delta+2)J^{-\Delta-3}+...
\end{align}
Taking expectations on both sides yields,
\begin{align}\notag
E\left(\hat{F}_{(\alpha)}\right) =& J^{-\Delta} - E\left(\hat{J}-J\right)\left(\Delta J^{-\Delta-1}\right) + \frac{E(\hat{J}-J)^2}{2}\Delta(\Delta+1)J^{-\Delta-2}+...\\\notag
=&J^{-\Delta}+\frac{Var\left(\hat{J}\right)}{2}\Delta(\Delta+1)J^{-\Delta-2}+...\\\notag
=&J^{-\Delta} + \frac{1}{2k}J^2(3-2\Delta)\Delta(\Delta+1)J^{-\Delta-2}+...\\\notag
=&J^{-\Delta} + J^{-\Delta}O\left(\frac{\Delta}{k}\right)=F_{(\alpha)}\left(1+O\left(\frac{\Delta}{k}\right)\right).
\end{align}
Evaluating the higher-order moments yields
\begin{align}\notag
E\left(\hat{F}_{(\alpha)}-J^{-\Delta}\right)^2 =& E\left[
- \left(\hat{J}-J\right)\left(\Delta J^{-\Delta-1}\right) + \frac{(\hat{J}-J)^2}{2}\Delta(\Delta+1)J^{-\Delta-2}+...
\right]^2\\\notag
=&E\left[\left(\hat{J}-J\right)^2\Delta^2 J^{-2\Delta-2}\right]-
E\left[
\frac{(\hat{J}-J)^3}{2}\Delta^2(\Delta+1)J^{-2\Delta-3}\right]+...\\\notag
=&\frac{F_{(\alpha)}^2}{k}\Delta^2\left(3-2\Delta + O\left(\frac{1}{k}\right)\right),
\end{align}
and
\begin{align}\notag
Var\left(\hat{F}_{(\alpha)}\right)
=&\frac{F_{(\alpha)}^2}{k}\Delta^2\left(3-2\Delta + O\left(\frac{1}{k}\right)\right).
\end{align}

\section{Proof of Lemma \ref{lem_hm}}\label{proof_lem_hm}

The task is to show that, as $\Delta = 1-\alpha\rightarrow 0$,
\begin{align}\notag
\frac{2\Gamma^2(1+\alpha)}{\Gamma(1+2\alpha)}-1 = \Delta + \Delta^2\left(2-\frac{\pi^2}{6}\right)+O\left(\Delta^3\right).
\end{align}

Using properties of Gamma functions, for example, $\Gamma(1+x)= x\Gamma(x)$, we obtain
\begin{align}\notag
\frac{2\Gamma^2(1+\alpha)}{\Gamma(1+2\alpha)} = \frac{2\alpha^2\Gamma^2(\alpha)}{2\alpha\Gamma(2\alpha)} = \frac{\alpha\Gamma^2(\alpha)}{\Gamma(2\alpha)} = \frac{\alpha\Gamma^2(1-\Delta)}{\Gamma(2-2\Delta)} =  \frac{\alpha(-\Delta)^2\Gamma^2(-\Delta)}{(1-2\Delta)(-2\Delta)\Gamma(-2\Delta)} = \frac{(1-\Delta)\Delta}{(-2)(1-2\Delta)}\frac{\Gamma^2(-\Delta)}{\Gamma(-2\Delta)}
\end{align}
Using the infinite product representation of the Gamma function\cite[8.322]{Book:Gradshteyn_94}, we obtain
\begin{align}\notag
\frac{\Gamma^2(-\Delta)}{\Gamma(-2\Delta)} =& \frac{ \frac{e^{2\gamma_e\Delta}}{(-\Delta)^2}}{\frac{e^{2\gamma_e\Delta}}{(-2\Delta) }} \frac{
\prod_{n=1}^\infty\left(1-\frac{\Delta}{n}\right)^{-2}e^{-2\Delta/n}
}{ \prod_{n=1}^\infty\left(1-\frac{2\Delta}{n}\right)^{-1}e^{-2\Delta/n} }, \hspace{0.5in} (\gamma_e \text{ is the Euler's constant})\\\notag
=&\frac{-2}{\Delta}\prod_{n=1}^\infty \left(1-\frac{\Delta}{n}\right)^2\left(1-\frac{2\Delta}{n}\right)\\\notag
=&\frac{-2}{\Delta}\prod_{n=1}^\infty \left(1+\frac{2\Delta}{n}+\frac{3\Delta^2}{n^2}+O\left(\Delta^3\right)\right)\left(1-\frac{2\Delta}{n}\right)\\\notag
=& \frac{-2}{\Delta}\prod_{n=1}^\infty \left(1-\frac{\Delta^2}{n^2}+O\left(\Delta^3\right)\right)\\\notag
=&\frac{-2}{\Delta}\exp\left(\sum_{n=1}^\infty \log\left(1-\frac{\Delta^2}{n^2}+O\left(\Delta^3\right)\right)\right)\\\notag
=& \frac{-2}{\Delta}\exp\left(-\Delta^2\sum_{n=1}^\infty \frac{1}{n^2}+O\left(\Delta^3\right)\right)\\\notag
=&\frac{-2}{\Delta}\exp\left(-\Delta^2\frac{\pi^2}{6} +O\left(\Delta^3\right)\right)\\\notag
=& \frac{-2}{\Delta}
\left(1-\frac{\Delta^2\pi^2}{6}+O\left(\Delta^3\right)\right)
\end{align}
Therefore,
\begin{align}\notag
\frac{2\Gamma^2(1+\alpha)}{\Gamma(1+2\alpha)}-1 =& \frac{1-\Delta}{1-2\Delta}\left(1-\frac{\Delta^2\pi^2}{6} +O\left(\Delta^3\right)\right)-1\\\notag
=&\left(1-\Delta\right)\left(1+2\Delta+4\Delta^2+O\left(\Delta^3\right)\right)\left(1-\frac{\Delta^2\pi^2}{6} +O\left(\Delta^3\right)\right)-1\\\notag
=&\left(1+\Delta+2\Delta^2+O\left(\Delta^3\right)\right)\left(1-\frac{\Delta^2\pi^2}{6} +O\left(\Delta^3\right)\right)-1\\\notag
=&\Delta + \Delta^2\left(2-\frac{\pi^2}{6}\right)+O\left(\Delta^3\right).
\end{align}

\section{Proof of Lemma \ref{lem_CDF}}\label{proof_lem_CDF}

Suppose a random variable $Z \sim S\left(\alpha<1,\beta=1,\cos\left(\frac{\pi}{2}\alpha\right)\right)$.
We can show that the cumulative distribution function is
\begin{align}\notag
&F_Z(t) =  \mathbf{Pr}\left(Z\leq t\right) =\frac{1}{\pi}\int_0^\pi  \exp\left(- \frac{  \left[\sin\left(\alpha \theta \right)\right]^{\alpha/\Delta} }{t^{\alpha/\Delta} \left[\sin \theta \right]^{1/\Delta}} \sin\left( \theta\Delta\right)\right) d\theta, \hspace{0.5in} (\Delta = 1-\alpha).
\end{align}
Recall $Z = \frac{  \sin\left(\alpha V \right)}{\left[\sin V
\right]^{1/\alpha}} \left[\frac{\sin\left( V\Delta\right)}{W}
\right]^{\frac{\Delta}{\alpha}}$. $V$ is uniform in $[0, \pi]$ and $W$ is exponential with mean 1. Therefore,
\begin{align}\notag
\mathbf{Pr}\left(Z\geq t\right) =&\mathbf{Pr}\left( \frac{  \sin\left(\alpha V \right)}{\left[\sin V
\right]^{1/\alpha}} \left[\frac{\sin\left( V\Delta\right)}{W}
\right]^{\frac{\Delta}{\alpha}}  \geq t \right)\\\notag
=&\mathbf{Pr}\left(  W \leq \frac{  \left[\sin\left(\alpha V \right)\right]^{\alpha/\Delta} }{t^{\alpha/\Delta} \left[\sin V
\right]^{1/\Delta}} \sin\left( V\Delta\right)\right)\\\notag
=&\text{E}\left(\mathbf{Pr}\left( \left. W \leq \frac{  \left[\sin\left(\alpha V \right)\right]^{\alpha/\Delta} }{t^{\alpha/\Delta} \left[\sin V
\right]^{1/\Delta}} \sin\left( V\Delta\right)\right|V\right)\right)\\\notag
=&1-\text{E}\left(\exp\left(- \frac{  \left[\sin\left(\alpha V \right)\right]^{\alpha/\Delta} }{t^{\alpha/\Delta} \left[\sin V \right]^{1/\Delta}} \sin\left( V\Delta\right)\right)\right)\\\notag
=&1-\frac{1}{\pi}\int_0^\pi  \exp\left(- \frac{  \left[\sin\left(\alpha \theta \right)\right]^{\alpha/\Delta} }{t^{\alpha/\Delta} \left[\sin \theta \right]^{1/\Delta}} \sin\left( \theta\Delta\right)\right) d\theta.
\end{align}

For $\theta\in(0,\pi)$, let
\begin{align}\notag
g(\theta;\Delta) = \frac{  \left[\sin\left(\alpha \theta \right)\right]^{\alpha/\Delta} }{\left[\sin \theta \right]^{1/\Delta}} \sin\left( \theta\Delta\right).
\end{align}
It is easy to show that, as $\theta\rightarrow 0+$,
\begin{align}\notag
\lim_{\theta\rightarrow 0+} g(\theta,\Delta) =& \lim_{\theta\rightarrow 0+}\frac{  \left[\sin\left(\alpha \theta \right)\right]^{\alpha/\Delta} }{\left[\sin \theta \right]^{1/\Delta}} \sin\left( \theta\Delta\right)\\\notag
=&\lim_{\theta\rightarrow 0+} \left(\frac{\sin\left(\alpha \theta \right) }{\sin \theta} \right)^{1/\Delta} \frac{\sin\left( \theta\Delta\right)}{\sin\left(\alpha \theta \right)}\\\notag
=&\alpha^{1/\Delta}\frac{\Delta}{\alpha} = \Delta\alpha^{1/\Delta-1}.
\end{align}

The proof of the monotonicity of $g(\theta,\Delta)$ is omitted, because it is can be inferred from the proof of the convexity.

To show $g(\theta;\Delta)$ is a convex function $\theta$, it suffices to show it is log-convex. Since
\begin{align}\notag
g(\theta;\Delta) = \sin(\theta\Delta)\frac{ [\sin(\alpha\theta)]^{\alpha/\Delta}}{[\sin(\theta)]^{1/\Delta}}
=\frac{\sin(\theta\Delta)}{\sin(\alpha\theta)}\left[\frac{\sin(\alpha\theta)}{\sin(\theta)}\right]^{1/\Delta}
\end{align}
it suffices to show that both $\frac{\sin(\theta\Delta)}{\sin(\alpha\theta)}$ and $\left[\frac{\sin(\alpha\theta)}{\sin(\theta)}\right]^{1/\Delta}$ are log-convex.

\begin{align}\notag
&\frac{\partial \left[ \log \sin(\theta\Delta) - \log \sin(\alpha\theta)\right]}{\partial \theta}
=\frac{\cos(\theta\Delta)}{\sin(\theta\Delta)}\Delta - \frac{\cos(\alpha\theta)}{\sin(\alpha\theta)}\alpha
\end{align}
\begin{align}\notag
&\frac{\partial^2\left[ \log \sin(\theta\Delta) - \log \sin(\alpha\theta)\right]}{\partial \theta^2}
=-\frac{\Delta^2}{\sin^2(\theta\Delta)} + \frac{\alpha^2}{\sin^2(\alpha\theta)}
=\left(\frac{\alpha}{\sin(\alpha\theta)}-\frac{\Delta}{\sin(\theta\Delta)}\right)
\left(\frac{\alpha}{\sin(\alpha\theta)}+\frac{\Delta}{\sin(\theta\Delta)}\right)
\end{align}

\begin{align}\notag
\frac{\partial\left[ \alpha\sin(\theta\Delta)-\Delta\sin(\alpha\theta)\right]}{\partial \theta} = \Delta\alpha(\cos(\theta\Delta)-\cos(\alpha\theta))\geq 0 \hspace{0.5in} (\text{because} \ \ \Delta<0.5)
\end{align}
Therefore, $\alpha\sin(\theta\Delta)-\Delta\sin(\alpha\theta)\geq 0$ and $\frac{\sin(\theta\Delta)}{\sin(\alpha\theta)}$ is convex.

\begin{align}\notag
&\frac{\partial \left[\log \sin(\alpha\theta) - \log \sin(\theta)\right]}{\partial \theta}
=\frac{\cos(\alpha\theta)}{\sin(\alpha\theta)}\alpha - \frac{\cos(\theta)}{\sin(\theta)}
\end{align}
\begin{align}\notag
&\frac{\partial^2\left[ \log \sin(\alpha\theta) - \log \sin(\theta)\right]}{\partial \theta^2}
=-\frac{\alpha^2}{\sin^2(\alpha\theta)} + \frac{1}{\sin^2(\theta)}
=\left(\frac{1}{\sin(\theta)}-\frac{\alpha}{\sin(\alpha\theta)}\right)
\left(\frac{1}{\sin(\theta)}+\frac{\alpha}{\sin(\alpha\theta)}\right)
\end{align}

\begin{align}\notag
\frac{\partial \left[\sin(\alpha\theta)-\alpha\sin(\theta)\right]}{\partial \theta} = \alpha(\cos(\alpha\theta)-\cos(\theta))\geq 0 \hspace{0.5in} (\text{because} \ \alpha = 1-\Delta>0.5)
\end{align}
Therefore,  we have proved the convexity of $g\left(\theta;\Delta\right)$.

\section{Proof of Theorem \ref{thm_MLE}}\label{proof_thm_MLE}

Given $k$ i.i.d. samples $x_j = cY_j$, the task is to estimate $c^\alpha$ using MLE. The CDF of $Y_j$ is given by
\begin{align}\notag
F_Y(t) = \mathbf{Pr}\left(Y\leq t\right) = \exp\left(-t^{-\alpha/\Delta}\Delta\alpha^{\alpha/\Delta}\right),  \hspace{0.2in} t \in [0, \infty).
\end{align}
By taking derivatives, the density function of $x_j$ is given by
\begin{align}\notag
f_X(t) = \frac{1}{c}f_Y(t/c) = c^{\alpha/\Delta}F_Z(t/c)\alpha^{1/\Delta} t^{-1/\Delta},
\end{align}
because
\begin{align}\notag
f_Z(t) = F_Z(t) (-\Delta)\left(-\alpha/\Delta\right)\alpha^{\alpha/\Delta} t^{-\alpha/\Delta-1}  = F_Z(t) \alpha^{1/\Delta} t^{-1/\Delta}.
\end{align}
Solving the MLE equation,
\begin{align}\notag
\sum_{j=1}^k \frac{\partial \log f_X(x_j)}{c^\alpha} = 0
\end{align}
we obtain
\begin{align}\notag
c^\alpha = \frac{1}{\Delta^\Delta\alpha^\alpha} \left[\frac{k}{\sum_{j=1}^k x_j^{-\alpha/\Delta}}\right]^\Delta
\end{align}

\section{Proof of Theorem \ref{thm_tail}}\label{proof_thm_tail}

From the previous results, we know
\begin{align}\notag
&\hat{F}_{(\alpha)} = \frac{1}{\Delta^\Delta}\left[\frac{k}{\sum_{j=1}^k x_j^{-\alpha/\Delta}}\right]^\Delta,\\\notag
&x_j \sim S\left(\alpha, \beta=1, \cos\left(\frac{\pi}{2}\alpha\right)F_{(\alpha)}\right),\\\notag
&E\left(x_j^\lambda\right) = F_{(\alpha)}^{\lambda/\alpha} \frac{ \Gamma\left(1-\frac{\lambda}{\alpha}\right) }
{\Gamma\left(1-\lambda\right)},\\\notag
&E\left(\frac{x_j^{-n\alpha/\Delta}}{F_{(\alpha)}^{-n/\Delta}}\right) = \frac{\Gamma\left(1+\frac{n}{\Delta}\right)}{\Gamma\left(1+\frac{n\alpha}{\Delta}\right)}.
\end{align}

We first study the right tail bound.
\begin{align}\notag
&\mathbf{Pr}\left(\hat{F}_{(\alpha)} \geq (1+\epsilon)F_{(\alpha)}\right)\\\notag
=& \mathbf{Pr}
\left( \frac{1}{\Delta^\Delta}\left[\frac{k}{\sum_{j=1}^k x_j^{-\alpha/\Delta}}\right]^\Delta     \geq (1+\epsilon)F_{(\alpha)}\right)\\\notag
=&\mathbf{Pr}\left(\sum_{j=1}^k x_j^{-\alpha/\Delta} \leq \frac{k}{(1+\epsilon)^{1/\Delta} \Delta F_{(\alpha)}^{1/\Delta}}\right)\\\notag
=&\mathbf{Pr}\left(-t\sum_{j=1}^k\frac{ x_j^{-\alpha/\Delta}}{F_{(\alpha)}^{-1/\Delta}} \geq -t\frac{k}{(1+\epsilon)^{1/\Delta} \Delta }\right) \hspace{0.5in} (\text{any  } t>0)\\\notag
\leq&E\left(\exp\left(-t\sum_{j=1}^k\frac{ x_j^{-\alpha/\Delta}}{F_{(\alpha)}^{-1/\Delta}}\right)\right)\exp\left(
t\frac{k}{(1+\epsilon)^{1/\Delta} \Delta }\right)\\\notag
=&E^k\left(\exp\left(-t\frac{ x_j^{-\alpha/\Delta}}{F_{(\alpha)}^{-1/\Delta}}\right)\right)\exp\left(
t\frac{k}{(1+\epsilon)^{1/\Delta} \Delta }\right)\\\notag
%=&E^k\left(\exp\left(-t\frac{ x_j^{-\alpha/\Delta}}{F_{(\alpha)}^{-1/\Delta}}\right)\right)\exp\left(
%t\frac{k}{(1+\epsilon)^{1/\Delta} \Delta }\right)\\\notag
=&E^k\left(\sum_{n=0}^\infty \frac{(-t)^n}{n!} \left(\frac{ x_j^{-\alpha/\Delta}}{F_{(\alpha)}^{-1/\Delta}}\right)^n \right)\exp\left(
t\frac{k}{(1+\epsilon)^{1/\Delta} \Delta }\right)\\\notag
=&\left(\sum_{n=0}^\infty \frac{(-t)^n}{n!} \frac{\Gamma\left(1+\frac{n}{\Delta}\right)}{\Gamma\left(1+\frac{n\alpha}{\Delta}\right)} \right)^k\exp\left(
t\frac{k}{(1+\epsilon)^{1/\Delta} \Delta }\right)\\\notag
=&\exp\left(k\left( \log \sum_{n=0}^\infty \frac{(-t)^n}{n!} \frac{\Gamma\left(1+\frac{n}{\Delta}\right)}{\Gamma\left(1+\frac{n\alpha}{\Delta}\right)} +
\frac{t}{(1+\epsilon)^{1/\Delta} \Delta }\right)\right)
\end{align}

We can choose the optimal $t$ to minimize this upper bound. Thus,
\begin{align}\notag
&\mathbf{Pr}\left(\hat{F}_{(\alpha)} \geq (1+\epsilon)F_{(\alpha)}\right) \leq \exp\left(-k\frac{\epsilon^2}{G_R}\right)\\\notag
&\frac{\epsilon^2}{G_R} = - \left( \log \sum_{n=0}^\infty \frac{(-t_R)^n}{n!} \frac{\Gamma\left(1+\frac{n}{\Delta}\right)}{\Gamma\left(1+\frac{n\alpha}{\Delta}\right)} +
\frac{t_R}{(1+\epsilon)^{1/\Delta} \Delta }\right)
\end{align}
where $t_R$ is the solution to
\begin{align}\notag
\frac{\sum_{n=1}^\infty \frac{(-1)^n(t_R)^{n-1}}{(n-1)!} \frac{\Gamma\left(1+\frac{n}{\Delta}\right)}{\Gamma\left(1+\frac{n\alpha}{\Delta}\right)}}
{\sum_{n=0}^\infty \frac{(-t_R)^n}{n!} \frac{\Gamma\left(1+\frac{n}{\Delta}\right)}{\Gamma\left(1+\frac{n\alpha}{\Delta}\right)}
} +
\frac{1}{(1+\epsilon)^{1/\Delta} \Delta } = 0
\end{align}

Now, we look into the left tail bound.
\begin{align}\notag
&\mathbf{Pr}\left(\hat{F}_{(\alpha)} \leq (1-\epsilon)F_{(\alpha)}\right)\\\notag
=& \mathbf{Pr}
\left( \frac{1}{\Delta^\Delta}\left[\frac{k}{\sum_{j=1}^k x_j^{-\alpha/\Delta}}\right]^\Delta     \leq (1-\epsilon)F_{(\alpha)}\right)\\\notag
=&\mathbf{Pr}\left(\sum_{j=1}^k x_j^{-\alpha/\Delta} \geq \frac{k}{(1-\epsilon)^{1/\Delta} \Delta F_{(\alpha)}^{1/\Delta}}\right)\\\notag
=&\mathbf{Pr}\left(t\sum_{j=1}^k\frac{ x_j^{-\alpha/\Delta}}{F_{(\alpha)}^{-1/\Delta}} \geq t\frac{k}{(1-\epsilon)^{1/\Delta} \Delta }\right) \hspace{0.5in} (\text{any  } t>0)\\\notag
\leq&E\left(\exp\left(t\sum_{j=1}^k\frac{ x_j^{-\alpha/\Delta}}{F_{(\alpha)}^{-1/\Delta}}\right)\right)\exp\left(
-t\frac{k}{(1-\epsilon)^{1/\Delta} \Delta }\right)\\\notag
=&E^k\left(\exp\left(t\frac{ x_j^{-\alpha/\Delta}}{F_{(\alpha)}^{-1/\Delta}}\right)\right)\exp\left(
-t\frac{k}{(1-\epsilon)^{1/\Delta} \Delta }\right)\\\notag
%=&E^k\left(\exp\left(t\frac{ x_j^{-\alpha/\Delta}}{F_{(\alpha)}^{-1/\Delta}}\right)\right)\exp\left(
%-t\frac{k}{(1-\epsilon)^{1/\Delta} \Delta }\right)\\\notag
=&E^k\left(\sum_{n=0}^\infty \frac{t^n}{n!} \left(\frac{ x_j^{-\alpha/\Delta}}{F_{(\alpha)}^{-1/\Delta}}\right)^n \right)\exp\left(
-t\frac{k}{(1-\epsilon)^{1/\Delta} \Delta }\right)\\\notag
=&\left(\sum_{n=0}^\infty \frac{t^n}{n!} \frac{\Gamma\left(1+\frac{n}{\Delta}\right)}{\Gamma\left(1+\frac{n\alpha}{\Delta}\right)} \right)^k\exp\left(
-t\frac{k}{(1-\epsilon)^{1/\Delta} \Delta }\right)\\\notag
=&\exp\left(k\left( \log \sum_{n=0}^\infty \frac{t^n}{n!} \frac{\Gamma\left(1+\frac{n}{\Delta}\right)}{\Gamma\left(1+\frac{n\alpha}{\Delta}\right)} -
\frac{t}{(1-\epsilon)^{1/\Delta} \Delta }\right)\right)
\end{align}

Again, we can choose the optimal $t = t_L$ to minimize this upper bound. Thus,
\begin{align}\notag
&\mathbf{Pr}\left(\hat{F}_{(\alpha)} \leq (1-\epsilon)F_{(\alpha)}\right) \leq \exp\left(-k\frac{\epsilon^2}{G_L}\right)\\\notag
&\frac{\epsilon^2}{G_L} = -  \log \sum_{n=0}^\infty \frac{(t_L)^n}{n!} \frac{\Gamma\left(1+\frac{n}{\Delta}\right)}{\Gamma\left(1+\frac{n\alpha}{\Delta}\right)}+
\frac{t_L}{(1-\epsilon)^{1/\Delta} \Delta}
\end{align}
where $t_L$ is the solution to
\begin{align}\notag
- \frac{\sum_{n=1}^\infty \frac{(t_L)^{n-1}}{(n-1)!} \frac{\Gamma\left(1+\frac{n}{\Delta}\right)}{\Gamma\left(1+\frac{n\alpha}{\Delta}\right)}}
{\sum_{n=0}^\infty \frac{(t_L)^n}{n!} \frac{\Gamma\left(1+\frac{n}{\Delta}\right)}{\Gamma\left(1+\frac{n\alpha}{\Delta}\right)}
}+
\frac{1}{(1-\epsilon)^{1/\Delta} \Delta } = 0
\end{align}

\section{Proof of Lemma \ref{lem_G->0}}\label{proof_lem_G->0}

We have derived $\frac{\epsilon^2}{G_R}$ and $\frac{\epsilon^2}{G_L}$ in Theorem \ref{thm_tail}. The task of this Lemma is to show that, as $\nu\rightarrow 0$,
\begin{align}\notag
\frac{G_R}{\Delta^2} \rightarrow 6-4\Delta, \hspace{0.5in} \ \  \frac{G_L}{\Delta^2} \rightarrow 6-4\Delta.
\end{align}
To proceed with the proof, we first assume that, as $\nu\rightarrow0$, we have
\begin{align}\notag
t_R\frac{\Delta}{e} = s_R = O\left(\nu\right), \hspace{0.5in} t_L\frac{\Delta}{e}=s_L = O\left(\nu\right),
\end{align}
which can be later verified.  With this assumption, we can expand $\frac{\epsilon^2}{G_L}$:
\begin{align}\notag
\frac{\epsilon^2}{G_L} =& - \log\left(1+\sum_{n=1}^\infty \left(s_L\right)^n \prod_{j=0}^{n-1} \frac{n-j\Delta}{(n-j)e}\right) + \frac{s_L}{e(1-\epsilon)^{1/\Delta}},\\\notag
=&-\log\left(1+s_L/e  + s_L^2 \frac{2-\Delta}{e^2}+...\right) +  \frac{s_L}{e(1-\epsilon)^{1/\Delta}}\\\notag
=&-\left( s_L/e  + s_L^2 \frac{2-\Delta}{e^2}+... - \frac{1}{2}\left(s_L/e  + s_L^2 \frac{2-\Delta}{e^2}+...\right)^2\right) + \frac{s_L}{e(1-\epsilon)^{1/\Delta}}\\\notag
=&-\left(s_L/e+ s_L^2 \frac{2-\Delta}{e^2} - \frac{s_L^2}{2e^2}+...\right)+ \frac{s_L}{e(1-\epsilon)^{1/\Delta}}\\\notag
=&-\left(s_L/e+ s_L^2 \frac{3-2\Delta}{2e^2}+...\right)+ \frac{s_L}{e(1-\epsilon)^{1/\Delta}}\\\notag
\end{align}
Setting the first derivative to zero,
\begin{align}\notag
-\frac{1}{e} - s_L\frac{3-2\Delta}{e^2}+ \frac{1}{e(1-\nu\Delta)^{1/\Delta}} = 0
\end{align}
we obtain
\begin{align}\notag
s_L = \frac{1}{e}\left(\frac{1}{(1-\nu\Delta)^{1/\Delta}} -1 \right) \frac{e^2}{3-2\Delta} = \left(\frac{1}{(1-\nu\Delta)^{1/\Delta}} -1 \right) \frac{e}{3-2\Delta} = \left(\nu + \frac{\nu^2}{2}(1+\Delta)\right) \frac{e}{3-2\Delta} + O\left(\nu^3\right),
\end{align}
which verifies that $s_L=t_L\frac{\Delta}{e}$ is indeed on the order of $\nu$. Therefore,
\begin{align}\notag
\frac{\epsilon^2}{G_L} =& - s_L/e  - s_L^2 \frac{3-2\Delta}{2e^2} + \frac{s_L}{e(1-\nu\Delta)^{1/\Delta}} + ...\\\notag
=&-\left(\nu + \frac{\nu^2}{2}(1+\Delta)\right) \frac{1}{3-2\Delta} - \left(\nu + \frac{\nu^2}{2}(1+\Delta)\right)^2 \frac{1}{2(3-2\Delta)}\\\notag
&+\left(\nu+\frac{\nu^2}{2}(1+\Delta) \right) \frac{1}{3-2\Delta} \left(1+\nu+\frac{\nu^2}{2}(1+\Delta)\right)+ O\left(\nu^3\right)\\\notag
=&\frac{1}{3-2\Delta} \left[
-\nu - \frac{\nu^2}{2}(1+\Delta) - \frac{\nu^2}{2} + \nu+\frac{\nu^2}{2}(1+\Delta)+\nu^2
\right] + O\left(\nu^3\right)\\\notag
=&\frac{\nu^2}{6-4\Delta} + O\left(\nu^3\right).
\end{align}
Thus, we have proved that $\frac{G_L}{\Delta^2}\rightarrow 6-4\Delta$ as $\nu\rightarrow 0$. A similar procedure can also prove  $\frac{G_R}{\Delta^2}\rightarrow 6-4\Delta$.

\clearpage

\section{Experiments}\label{app_exp}

This section demonstrates that the proposed estimator $\hat{F}_{(\alpha)}$ in (\ref{eqn_F}) for Compressed Counting (CC) is a truly practical algorithm, while the previously proposed {\em geometric mean} algorithm\cite{Proc:Li_SODA09} for CC is inadequate for entropy estimation. We also demonstrate that algorithms based on {\em symmetric stable random projections} \cite{Article:Indyk_JACM06,Proc:Li_SODA09} are not suitable for entropy estimation.

\subsection{Data}

Since the estimation accuracy is what we are interested in, we can simply use static data instead of real data streams. This is because the projected data vector $X = \mathbf{R}^\text{T} A_t$ is the same at the end of the stream (i.e., time $t$), regardless whether it is computed at once  (i.e., static) or incrementally (i.e., dynamic).

Eight English words are selected from a chunk of Web crawl data. The words are selected fairly randomly, although we make sure they cover a whole range of data sparsity, from function words (e.g., ``A''), to  common words (e.g., ``FRIDAY'') to rare words (e.g., ``TWIST'').  Thus, as summarized in Table \ref{tab_data},  our data set consists of 8 vectors and the entries are the numbers of word occurrences in each document.

\begin{table}[h]
\caption{\small  The data set consists of 8 English words selected from a corpus of Web pages, forming 8 vectors  whose values are the word occurrences. The table lists their fractions of non-zeros (sparsity) and the Shannon entropies ($H$).
 }
\begin{center}{
\begin{tabular}{l l l l l l l}
\hline \hline\\
Word &Sparsity  & Entropy $H$\\  \\\hline
TWIST &0.004 &5.4873 \\
%RICE &490 &5.4474 &5.4997 &5.3937 &6.3302 &4.7276\\
FRIDAY &0.034 &7.0487\\
FUN &0.047 & 7.6519\\
BUSINESS &0.126 &8.3995\\
NAME & 0.144 &8.5162 \\
HAVE & 0.267  &8.9782\\
THIS & 0.423  &9.3893 \\
A    & 0.596  &9.5463\\
%THE  & 42754  & 9.4231 &9.4828  &9.3641  &12.133  &7.4775\\
\hline\hline
\end{tabular}
}
\end{center}
\label{tab_data}
\end{table}

\vspace{-0.3in}
\subsection{Estimating Frequency Moments}

We estimate the $\alpha$th frequency moments , for $\Delta = 1-\alpha= 0.2$,\ 0.1,\  ..., \ $10^{-16}$, using the proposed new estimator $\hat{F}_{(\alpha)}$ and the {\em geometric mean} estimator $\hat{F}_{(\alpha),gm}$, as well as the {\em geometric mean} estimator for {\em symmetric stable random projections} proposed in \cite{Proc:Li_SODA08}.  Recall
\begin{align}\notag
&\hat{F}_{(\alpha)} = \frac{1}{\Delta^\Delta} \left[\frac{k}{\sum_{j=1}^k x_j^{-\alpha/\Delta}}\right]^\Delta,\hspace{0.5in}
\hat{F}_{(\alpha),gm} = \left[\frac{\Gamma\left(1-\frac{\alpha}{k}\right)}{\Gamma\left(1-\frac{1}{k}\right)}\right]^k \prod_{j=1}^k x_j^{\alpha/k}.
\end{align}

\begin{figure}[h]
\begin{center}\mbox{
\includegraphics[width=2.25in]{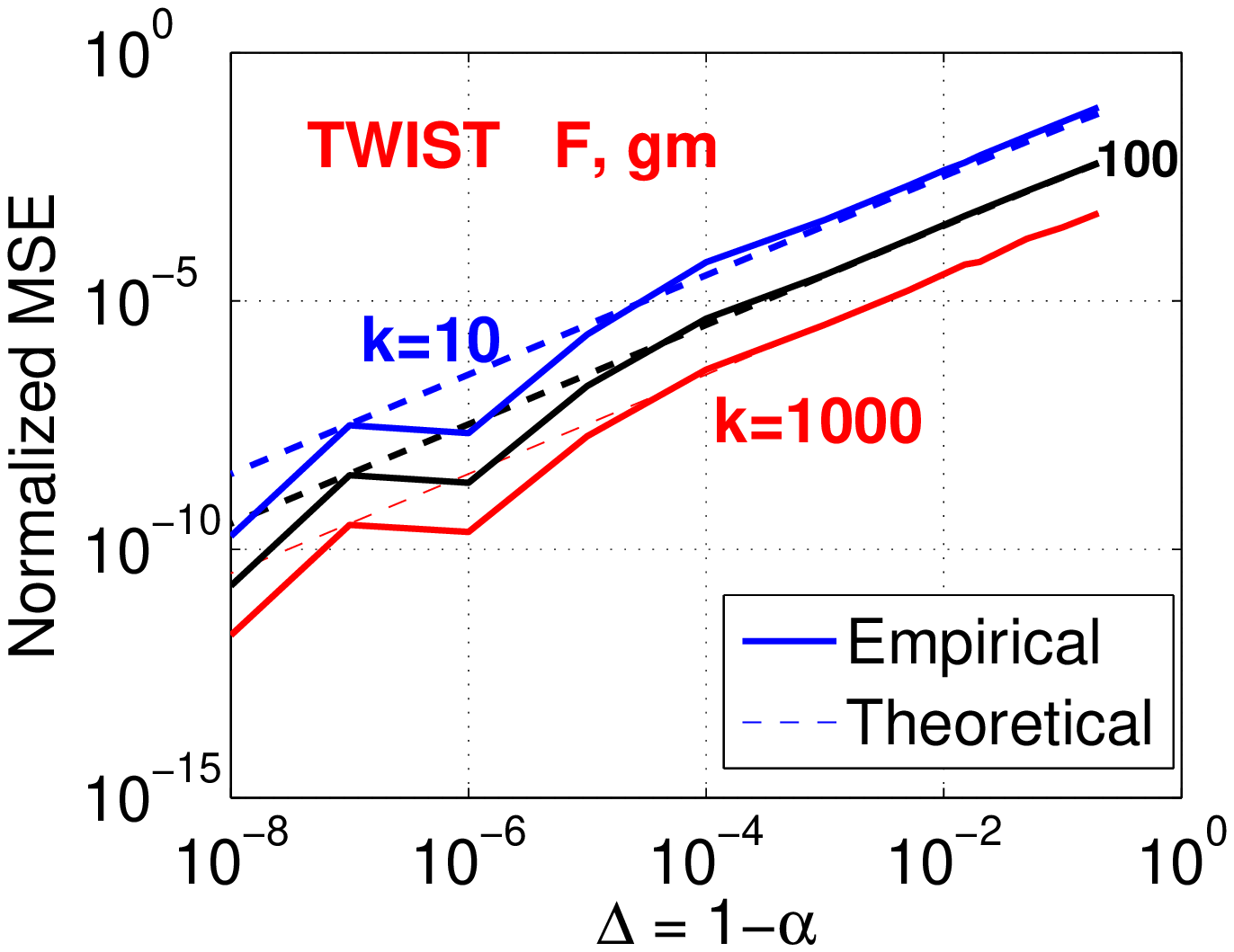}\hspace{-0.16in}
\includegraphics[width=2.25in]{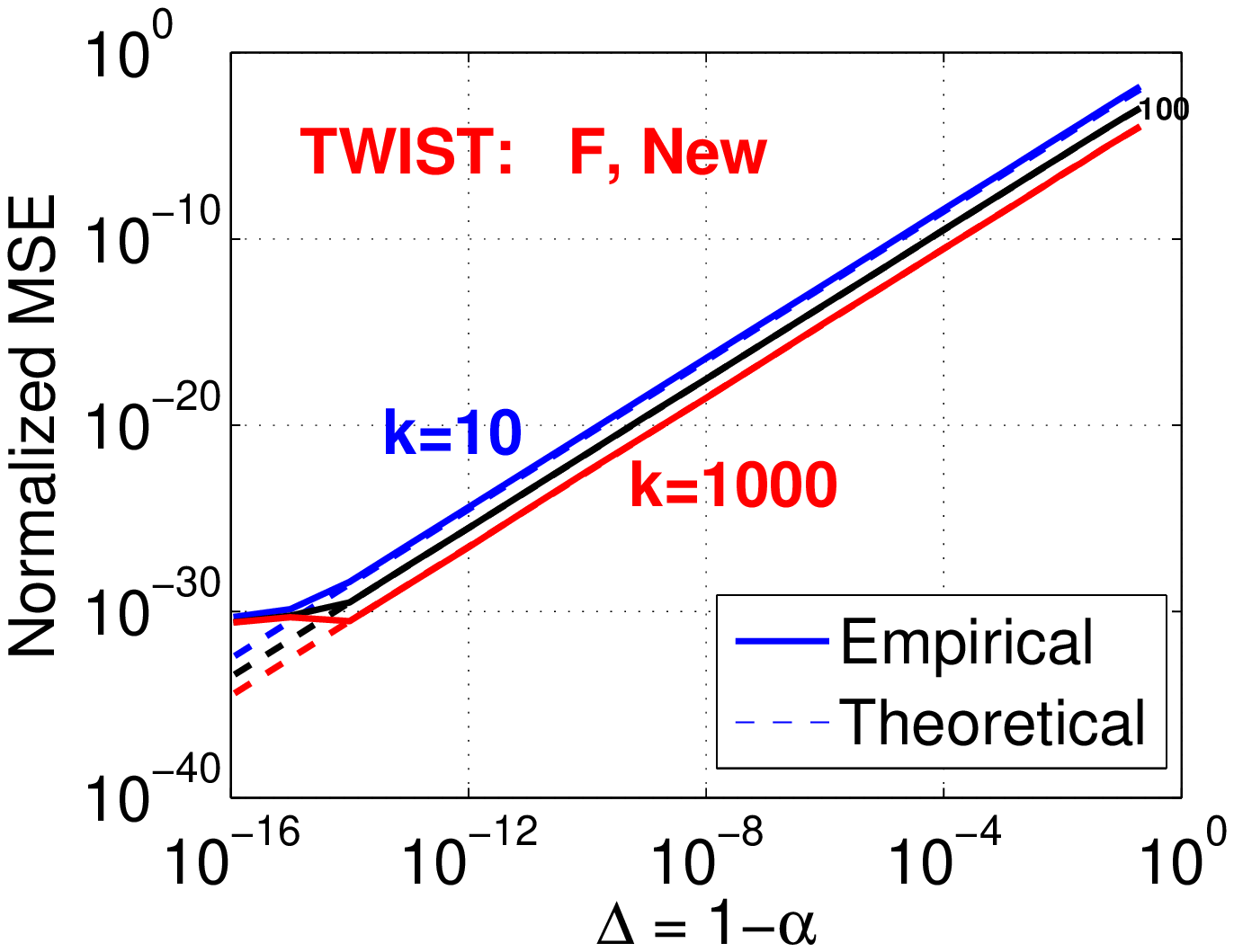}\hspace{-0.16in}
\includegraphics[width=2.25in]{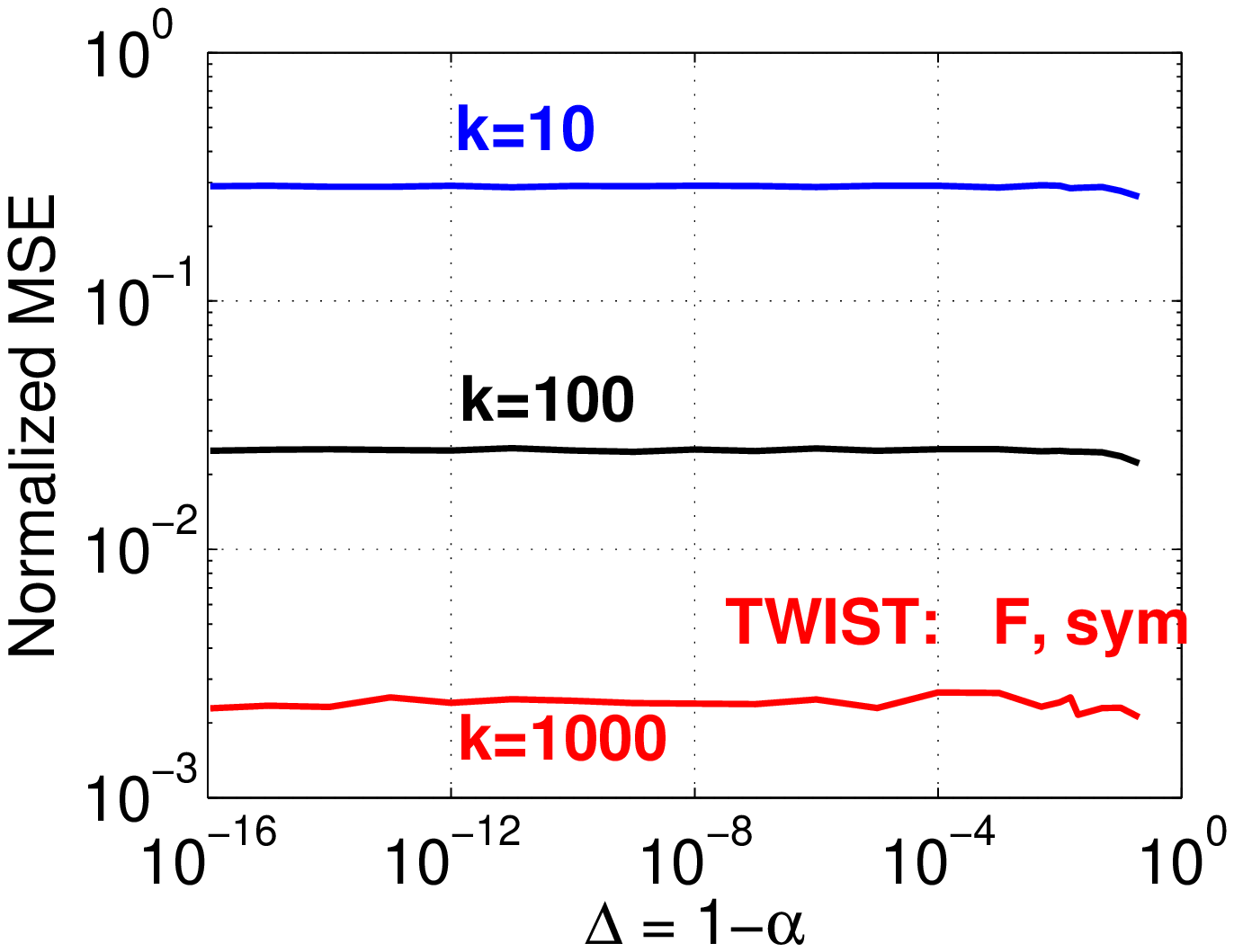}
}
\end{center}
\vspace{-0.2in}
\caption{Normalized MSEs for estimating the  $\alpha$th frequency moments using the {\em geometric mean} estimator $\hat{F}_{(\alpha),gm}$ (left panel) and the proposed new estimator $\hat{F}_{(\alpha)}$ (middle panel) for CC, together with the {\em geometric mean} estimator (right panel) for {\em symmetric stable random projections}. For  $\hat{F}_{(\alpha),gm}$ and $\hat{F}_{(\alpha)}$, we also plot their theoretical variances (dashed curves), which largely overlap the empirical MSEs whenever the algorithms are numerically stable. \hspace{0.1in} The proposed new estimator $\hat{F}_{(\alpha)}$ is numerically very stable even when $\Delta=10^{-14}$. In comparison,  $\hat{F}_{(\alpha),gm}$ is not stable  if $\Delta<10^{-5}$.  We present results at the sample sizes $k=10$, 100, and 1000. }\label{fig_TWIST_F}
\end{figure}

We find $\hat{F}_{(\alpha)}$ is numerically very stable, if we express it as
\begin{align}\notag
\hat{F}_{(\alpha)} = \frac{1}{\Delta^\Delta} \left[\frac{k}{\sum_{j=1}^k \exp\left(-\frac{\alpha}{\Delta} \log\frac{x_j}{F_{(1)}}\right)}\right]^\Delta \times F_{(1)}^\alpha,
\end{align}
where $F_{(1)}$, the first moment, can be computed exactly.  Using Matlab (the 32-bit version), we find no numerical problems with $\hat{F}_{(\alpha})$ even for very small $\Delta$ (e.g., $\Delta = 10^{-14}$; see Figure \ref{fig_TWIST_F}).\\

However, we could not find a numerically very stable implementation of the {\em geometric mean} estimator $\hat{F}_{(\alpha),gm}$, when $\Delta<10^{-5}$. We  tried a variety of ways (including the tricks in implementing $\hat{F}_{(\alpha)}$) to implement $\hat{F}_{(\alpha),gm}$ and the Gamma functions (e.g., using ``gammaln'' instead of ``gamma'' in Matlab). Fortunately, we believe $\Delta=10^{-5}$ is sufficiently small for comparing the two estimators.

We experiment with three $k$ values: 10, 100, and 1000; and we present the estimation errors in terms of the normalized  mean square errors (MSE, normalized by the square of the true values). As $\Delta$ decreases, the MSEs  for the {\em symmetric stable random projections} (in the right panel of Figure \ref{fig_TWIST_F}) are roughly flat, verifying that algorithms based on {\em symmetric stable random projections} do not capture the fact that the first moment ($\alpha=1$) should be a trivial problem.\\

Using Compressed Counting (CC), the {\em geometric mean} estimator, $\hat{F}_{(\alpha),gm}$ (in the left panel of Figure \ref{fig_TWIST_F}), and proposed new estimator, $\hat{F}_{(\alpha)}$ (in the middle panel), clearly exhibit the desired property that the MSEs decrease as $\Delta$ decreases. Of course, as expected,  $\hat{F}_{(\alpha)}$, has a much faster  rate of decreasing than $\hat{F}_{(\alpha),gm}$; the latter is also numerically much less stable when $\Delta<10^{-5}$.

\subsection{Estimating Shannon Entropies}

\begin{figure}[h]
\begin{center}

\mbox{
\includegraphics[width=2.25in]{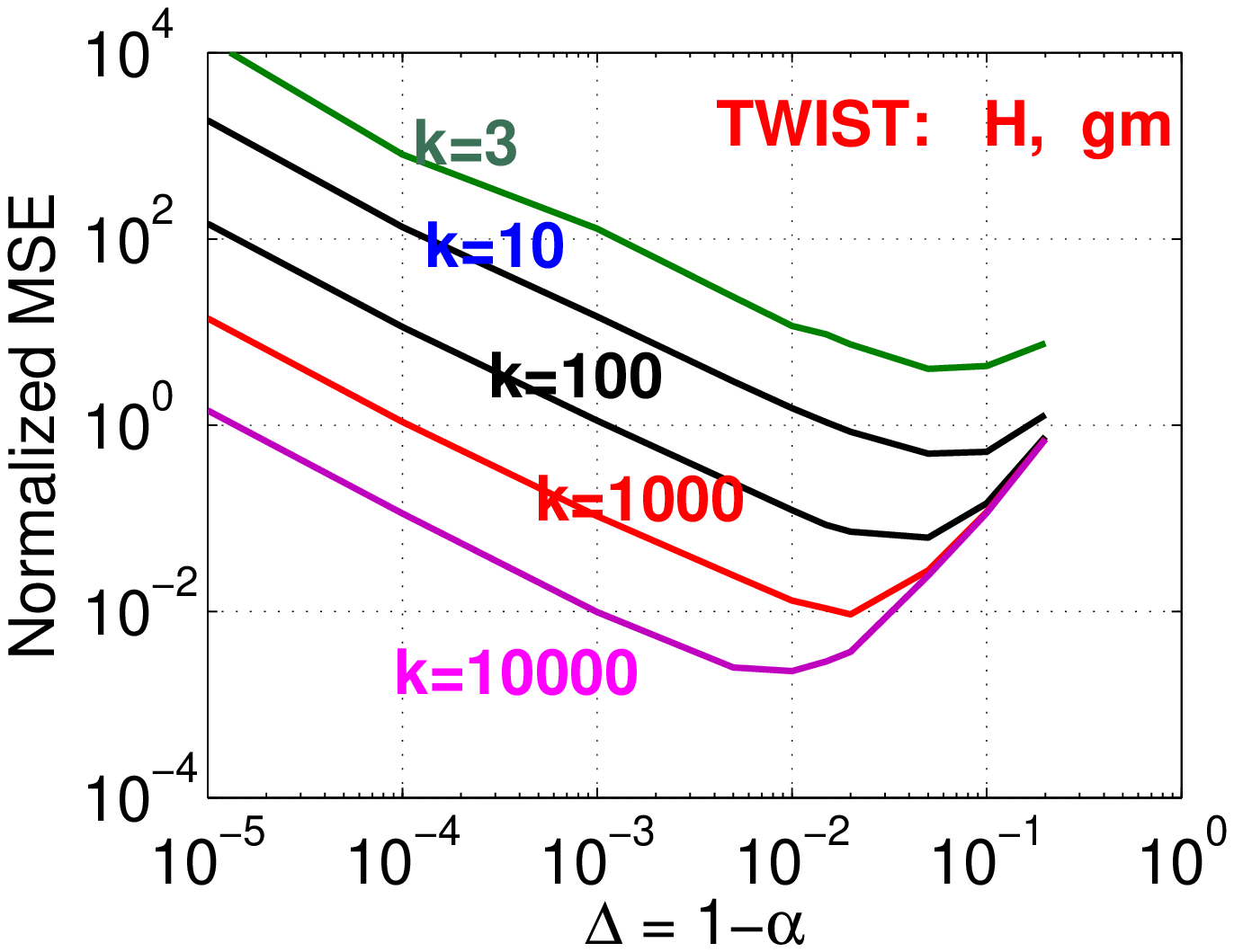}\hspace{-0.16in}
\includegraphics[width=2.25in]{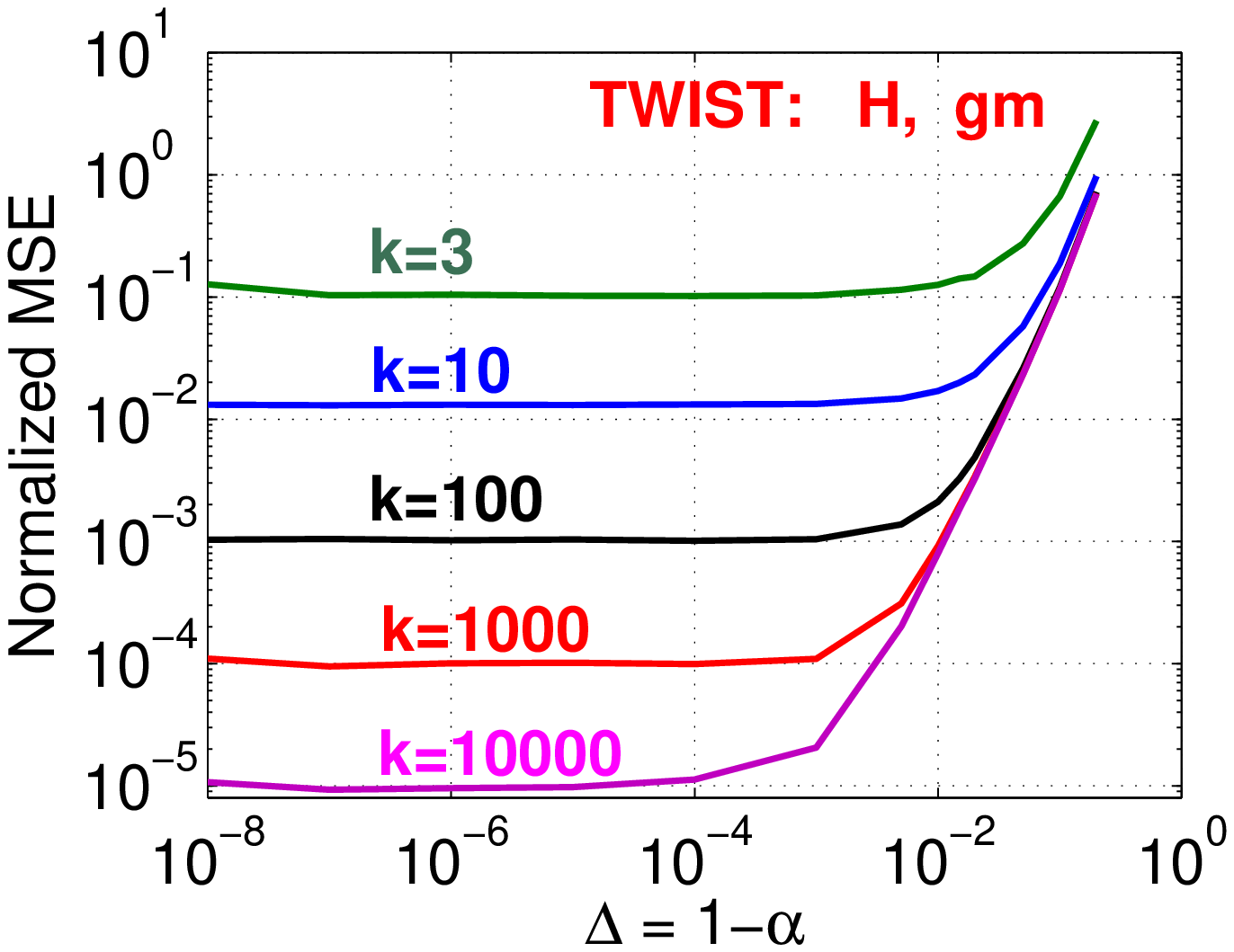}\hspace{-0.16in}
\includegraphics[width=2.25in]{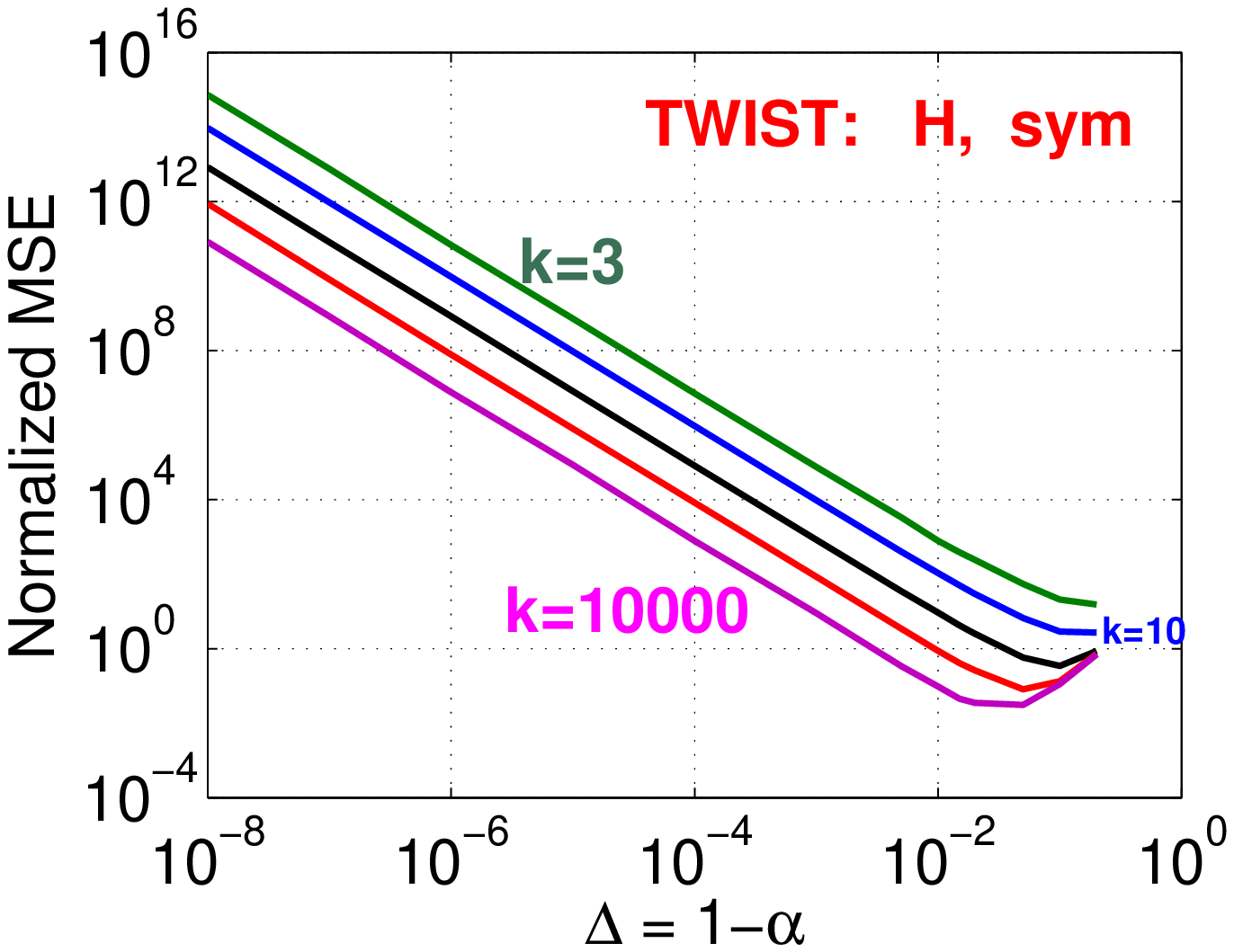}}

\mbox{
\includegraphics[width=2.25in]{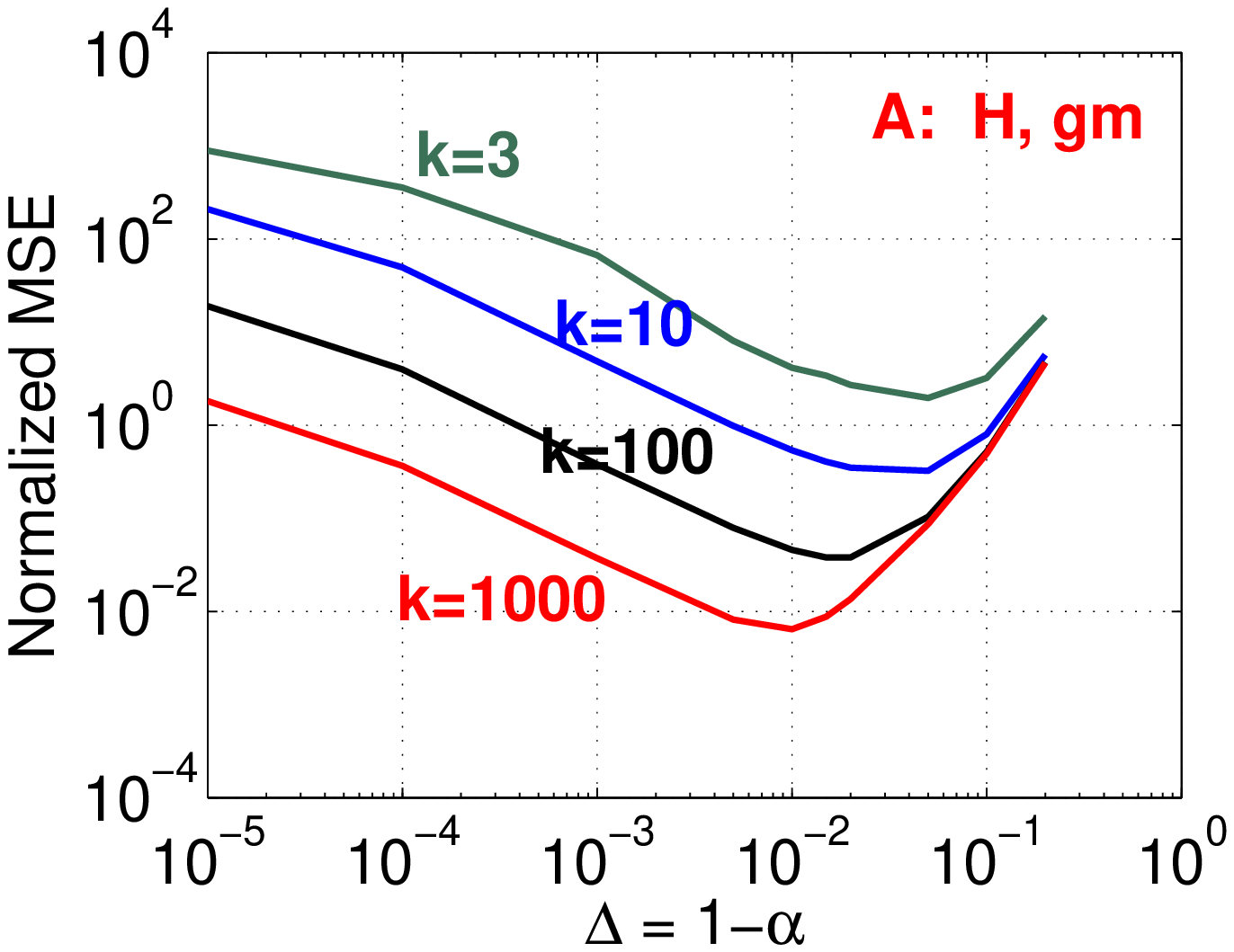}\hspace{-0.16in}
\includegraphics[width=2.25in]{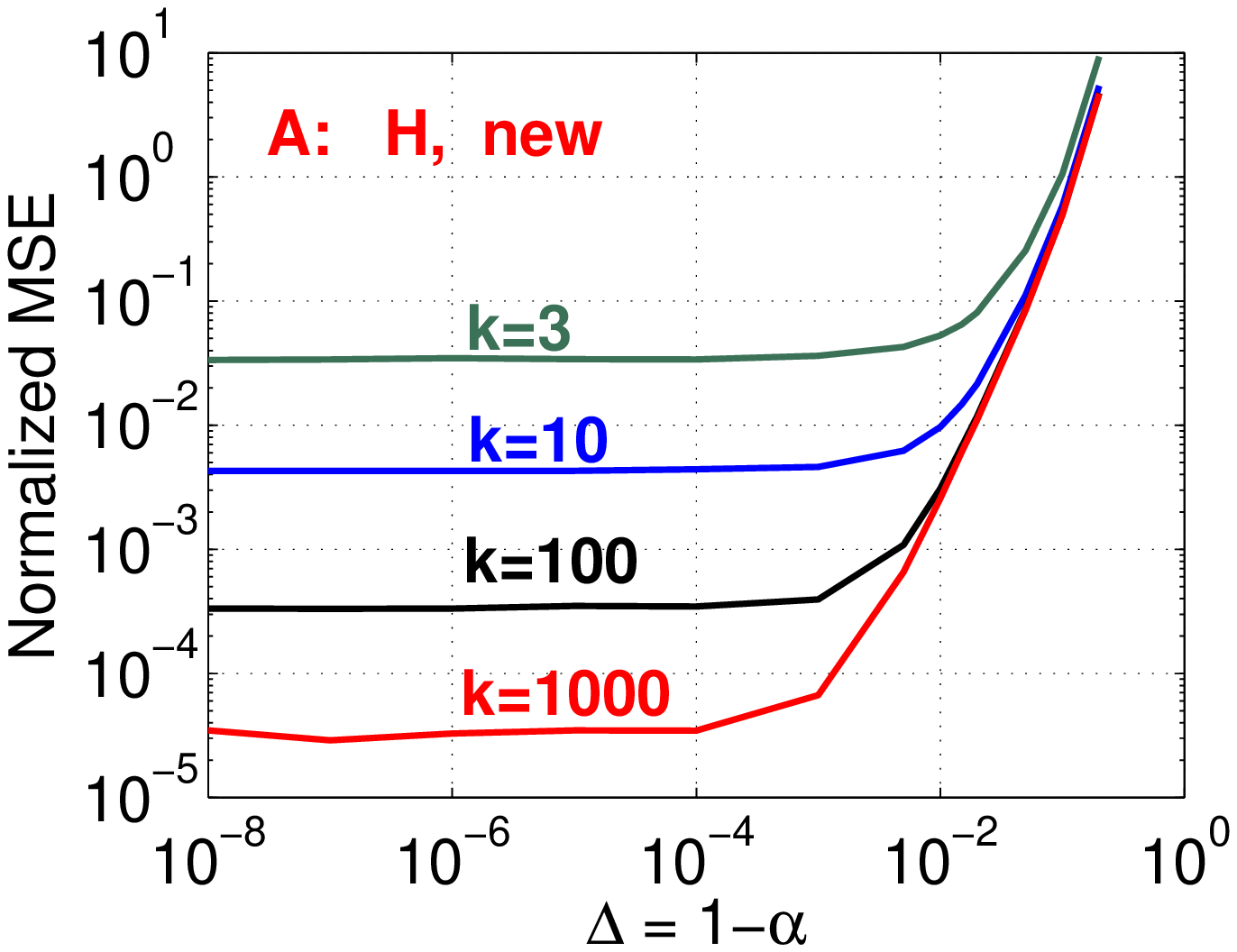}\hspace{-0.16in}
\includegraphics[width=2.25in]{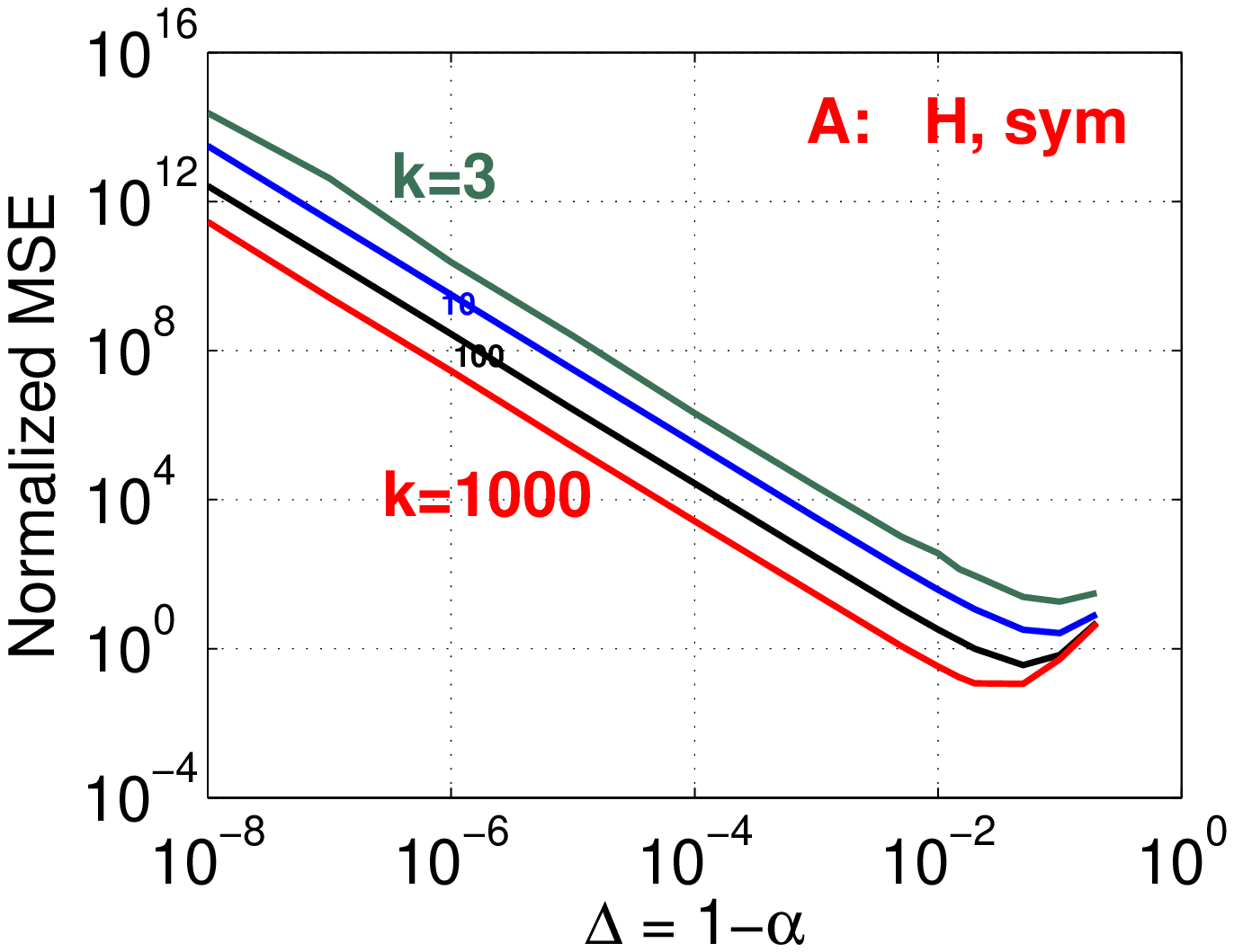}}

\mbox{
\includegraphics[width=2.25in]{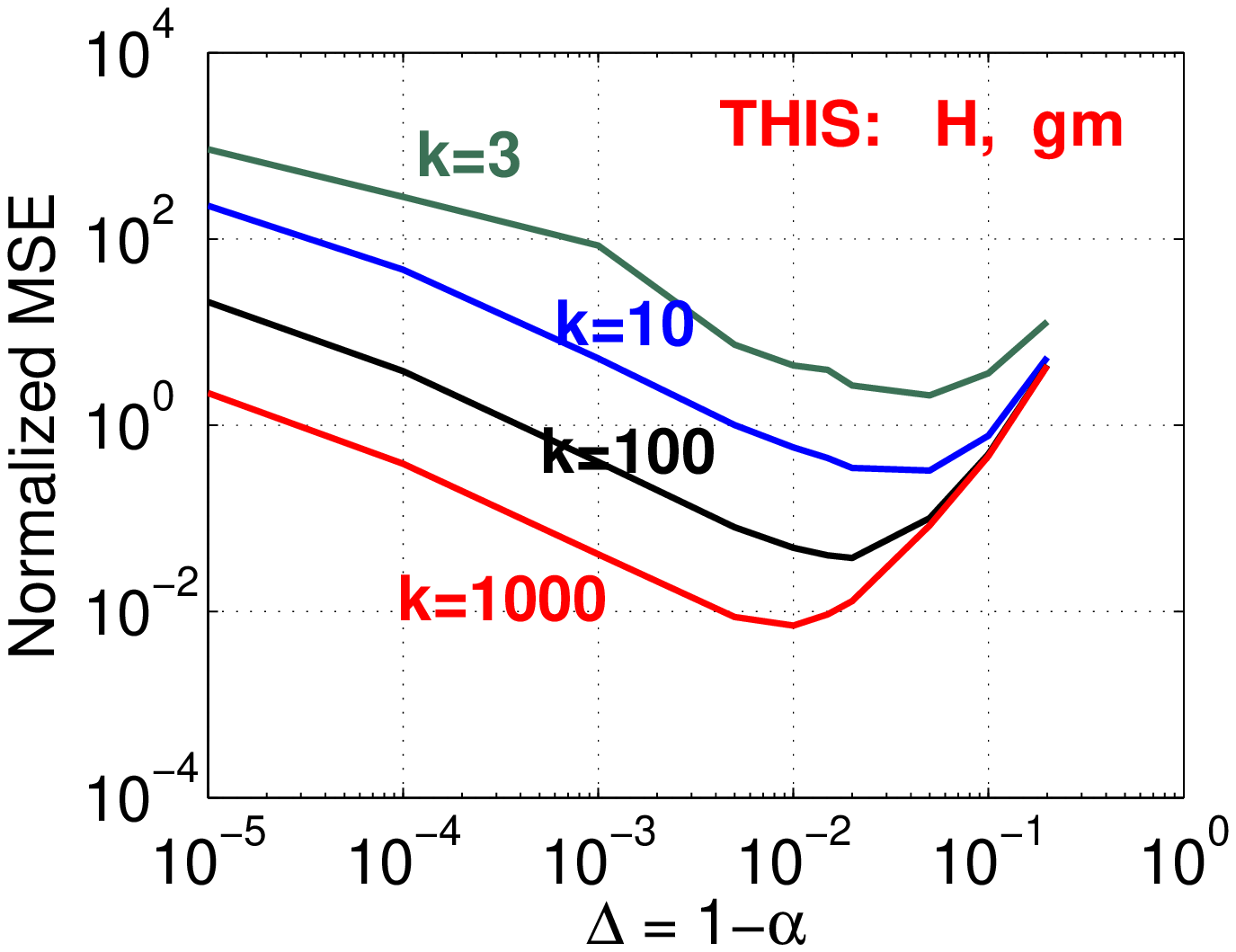}\hspace{-0.16in}
\includegraphics[width=2.25in]{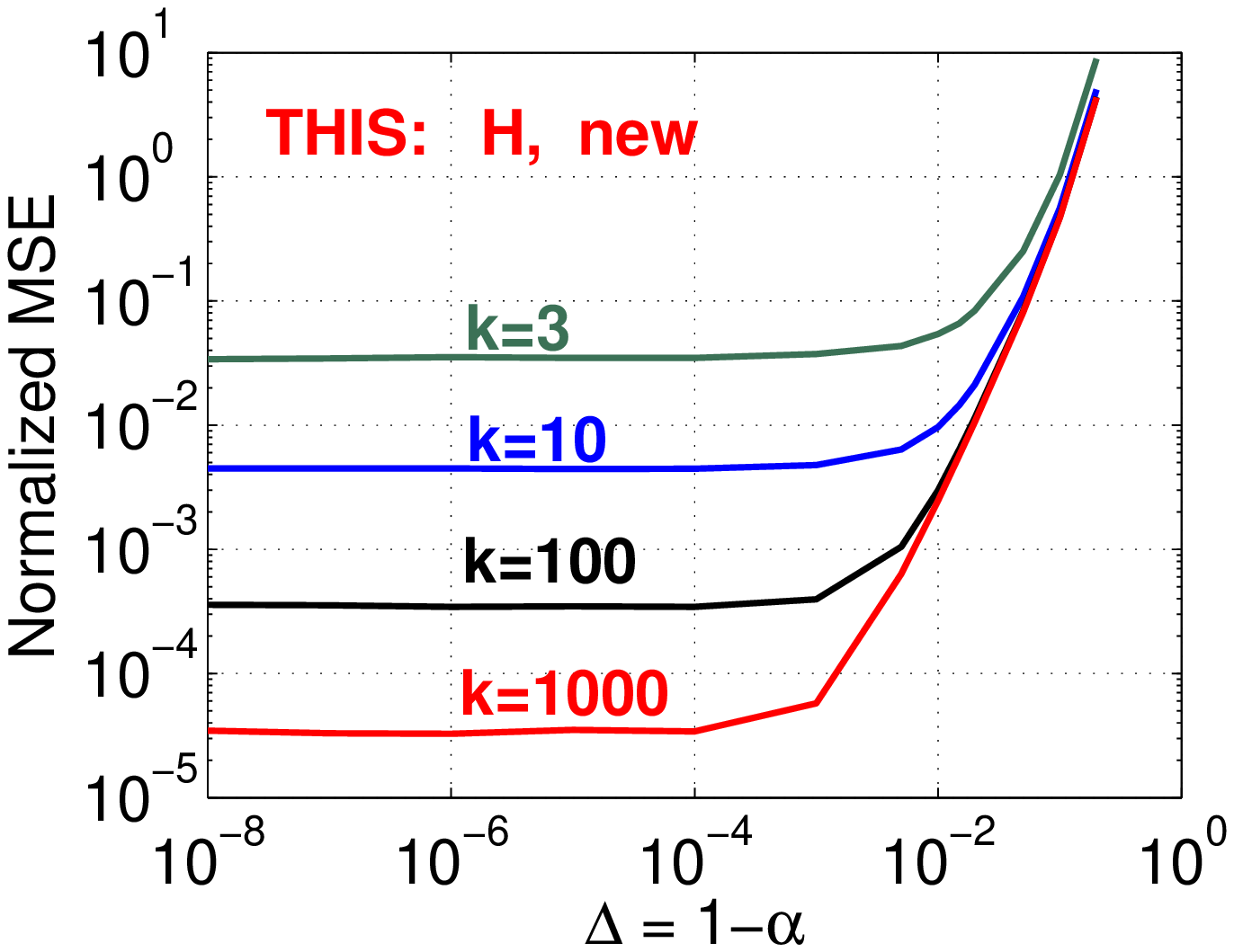}\hspace{-0.16in}
\includegraphics[width=2.25in]{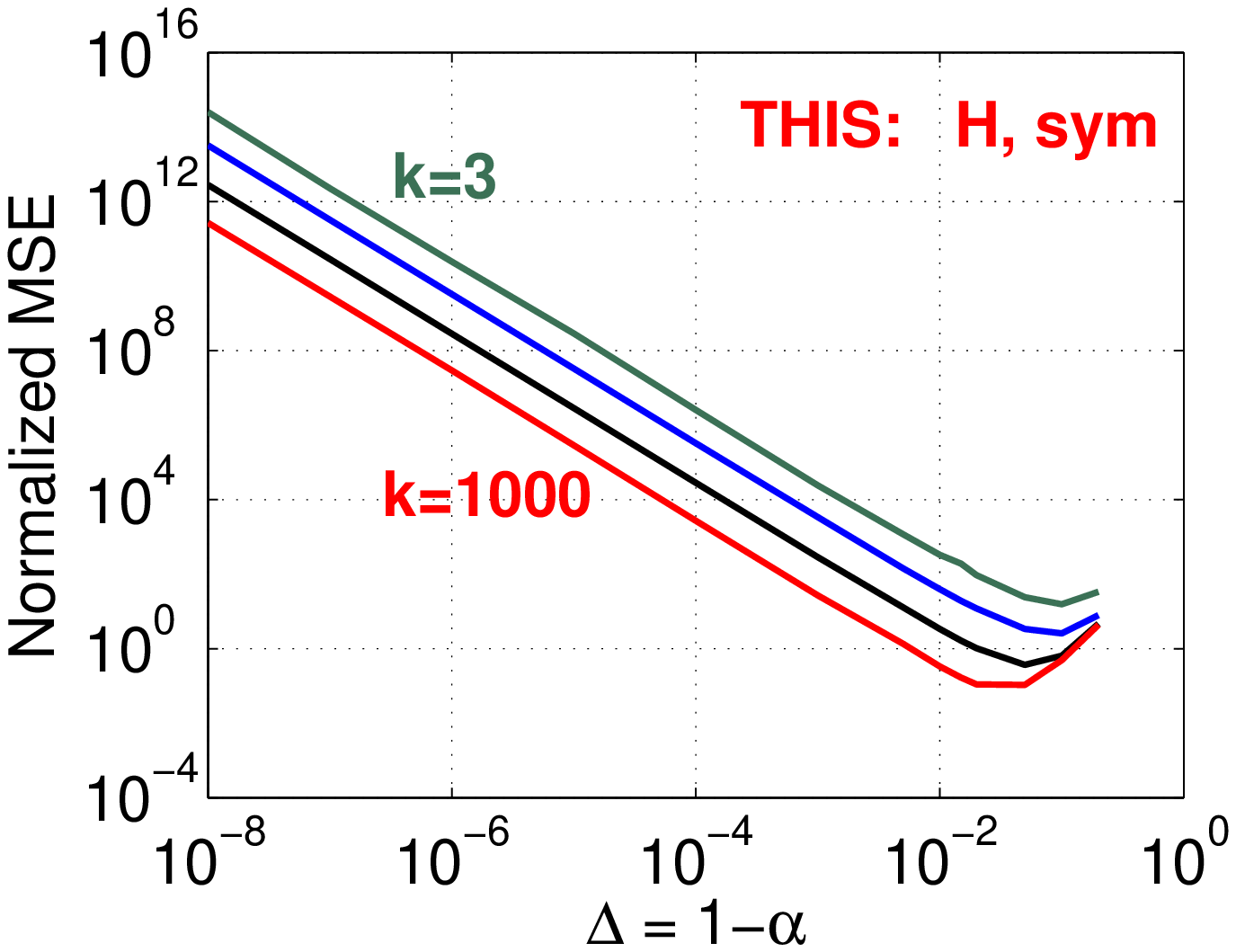}}

\mbox{
\includegraphics[width=2.25in]{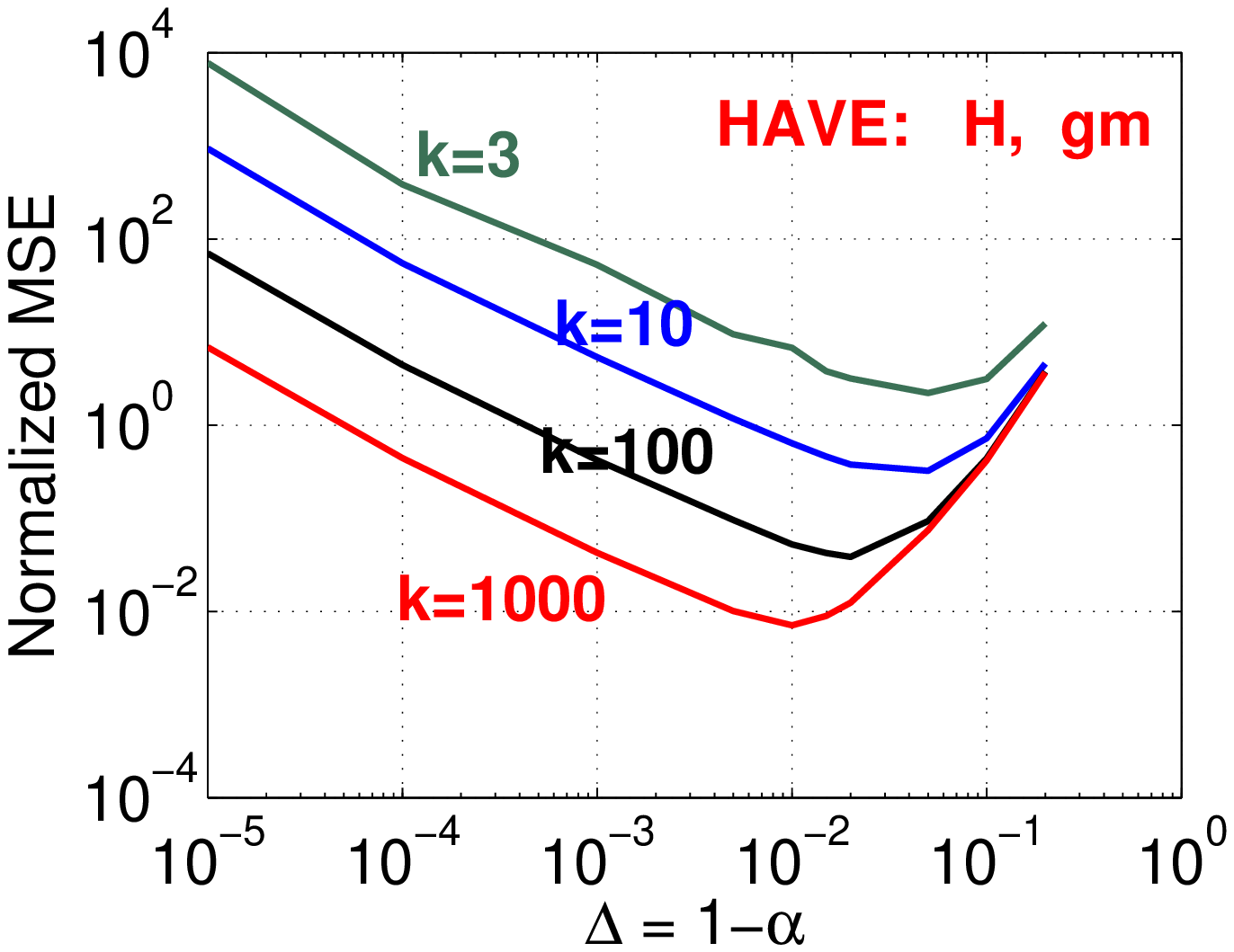}\hspace{-0.16in}
\includegraphics[width=2.25in]{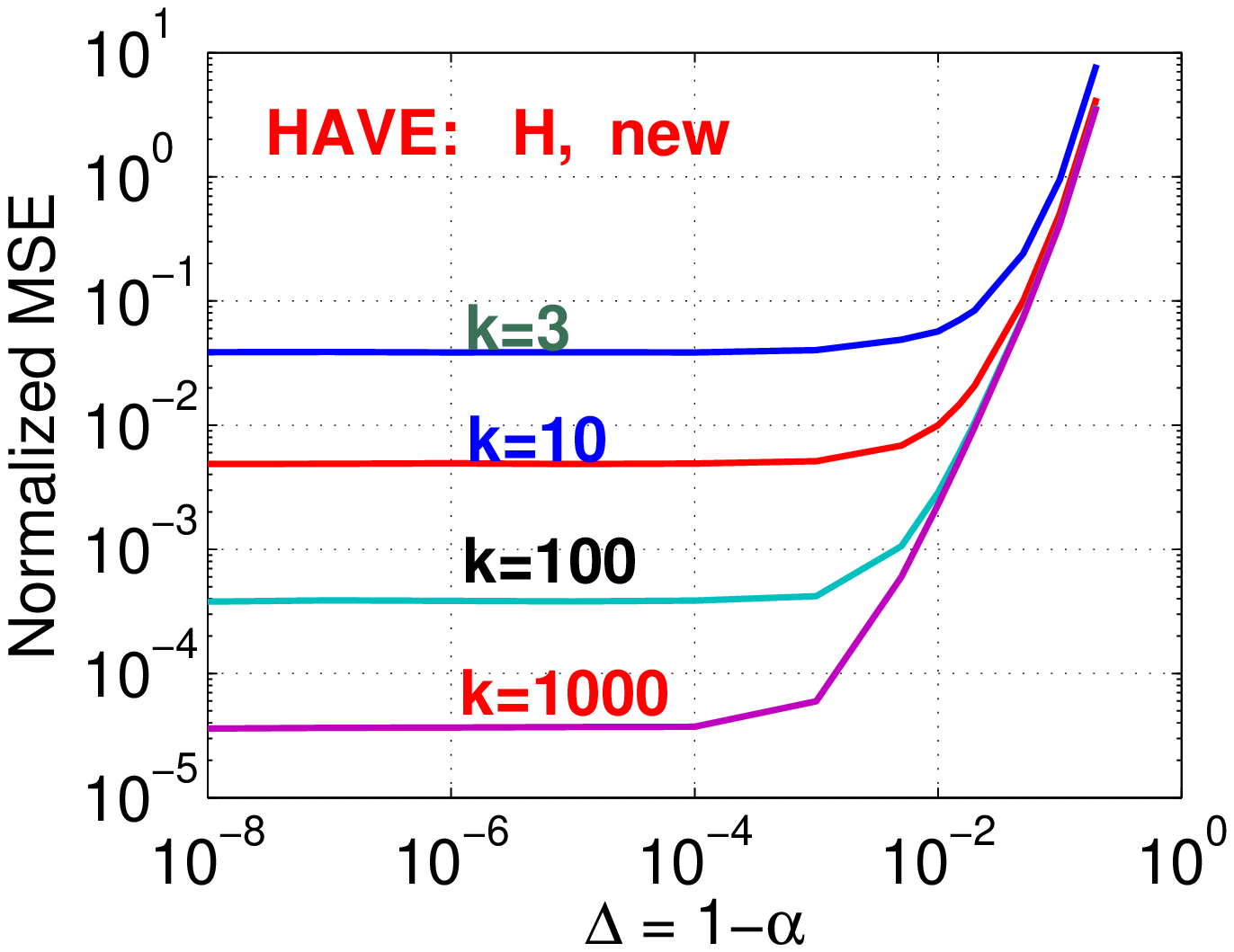}\hspace{-0.16in}
\includegraphics[width=2.25in]{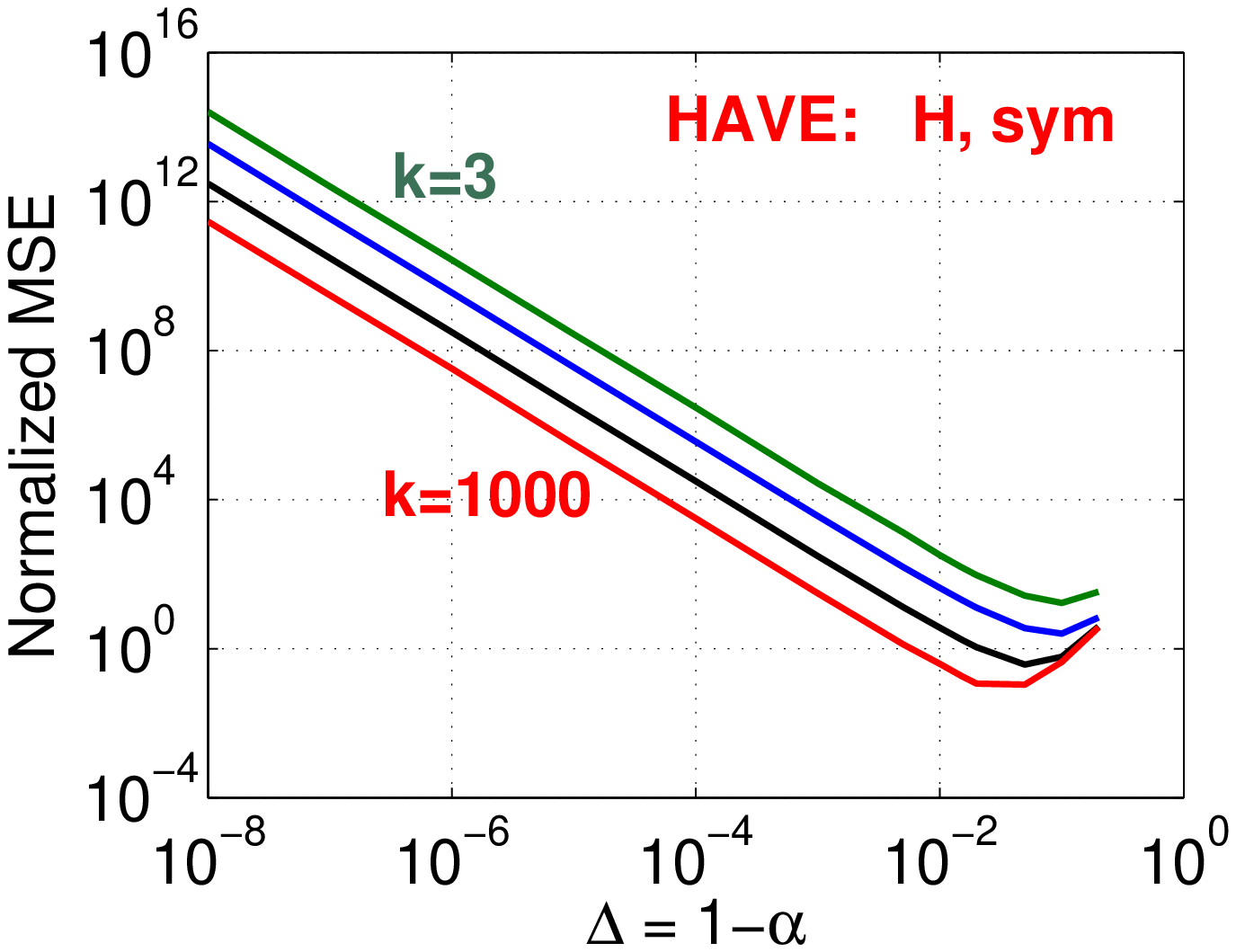}}

\end{center}
\vspace{-0.2in}
\caption{Normalized MSEs for estimating Shannon entropies using $\hat{F}_{(\alpha),gm}$ (left panels) and $\hat{F}_{(\alpha)}$ (middle panels) for CC, and the {\em geometric mean} estimator for {\em symmetric stable random projections} (right panels).
}\label{fig_H1}
\end{figure}

\begin{figure}[h]
\begin{center}
\mbox{
\includegraphics[width=2.25in]{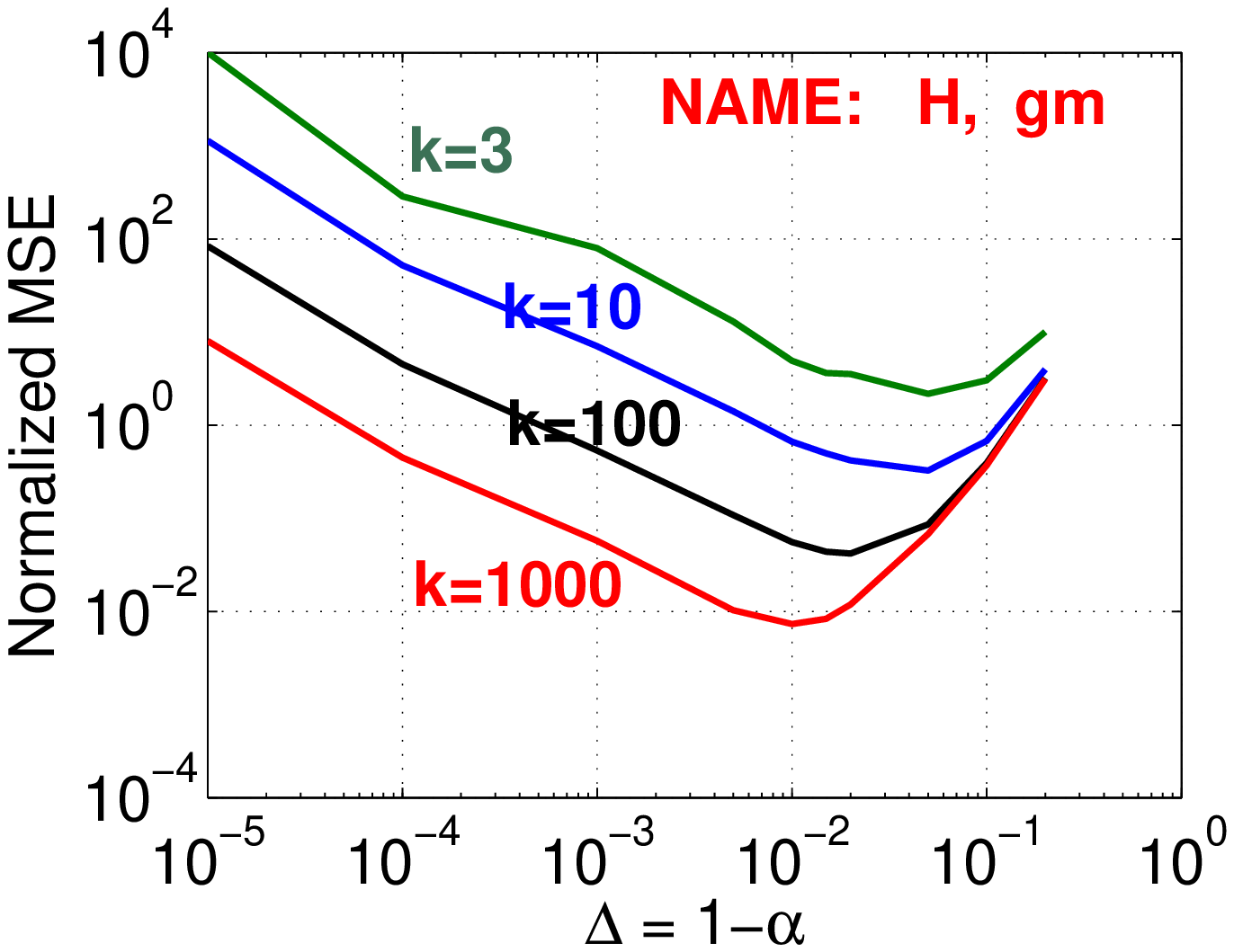}\hspace{-0.16in}
\includegraphics[width=2.25in]{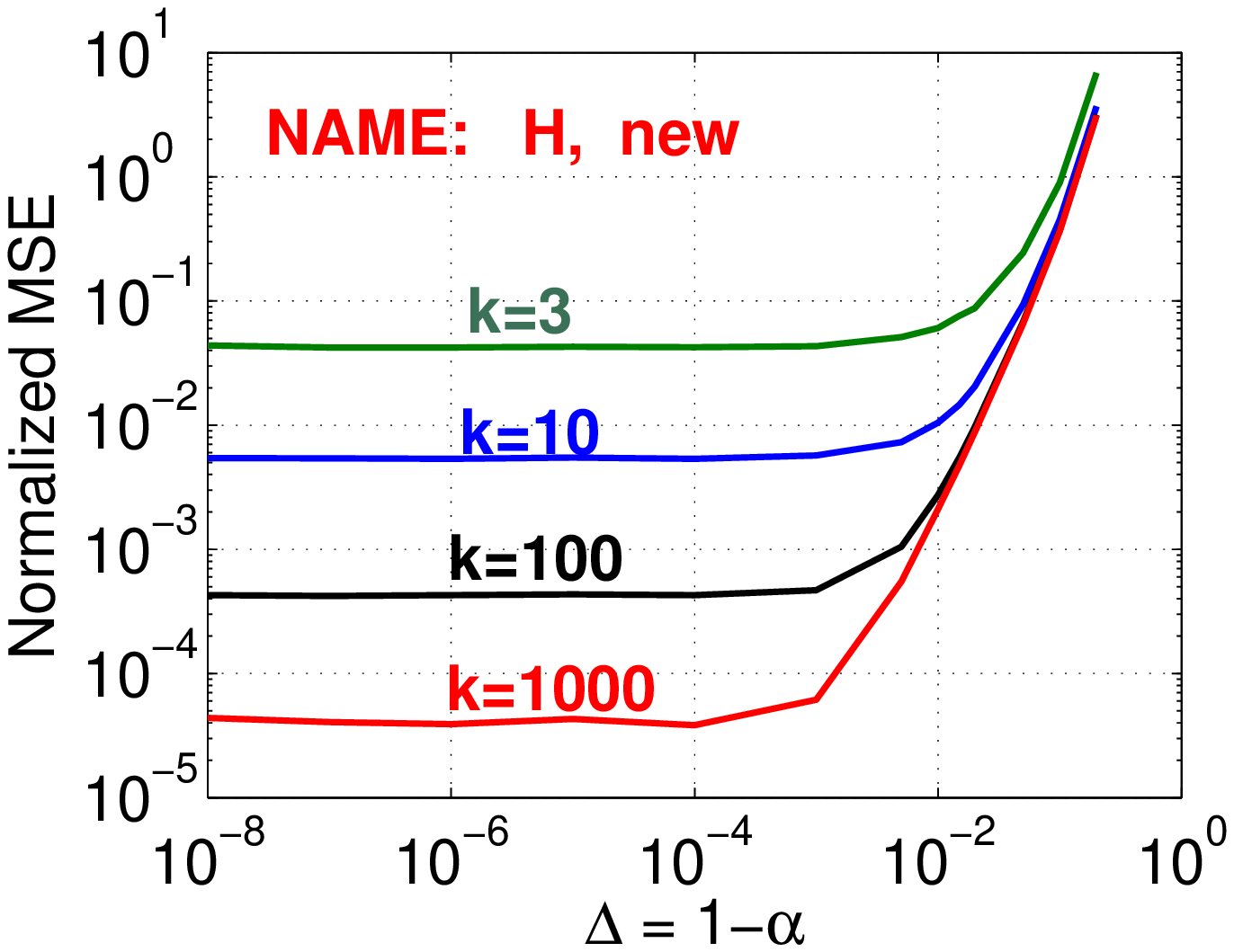}\hspace{-0.16in}
\includegraphics[width=2.25in]{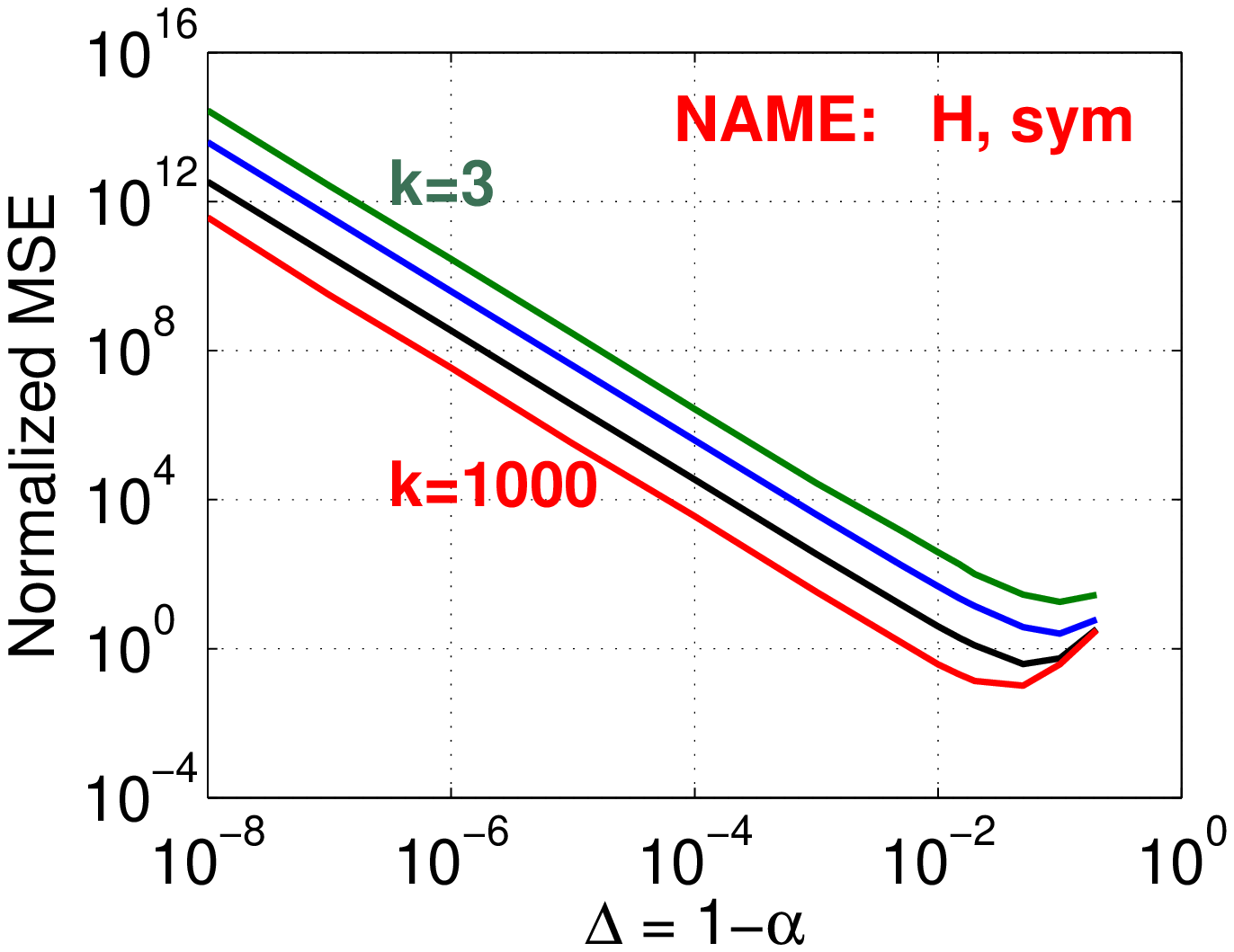}}

\mbox{
\includegraphics[width=2.25in]{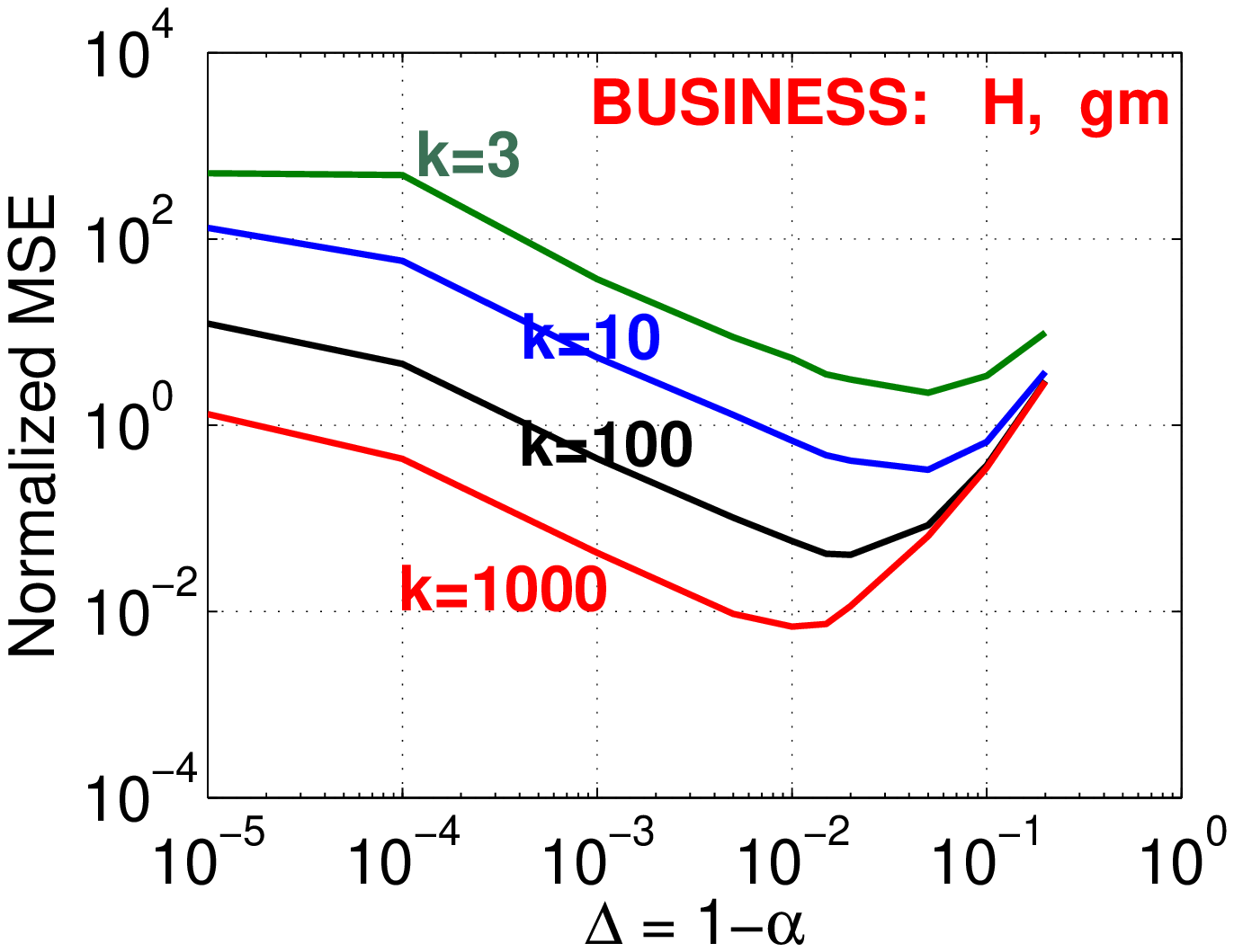}\hspace{-0.16in}
\includegraphics[width=2.25in]{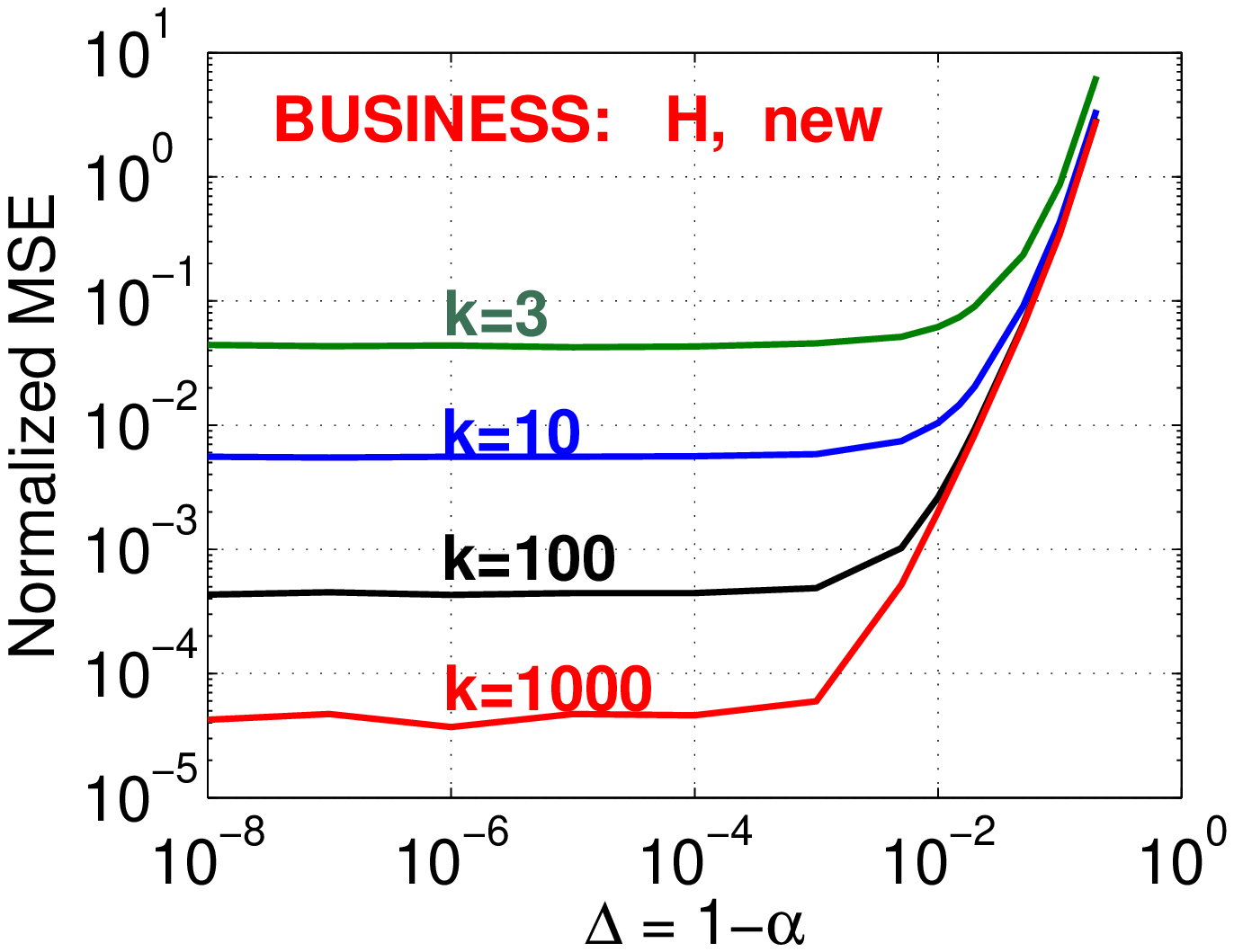}\hspace{-0.16in}
\includegraphics[width=2.25in]{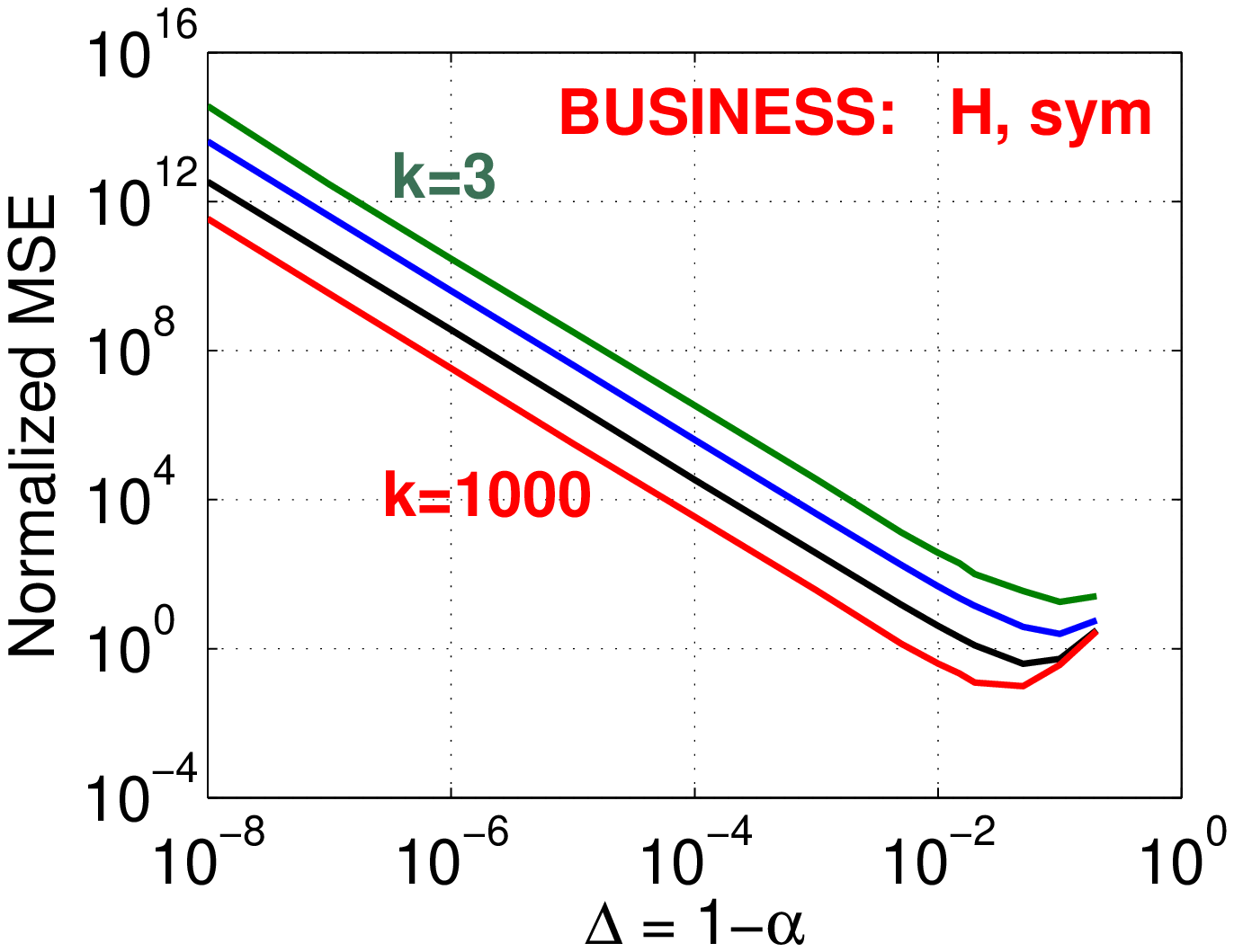}}
\mbox{
\includegraphics[width=2.25in]{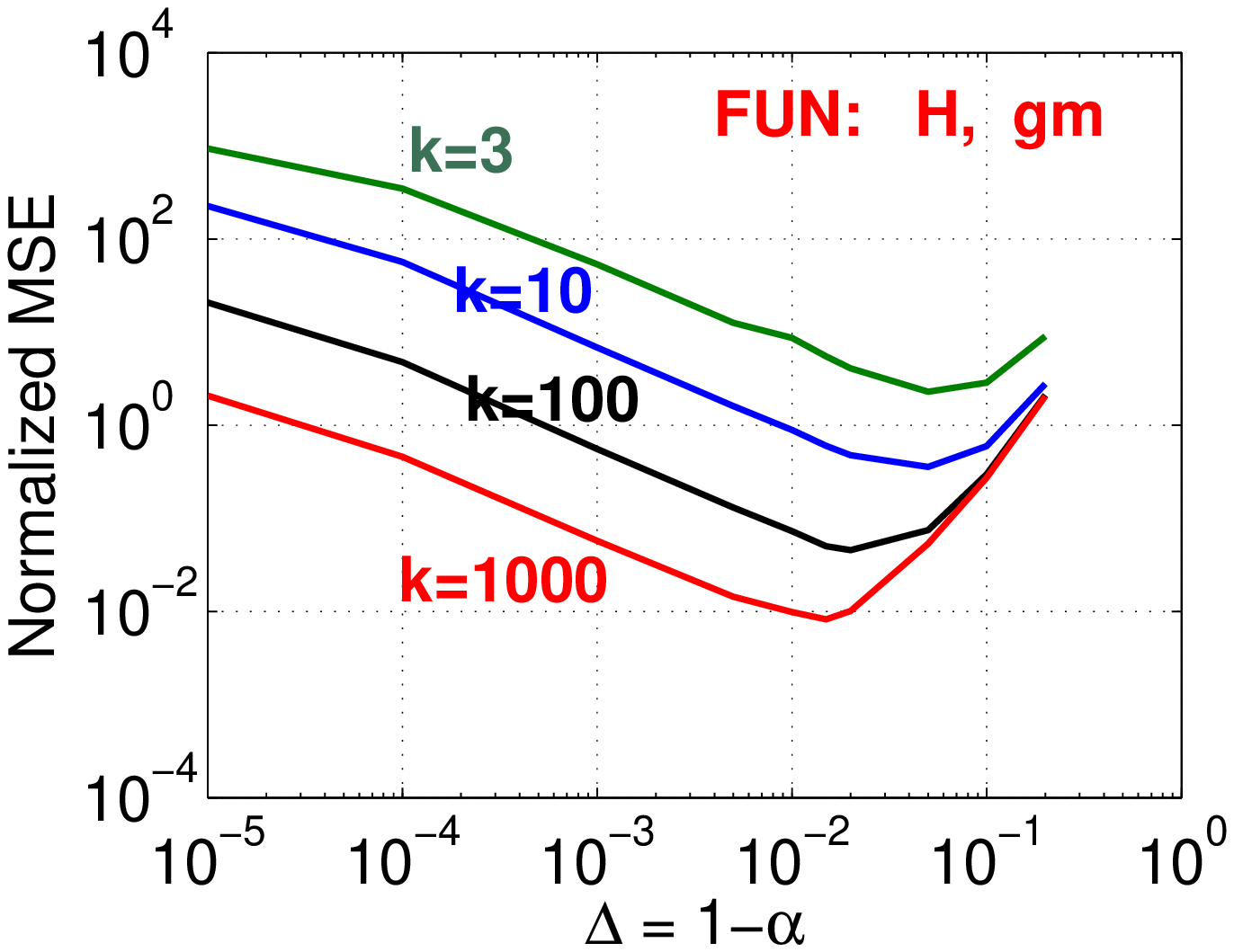}\hspace{-0.16in}
\includegraphics[width=2.25in]{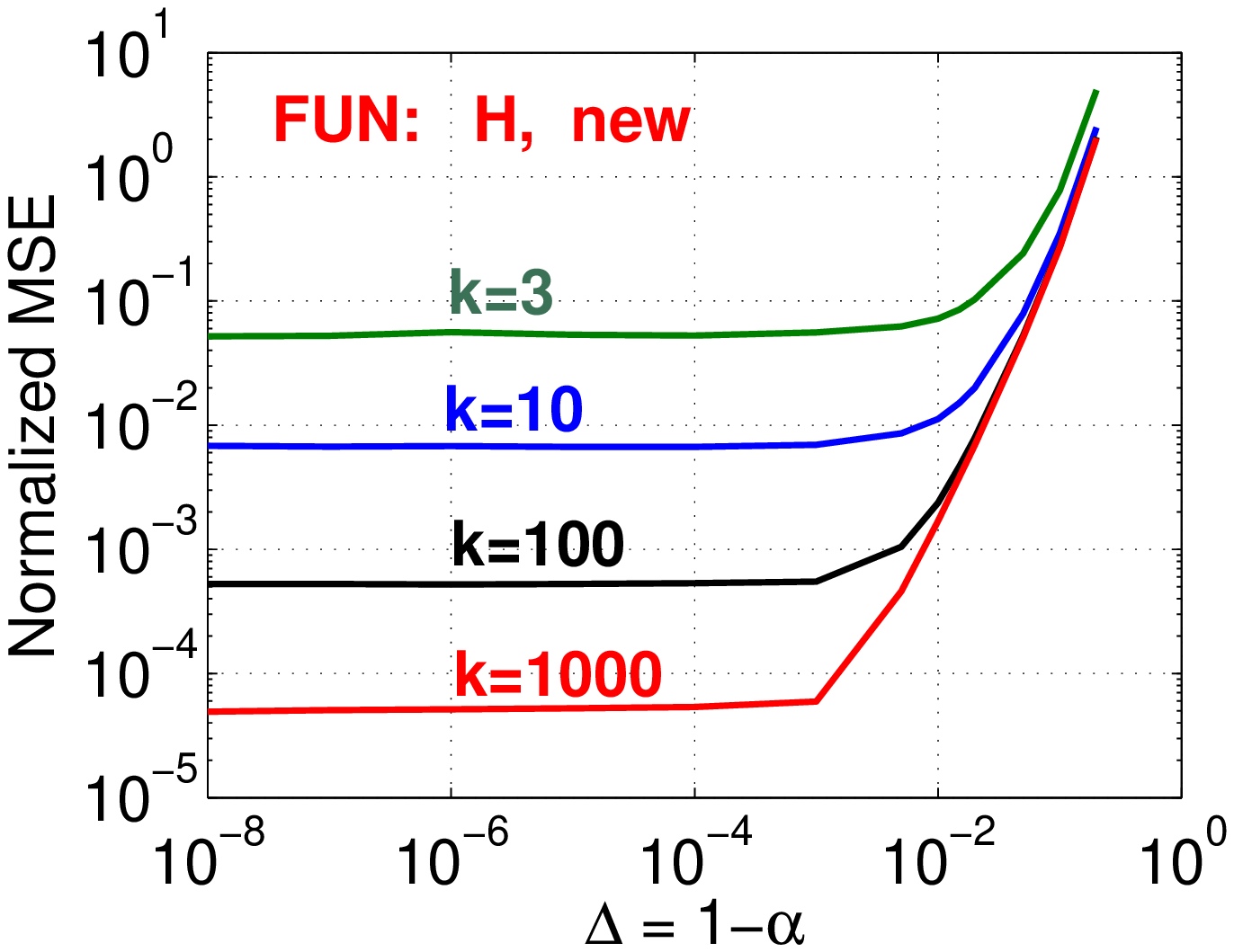}\hspace{-0.16in}
\includegraphics[width=2.25in]{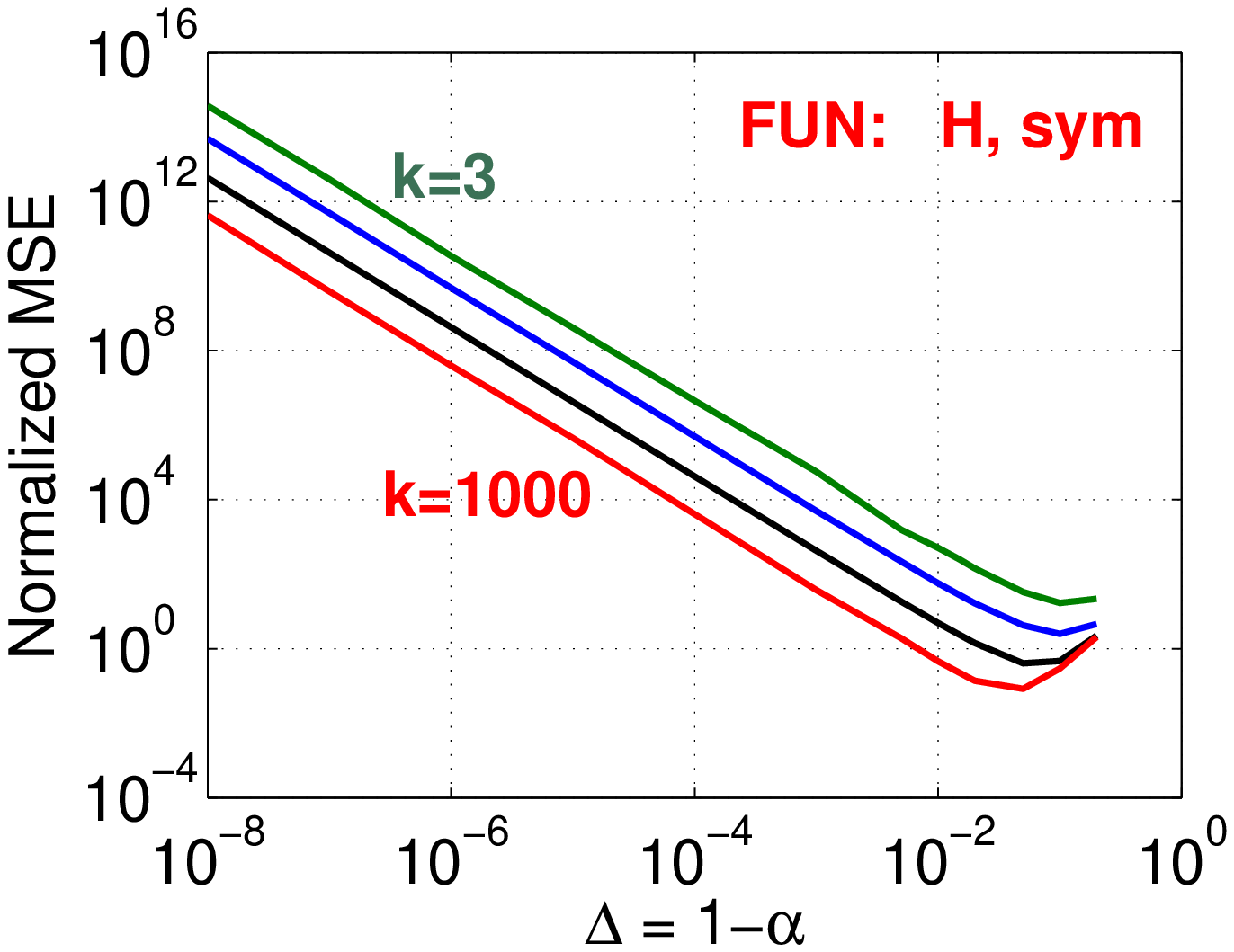}}

\mbox{
\includegraphics[width=2.25in]{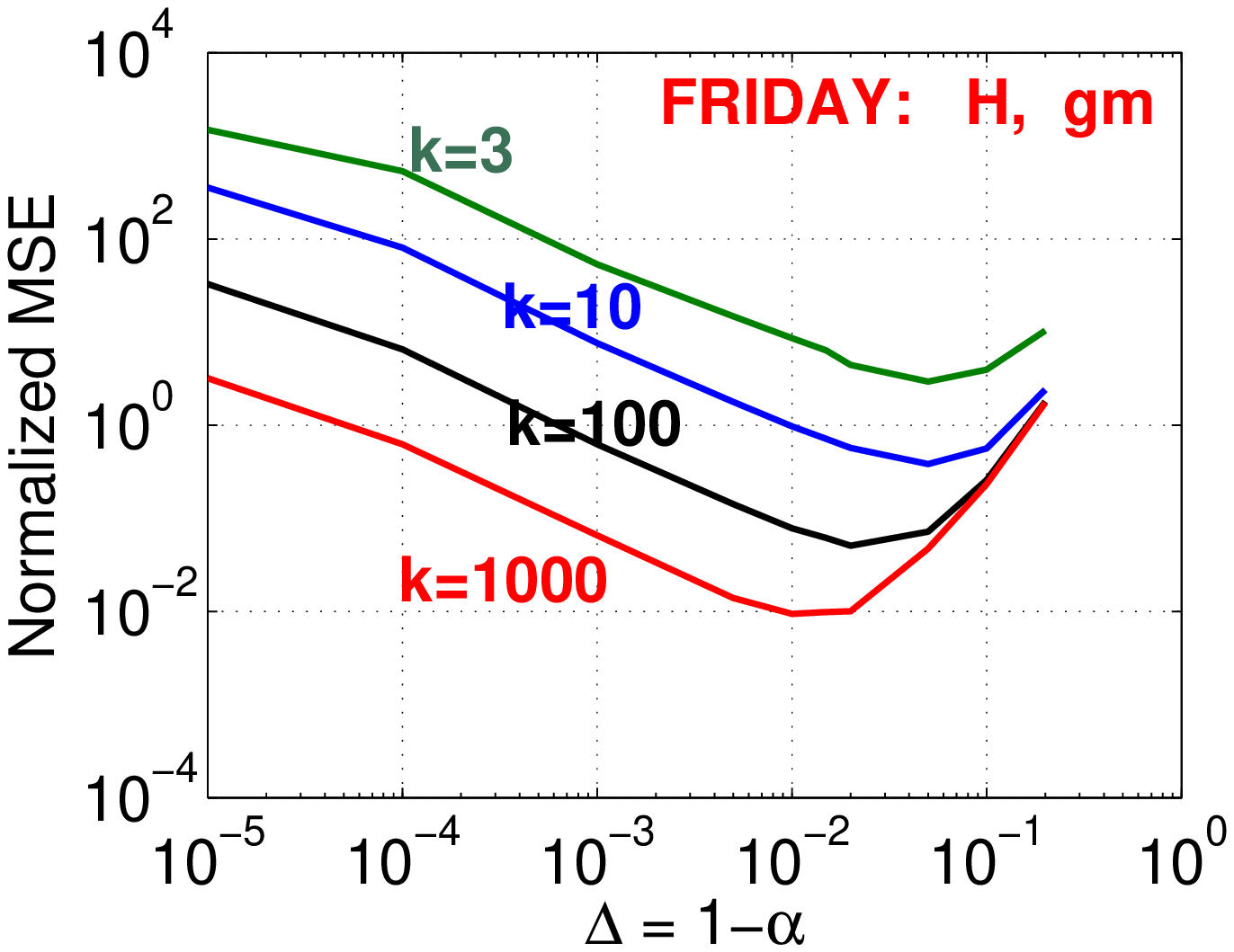}\hspace{-0.16in}
\includegraphics[width=2.25in]{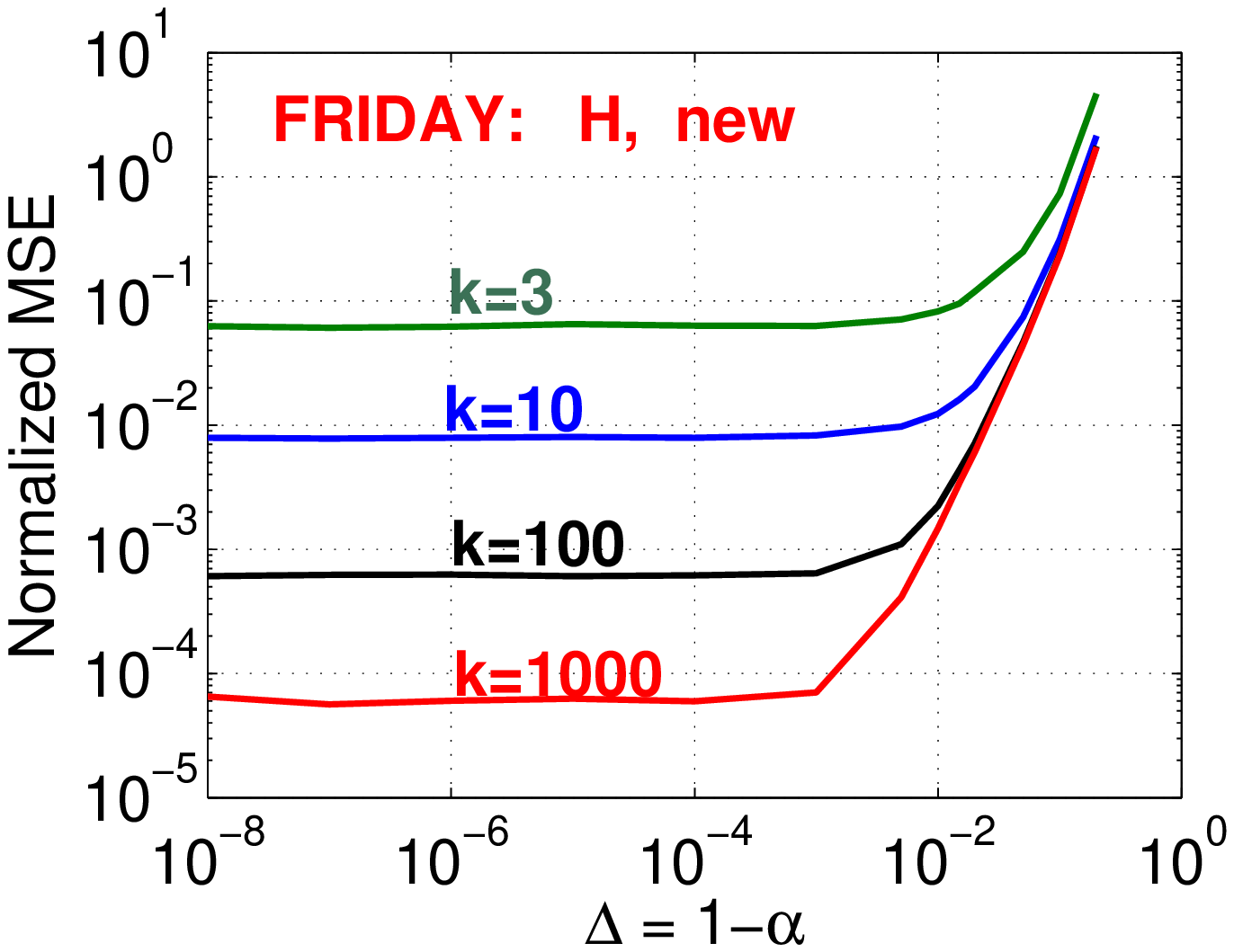}\hspace{-0.16in}
\includegraphics[width=2.25in]{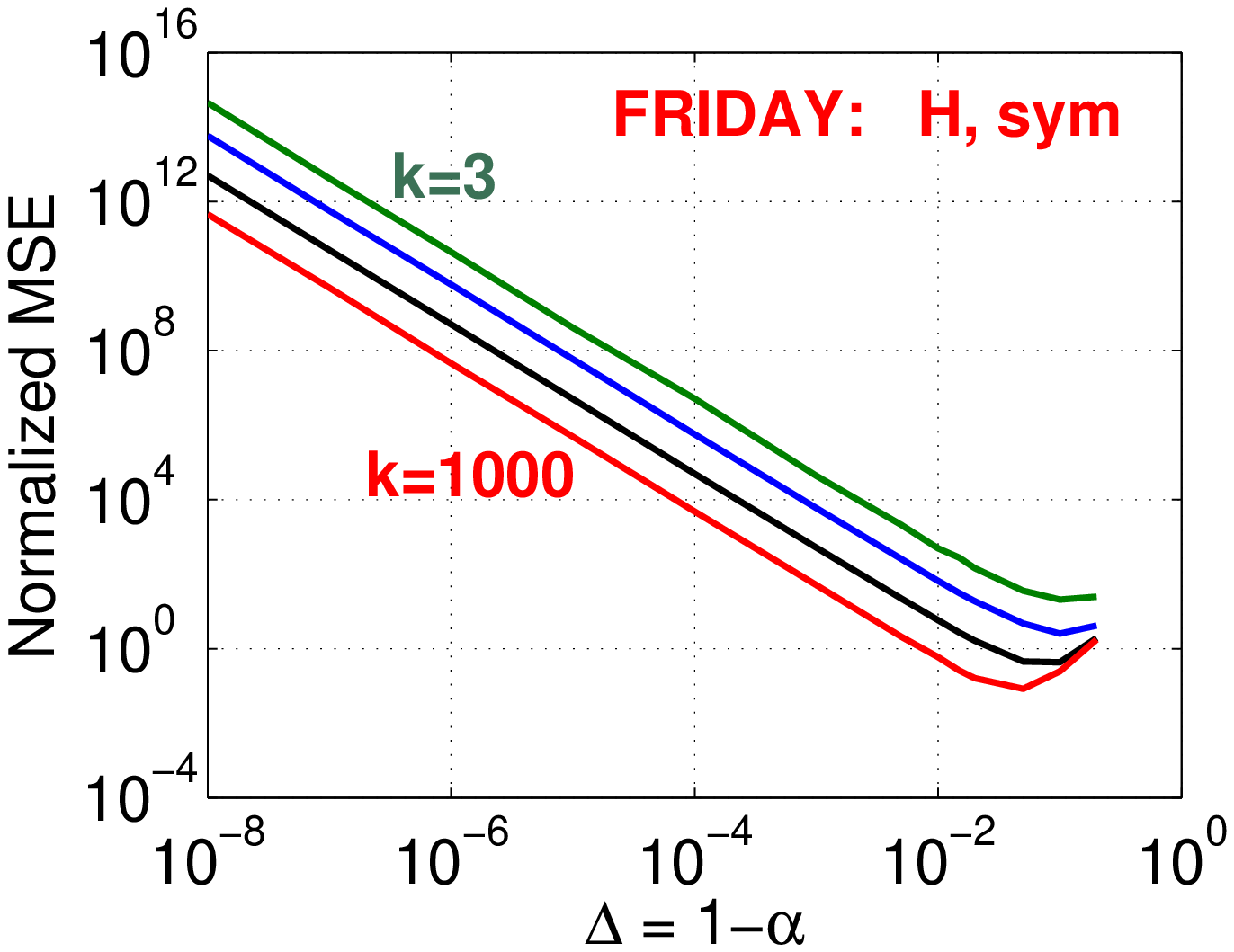}}

\end{center}
\vspace{-0.2in}
\caption{Normalized MSEs for estimating Shannon entropies using $\hat{F}_{(\alpha),gm}$ (left panels) and $\hat{F}_{(\alpha)}$ (middle panels) for CC, and the {\em geometric mean} estimator for {\em symmetric stable random projections} (right panels).}\label{fig_H2}
\end{figure}

After we have estimated the  frequency moments, we use them to estimate the Shannon entropies using Tsallis entropies. For the data vector ``TWIST'', we present results at sample sizes $k=3, 10, 100, 1000$, and 10000. For all other vectors, we do not experiment with $k=10000$.  Figure \ref{fig_H1} and Figure \ref{fig_H2} present the normalized MSEs.\\

Using CC and the proposed estimator $\hat{F}_{(\alpha)}$ (middle panels), only $k=10$ samples already produces fairly accurate estimates. In fact, for some vectors (such as ``A''), even $k=3$ may provide reasonable estimates. We believe the performance of the new estimator is remarkable.  Another nice property is that the estimation errors (MSEs) become stable after (e.g.,) $\Delta<10^{-3}$ (or $10^{-4}$).\\

In comparisons, the performance of the {\em geometric mean} estimator (left panels) for CC is not satisfactory. This is because  its variance only decreases only at the rate of $O(\Delta)$, not $O(\Delta^2)$.

Also clearly, using {\em symmetric stable random projections} (right panels) would not provide good estimates of the Shannon entropy (unless the sample size is extremely large ($\gg 10000$) and one could carefully choose a good $\Delta$ to exploit the bias-variance trade-off).

\clearpage

{%\small
%\bibliographystyle{plain}
%\bibliography{../bib/IEEEabrv,../bib/mybibfile}
}

\end{document}